\documentstyle[11pt,epsf,psfig]{article}
\def\zth{z_{\theta^*}}
\def\gamhtot{\Gamma_{\h}^{\rm tot}}
\def\sig{\sigma}
\def\hhat{\what h}
\def\mhhat{m_{\hhat}}

\def\Eq#1{Eq.~(\ref{#1})}

\def\Fig#1{Fig.~\ref{#1}}
\def\gamres{\Gamma_{\rm res}}
\def\call{{\cal L}}
\def\cala{{\cal A}}
\def\anti{\overline}
\def\gam{\gamma}
\def\br{B}
\def\ifmath#1{\relax\ifmmode #1\else $#1$\fi}

\def\lsim{\mathrel{\raise.3ex\hbox{$<$\kern-.75em\lower1ex\hbox{$\sim$}}}}
\def\gsim{\mathrel{\raise.3ex\hbox{$>$\kern-.75em\lower1ex\hbox{$\sim$}}}}
\def\half{\ifmath{{\textstyle{1 \over 2}}}}

\def\third{\ifmath{{\textstyle{1 \over 3}}}}

    \def\fillboxx#1#2{\hbox to #1{\vbox to #2{\vfil}\hfil}    }

\def\ie{{\it i.e.}}

\def\vev#1{\langle #1 \rangle}

\def\Eq#1{Eq.~(\ref{#1})}

\def\etc{{\it etc.}}

\def\eg{{\it e.g.}}

\def\msusy{m_{\rm SUSY}}

\def\tanb{\tan\beta}

\def\mt{m_t}

\def\mz{m_Z}
\def\mw{m_W}

\def\h{h}
\def\mh{m_{\h}}

\def\hpm{H^{\pm}}

\def\call{{\cal L}}

\def\what{\widehat}

\def\lam{\lambda}
\def\br{BR}
\def\tauptaum{\tau^+\tau^-}

\def\gam{\gamma}

\def\anti{\overline}
\def\epem{e^+e^-}
\def\emem{e^-e^-}

\def\rts{\sqrt s}
\def\ie{{\it i.e.}}
\def\eg{{\it e.g.}}
\def\eps{\epsilon}
\def\anti{\overline}

\def\mw{m_W}
\def\mz{m_Z}
\def\h{h}
\def\mh{m_{\h}}

\def\hsm{h_{SM}}
\def\mhsm{m_{\hsm}}

\def\hl{h^0}
\def\mhl{m_{\hl}}

\def\ha{A^0}
\def\mha{m_{\ha}}

\def\hh{H^0}
\def\mhh{m_{\hh}}
\def\hpm{H^{\pm}}
\def\mhpm{m_{\hpm}}

\def\fbi{~{\rm fb}^{-1}}

\def\abi{~{\rm ab}^{-1}}

\def\gev{~{\rm GeV}}
\def\tev{~{\rm TeV}}

\def\mt{m_t}

\def\MPL #1 #2 #3 {{ Mod.~Phys.~Lett.}~{\bf#1} (#3) #2}
\def\NPB #1 #2 #3 {{ Nucl.~Phys.}~{\bf #1} (#3) #2}
\def\PLB #1 #2 #3 {{ Phys.~Lett.}~{\bf #1} (#3) #2}
\def\PR #1 #2 #3 {{ Phys.~Rep.}~{\bf#1} (#3) #2}
\def\PRD #1 #2 #3 {{ Phys.~Rev.}~{\bf #1} (#3) #2}
\def\PRL #1 #2 #3 {{ Phys.~Rev.~Lett.}~{\bf#1} (#3) #2}
\def\RMP #1 #2 #3 {{ Rev.~Mod.~Phys.}~{\bf#1} (#3) #2}
\def\ZPC #1 #2 #3 {{ Z.~Phys.}~{\bf #1} (#3) #2}
\def\IJMP #1 #2 #3 {{ Int.~J.~Mod.~Phys.}~{\bf#1} (#3) #2}
\def\NIM #1 #2 #3 {{ Nucl.~Inst.~and~Meth.}~{\bf#1} {#3} #2}
\def\JHEP #1 #2 #3 {{ JHEP}~{\bf#1} (#3) #2}

\newcommand{\nc}{\newcommand}
\nc{\beq}{\begin{equation}}   \nc{\eeq}{\end{equation}}
\nc{\bea}{\begin{eqnarray}}   \nc{\eea}{\end{eqnarray}}
\nc{\baa}{\begin{array}}      \nc{\eaa}{\end{array}}
\nc{\bit}{\begin{itemize}}    \nc{\eit}{\end{itemize}}
\nc{\bed}{\begin{description}}    \nc{\eed}{\end{description}}
\nc{\ben}{\begin{enumerate}}  \nc{\een}{\end{enumerate}}
\nc{\bce}{\begin{center}}     \nc{\ece}{\end{center}}
\def\beqa{\begin{eqnarray}}
\def\eeqa{\end{eqnarray}}

\def\tightenlines{\def\baselinestretch{1.3}\small\normalsize}
\tightenlines
\begin{document}
\font\fortssbx=cmssbx10 scaled \magstep2
\hbox to \hsize{
%
%
\hfill
$\vcenter{\normalsize
\hbox{\bf LC Note: LCC-0074}
\hbox{\bf UCD-2001-8} 
\hbox{\bf UCRL-ID-143967}
\hbox{\bf hep-ph/0110320}
\hbox{October, 2001}
}$
}
\begin{center}
{ \Large \bf  Detecting and Studying Higgs Bosons at a Photon-Photon Collider}
\rm
\vskip1pc
{\large\bf David M. Asner$^1$, Jeffrey B.  Gronberg$^1$, and John F. Gunion$^2$}\\
\medskip
{\it 1. Lawrence Livermore National Laboratory, Livermore, CA 94550}\\
{\it 2. Davis Institute for High Energy Physics, University of California, Davis, CA 95616}\\
\end{center}

\begin{abstract}
We examine the potential for detecting
and studying Higgs bosons at a photon-photon collider
facility associated with a future linear
collider. Our study incorporates realistic $\gam\gam$ luminosity spectra
based on the most probable available laser technology. Results
include detector simulations. We  study the cases of: a) a SM-like
Higgs boson; b) the heavy MSSM Higgs bosons; c) a Higgs boson
with no $WW/ZZ$ couplings from a general two Higgs doublet model. 
\end{abstract}

\section{Introduction}
Higgs production in $\gamma\gamma$ collisions,
first studied in  \cite{Gunion:1993ce,Borden:1993cw}, 
 offers a unique 
capability to measure the two-photon width of the Higgs and
to determine its charge conjugation and parity (CP) 
composition through control of the photon
polarization. Both measurements have unique value in understanding
the nature of a Higgs boson eigenstate.  Photon-photon collisions
also offer one of the best means for producing a heavy Higgs
boson singly, implying significantly greater mass reach than
$\epem$ production of a pair of Higgs bosons.  In this paper, we
present a realistic assessment of the prospects for these studies
based on the current Next Linear Collider (NLC) machine and detector 
designs~\cite{Abe:2001nr,Abe:2001gc,Abe:2001wn}, but we will
also comment on changes in our results based on the 
TeV-Energy Superconducting Linear Accelerator (TESLA) 
design~\cite{Aguilar-Saavedra:2001rg}. When referring to either
of these machines in a generic context, we will use the phrase
``Linear Collider'' (LC). Summaries of and 
references to other recent work on $\gam\gam$
Higgs production at the LC appear 
in \cite{Abe:2001nr,Abe:2001gc,Abe:2001wn,Aguilar-Saavedra:2001rg}.
In our work, we attempt to assess the potential
of $\gam\gam$ Higgs production using a realistic computation of
the luminosity and polarizations of the colliding back-scattered
photons and of the resulting backgrounds, including detector
simulation and appropriate cuts. We will particularly focus
on: a) studying a light Standard-Model-like Higgs boson,
including a determination of its CP; and b)
determining the best strategy for detecting the heavy Higgs
bosons of the MSSM for model parameter choices such that they
will not be seen either at the LHC or in $\epem$ collision
operation of the LC.

There are many important reasons for measuring 
the $\gam\gam$ coupling of a Higgs boson, generically denoted
$\h$.  In the Standard Model (SM), the coupling
of the Higgs boson, $\hsm$, to two photons 
receives contributions from loops containing any charged particle
whose mass arises in whole or part from the vacuum expectation
value (vev) of the neutral Higgs field.
In the limit of infinite mass for the charged particle in the loop, the
contribution asymptotes to a  value  that depends 
on the particle's spin (\ie\ the contribution does not decouple).
Thus, a measurement of $\Gamma(\hsm\to\gam\gam)$ 
provides the possibility of revealing the presence of arbitrarily
heavy charged particles, since in the SM context all particles
acquire mass via the Higgs 
mechanism.\footnote{Loop contributions from charged particles that acquire
a large mass from some other mechanism, beyond
the SM context, will decouple as $({\rm mass})^{-2}$
and, if there is a SM-like Higgs boson $\h$,
$\Gamma(\h\to\gam\gam)$ will not be sensitive to their presence.}
Of course, since such masses are basically proportional to
some coupling times $v=174\gev$ (the Higgs field vacuum expectation value), 
if the coupling is perturbative
the masses of these heavy particles are unlikely to be much 
larger than $0.5-1\tev$. Since
$\br(\hsm\to X)$ is entirely determined by the spectrum
of light particles, and is thus not affected by heavy states,
$N(\gam\gam\to\hsm\to X)\propto \Gamma(\hsm\to\gam\gam)\br(\hsm\to X)$
will then provide an extraordinary probe for such heavy states.

Even if there are no new particles that acquire mass via the Higgs
mechanism, a precision measurement of $N(\gam\gam\to\h\to X)$
for specific final states $X$ ($X=b\anti b,WW^*,\ldots$)
can allow one to distinguish between a $\h$ that is part
of a larger Higgs sector and the SM $\hsm$. The ability to detect
deviations from SM expectations will be enhanced by combining this 
with other types of precision measurements for the SM-like Higgs boson. 
Observation of small deviations would be typical 
for an extended Higgs sector as one approaches the decoupling limit
in which all other Higgs bosons are fairly heavy, leaving
behind one SM-like light Higgs boson. In such models,
the observed small deviations could then be interpreted as implying the
presence of heavier Higgs bosons. 
Typically,\footnote{But there are exceptional regions
of parameter space for which this is not true \cite{gminprogress}.}
deviations exceed 5\% if the other heavier Higgs bosons
have masses below about 400 to 500 GeV. A precise measurement
of the deviations, coupled with enough other information
about the model, might then allow one to constrain the masses of the heavier
Higgs bosons, thereby allowing one to understand how to go about
detecting them directly. For example, in the case of the two-doublet 
minimal supersymmetric Standard Model (MSSM)
Higgs sector there are five physical Higgs bosons (two CP-even, $\hl$
and $\mhh$ with $\mhl<\mhh$; one CP-odd, $\ha$; and a charged
Higgs pair, $\hpm$). In this model, significant deviations
of the $\hl$ properties from those of the $\hsm$ would
indicate that $\mha$ might well be sufficiently small that
the approximately degenerate $\hh$ and $\ha$ could be discovered in
$\gam\gam\to \hh,\ha$ production at a LC collider with energy
of order $\rts=500-600\gev$.

Of course, the ability to detect $\gam\gam\to \hh,\ha$ will be of
greatest importance if the $\hh$ and $\ha$ cannot be detected either
at the Large Hadron Collider (LHC) 
or in $\epem$ collisions at the LC.  In fact, there is
a very significant section of parameter space in the MSSM for which
this is the case, often referred to as the `wedge' region.  The wedge
basically occupies the following region of $\mha$--$\tanb$ parameter space.
\bit
\item
 $\mha\sim\mhh\gsim \rts/2$, for which $\epem\to\hh\ha$
pair production is impossible --- we will be focusing on
an LC with $\rts=630\gev$, implying that 
the wedge begins at $\mha< 315\gev$.
\item
$\tanb>3$ --- below this, the LHC will be able to detect the $\hh,\ha$
in a variety of modes such as $\hh\to \hl\hl$ and $\ha\to Z\hl$
for $\mha\lsim 2\mt$ and $\hh,\ha\to t\anti t$ for $\mha\gsim 2\mt$.
In some versions of the MSSM (e.g. the maximal mixing scenario),
most of this region is already eliminated by constraints from 
the Large Electron Positron Collider (LEP) data.
\item
$\tanb< \tanb^{\rm min}(\mha)$, where $\tanb^{\rm min}(\mha)$ is the
minimum value of $\tanb$ for which the LHC can detect 
$b\anti b \hh+b\anti b\ha$ production in the $\ha,\hh\to \tauptaum$ decay
modes (currently deemed the most accessible) --- $\tanb^{\rm min}(\mha)$
rises from $\sim 12$ at $\mha=315\gev$ to $\sim 18$ at $\mha=500\gev$.
\item In this wedge, 
the LC alternatives of $\epem\to b\anti b\hh$ and 
$\epem\to b\anti b\ha$ production also have such extremely small rates 
as to be undetectable --- see, \eg\ \cite{Grzadkowski:2000wj}.
\eit
This wedge will be discussed in greater detail later in the paper.
A LC for which the maximum  $\epem$ center of mass energy is
$\rts=630\gev$ can potentially probe Higgs masses in $\gam\gam$ collisions as
high as $\sim 500\gev$, the point at which the $\gam\gam$ luminosity
spectrum runs out.  An important goal of this paper is to determine
the portion of the `wedge' $[\mha,\tanb]$ parameter region for
which $\hh,\ha$ will be detectable via $\gam\gam$ collisions.
We find the following. 
\bit
\item If $\mhh$ and $\mha$ are known to within roughly $50\gev$
on the basis of precision $\hl$ data (and there is sufficient knowledge
of other MSSM parameters from the LHC to know how
to interpret these data), 
then we find that it is almost certain that we can detect the $\hh$ and $\ha$ 
by employing just one or two $\rts$ settings
and electron-laser-photon polarizations such as
to produce a $\gam\gam$ spectrum peaked in the region of interest. 
\item
However, it is very possible that
there will be no fully reliable constraints
on the $\hh,\ha$ masses (other
than $\mha \sim\mhh>\rts/2$ from LC running in the $\epem$
collision mode). In this case, for expected luminosities,
the simplest, and probably also the most efficient, procedure
will be to simply operate the machine at a single (high) energy,
roughly 2/3 to 3/4 
of the time using electron-laser-photon polarization configurations
that produce a broad spectrum $E_{\gam\gam}$
spectrum and 1/3 to 1/4 of the time using configurations
that yield a spectrum peaked at high $E_{\gam\gam}$.
We will find that after three to four years of operation this procedure will
yield a visible signal for $\hh,\ha$ production
for most of the wedge parameter space,
and, more generally, for many $[\mha,\tanb]$ parameter choices. 
\eit
Earlier work on detecting the heavy MSSM Higgs bosons in $\gam\gam$
collisions appears in (\cite{Muhlleitner:2001kw,Muhlleitner:2000jj}.
Our study employs the best available predictions for
the $\gam\gam$ luminosity spectrum and polarizations using
the realistic assumption of 80\% polarization for the colliding
electron beams.

The $\gam\gam$ collider would also play a very important role
in exploring a non-supersymmetric general two-Higgs-doublet model
(2HDM).  In this paper, we will explore the role
of a $\gam\gam$ collider in the context of a CP-conserving (CPC)
type-II~\footnote{In a type-II 2HDM, at tree-level the vacuum
expectation value of the neutral field of one doublet
gives rise to up-type quark masses while the vev of
the neutral field of the second doublet gives rise
to down-type quark masses and lepton masses.}
2HDM (of which the MSSM Higgs sector is a special case).
In particular, there are CPC type-II 2HDM's
with Higgs sector potentials for which the lightest Higgs
boson is not at all SM-like, despite the fact that the other
Higgs bosons are fairly heavy. Several such models were
considered in Ref.~\cite{Chankowski:2000an}. 
In the models considered, there is a light Higgs boson
with no $WW,ZZ$ coupling, generically denoted $\hhat$,
while all other Higgs bosons (including
a heavy neutral Higgs boson with SM-like couplings) are heavier than $\rts$.
Further, there is a wedge (somewhat analogous to, but larger than,
that of the MSSM) of moderate $\tanb$ values
in which the $\epem\to b\anti b \hhat$ and $\epem\to t\anti t\hhat$ 
production processes both yield fewer than 20 events for $L=1\abi$
and in which LHC detection will also be impossible. 
If $\mhhat$ is also so heavy ($\mhhat>150\gev,250\gev$ 
for $\rts=500\gev,800\gev$,
respectively) as to yield few or no events in 
$\epem\to Z\hhat\hhat$ or $\epem\to \nu\anti\nu\hhat\hhat$ production,
then only $\gam\gam\to \hhat\to b\anti b$ 
might allow detection of the $\hhat$. We again find that such detection would
be possible for a significant fraction of the $[\mhhat,\tanb]$
parameter space that is not accessible
at the LHC or in $\epem$ LC operation, the precise values
depending upon the luminosity expended for the search.

Once one or several Higgs bosons have been detected, 
precision studies can be
performed. Primary on the list would be
the determination of the CP nature of any observed Higgs boson.
This and other types of measurements
become especially important if one is in the decoupling limit of a 2HDM.
The decoupling limit is defined by the situation in which 
there is a light SM-like Higgs boson, while the other
Higgs bosons ($\hh,\ha,\hpm$) are heavy and quite degenerate.
In the MSSM context, such decoupling is automatic
in the limit of large $\mha$.
In this situation, a detailed scan to
separate the $\hh$ and $\ha$ would be very important and
entirely possible at the $\gam\gam$ collider. Further,
measurements of relative branching fractions for the $\hh$ and
$\ha$ to various possible final states would also be possible
and reveal much about the Higgs sector model. In the MSSM
context, the branching ratios for supersymmetric final states
would be measurable; these are
especially important for determining the basic supersymmetry breaking 
parameters \cite{Gunion:1995bh,Gunion:1997cc,Gunion:1996qd,Feng:1997xv,Muhlleitner:2001kw,Muhlleitner:2000jj}.

\section{Production Cross Sections and Luminosity Spectra}

The rate for $\gam\gam\to\h\to X$ production of any final state $X$ 
consisting of two jets is given by
\bea
N(\gam\gam\to \h \to X)&=&\sum_{\lam=\pm 1,\lam'=\pm 1}
\int dz dz' d\zth 
{d\call_{\gam}^{\lam}(\lam_e,P,z)\over dz}{d\call_{\gam}^{\lam'}(\lam_e^\prime,P',z') \over dz'}
A(z,z',\zth)\times\nonumber\\
&&\qquad \left\{ {1+\lam\lam'\over 2} {d\sig_{J_z=0}\over d\zth}(zz's,\zth) +
{1-\lam\lam'\over 2} {d\sig_{J_z=\pm 2}\over d\zth}(zz's,\zth)\right\}\,.
\eea
Here $\call_{\gam}^\lam(\lam_e,P,z)$ is the luminosity distribution for a back-scattered
photon of polarization $\lam$. It depends upon the initial electron beam
polarization $\lam_e$ ($|\lam_e|\leq 0.5$), the polarization 
of the laser beam ($P=\pm 1$), assumed temporarily to be entirely circular,
and the fraction $z$ of the $e$ beam momentum, $\half\rts$,
carried by the photon. The quantity $A(z,z',\zth)$ denotes
the acceptance of the event, including cuts, as a function
of the photon momentum fractions, $z$ and $z'$, and $\zth=\cos\theta^*$,
where  $\theta^*$ is the scattering angle
of the two jets in their center of mass frame.
The cross section for the two-jet final state is written
in terms of its $J_z=0$ component ($\lam\lam'=1$) and its $J_z=\pm 2$
component ($\lam\lam'=-1$).  Each component depends upon the subprocess
energy $zz's$ and $\zth$.
For the Higgs signal, $d\sig_{J_z=0}/d\zth$ is non-zero, but independent
of $\zth$, while $d\sig_{J_z=\pm2}/d\zth=0$:
\beq
{d\sig_{J_z=0}\over d\zth}(s',\zth)=
{8\pi \Gamma(\h\to \gam\gam)\Gamma(\h\to X)\over
(s'-\mh^2)^2+[\gamhtot]^2\mh^2}\,,
\label{hresform}
\eeq
where $s'=E_{\gam\gam}^2=zz's$.
This is the usual resonance form for the Higgs cross section.  
For the background, the tree level cross sections may be written
\bea
{d\sigma_{J_z=0}\over dt'}(s',t',u')&=& {12\pi\alpha^2 Q_q^4\over  
s^{\prime\,2}} \, {m_q^2( s'-2m_q^2)\over \what t^{2} \what u^{2}}
\label{jz0bkgnd}\\
{d\sigma_{J_z=\pm2}\over dt'}(s',t',u')&=& {12\pi\alpha^2 Q_q^4\over \what s^{\prime\,2}} \, 
{(\what t\what u-m_q^2  s')(\what t^{2}+\what u^{2}-2m_q^2  s')\over
\what t^{2} \what u^{2}}
\label{jz2bkgnd}
\eea
where $s',t',u'$ are the invariants of the subprocess,
with $s'=zz's$, $\what t=t'-m_q^2=-\half  s'(1-\beta_q\zth)$, $\what u=u'-m_q^2=-\half s'(1+\beta_q\zth)$,
$dt'=\half s'\beta_q d\zth$, and $Q_q$ and $m_q$ are the charge
and mass of the quark produced. As is well known, the $J_z=0$
portion of the background is suppressed by a factor of $m_q^2/s$
relative to the $J_z=\pm 2$ part of the background, implying
that choices yielding $\lam\lam'$ near 1 will suppress the background
while at the same time enhancing the signal.
In a common approximation, the dependence of the acceptance and cuts
on $z$ and $z'$ is ignored and one writes
\beq
\sum_{\lam,\lam'}\int dzdz' {d\call_{\gam}^{\lam}(\lam_e,P,z)\over dz}{d\call_{\gam}^{\lam'}(\lam_e^\prime,P',z') \over dz'}[1,\lam\lam']=\int dy {d\call_{\gam\gam}(\lam_e,\lam_e^\prime,P,P',y)\over dy}[1,\vev{\lam\lam'}(y)]\,,
\label{dldydef}
\eeq
where $y=E_{\gam\gam}/\rts=\sqrt{s'}/\rts=zz'$.
In this approximation, one obtains \cite{Gunion:1993ce}
\bea
N(\gam\gam\to{\h}\to X)&=&{4\pi^2\Gamma({\h}\to\gam\gam)
\br({\h}\to X)(1+\vev{\lam\lam'}(y))\over \rts \mh^2}
\left.{d\call_{\gam\gam}\over dy}\right|_{y=\mh/\rts}
{\int_{-1}^1 d\zth A(\zth)\over 2}\nonumber\\ 
&\equiv& I_\sigma(\gam\gam\to\h\to X)\left[(1+\vev{\lam\lam'})
{d\call_{\gam\gam}\over dE_{\gam\gam}}\right]_{E_{\gam\gam}=\mh}{\int_{-1}^1 d\zth A(\zth)\over 2}\,,
\label{ngamgam}
\eea
where we have assumed that the resolution, $\gamres$, in the final state
invariant mass $m_X$ is such that $\gamres\gg \gamhtot$
and that ${d\call \over dE_{\gam\gam}}$ does not change significantly
over an interval of size $\gamhtot$.
The first line reduces to the usual form if $A(\zth)=1$,
implying $\int_{-1}^1 A(\zth)=2$.
The maximum value of $y$ is given by $y_{\rm max}=x/(1+x)$,
where $x\simeq {4 E_{\rm beam}\omega_{\rm laser}\over m^2c^4}$.

Whether or not one-loop and higher-order corrections (generically
referred to here as NLO corrections) to the above tree-level cross
sections will be large and important depends on many factors.  In this
paper, we will employ tree-level predictions inserted into a Monte
Carlo framework that generates radiative corrections in the leading
logarithmic approximation. We argue in Appendix B that, for expected
luminosity and polarizations of the colliding photons and for suitable
cuts, our procedure yields a realistic assessment of the prospects for
Higgs study and detection via $\gam\gam$ collisions for the various
SM, MSSM and 2HDM scenarios we consider.  The basic point is that the
luminosity spectra and polarizations we employ predict that the
$J_z=2$ background is far larger than the $J_z=0$ background after
cuts. Consequently, even if NLO corrections enhance the $J_z=0$ background
by a factor of 5 to 10 (as is possible), the $J_z=0$ background will
still yield at most a 10\%-20\% correction to the $J_z=2$ background
at low Higgs masses ($\sim 120\gev$) and a 5\%-10\% correction
at high Higgs masses ($>300\gev$).  Such corrections are
well within the other uncertainties implicit in this study.
Further, the NLO corrections do not significantly alter the shape of
the kinematical distributions of the $J_z=0$
background~\cite{srprivcom}.  In other words, the NLO corrections act
mainly to change the overall normalization of the $J_z=0$ background,
implying that the cuts employed do not cause additional enhancement
(or suppression) of this background.

\begin{figure}[h!]
\centerline{\psfig{file=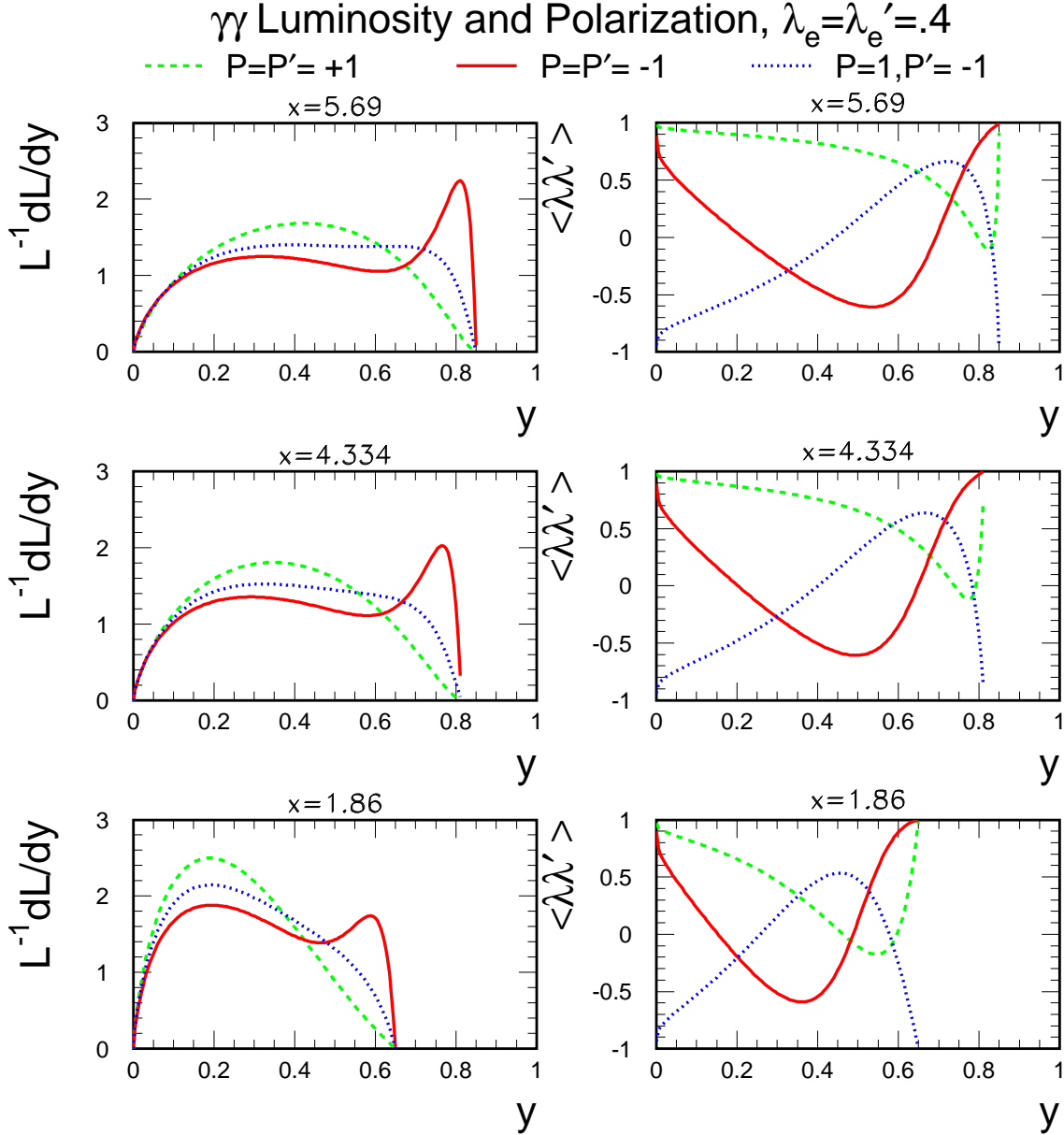,width=17cm}}
\caption[0]{The normalized differential luminosity
  ${1\over \call_{\gam\gam}}{d\call_{\gam\gam}\over dy}$ and the corresponding
  $\protect\vev{\lam\lam'}$ for $\lam_e=\lam'_e=.4$ (80\%
  polarization)
and three different choices of the initial laser photon polarizations
$P$ and $P'$. The distributions shown are for 
$\rho^2\ll 1$ \cite{Ginzburg:1983vm,Ginzburg:1984yr}. Results for  $x=5.69$,
$x=4.334$ and $x=1.86$ are compared.
}
\label{f:gamgamlum_plot}
\end{figure}

The computation of $d\call_{\gam\gam}/dy$ was first considered in 
\cite{Ginzburg:1983vm,Ginzburg:1984yr}.  We review results
based on their formulae assuming $\rho^2\ll 1$, where $\rho$
characterizes the distance from the electron laser collision to
the $\gam\gam$ interaction point. 
(See \cite{Ginzburg:1983vm,Ginzburg:1984yr}. When
$\rho$ is substantial in size, the low $E_{\gam\gam}$ part of the spectrum 
predicted by their formulae is
suppressed. However, beamstrahlung greatly enhances the luminosity
in this region, as we shall discuss.)
There are three independent choices for $\lam_e$, $\lam'_e$, $P$ and
$P'$. Assuming 80\% polarization is possible for the $e$ beams,
the values of $F(y)={1\over \call_{\gam\gam}}{d\call_{\gam\gam}\over dy}$ 
and $\vev{\lam\lam'}$
are plotted as a function of $y$ in Fig.~\ref{f:gamgamlum_plot}
for the three independent choices of relative electron
and laser polarization orientations, and for $x=5.69$, $x=4.334$ and $x=1.86$.
(The relevance of these particular $x$ values will emerge very shortly).
We observe that the choice (I) of $\lam_e=\lam'_e=.4$, $P=P'=1$ gives
large $\vev{\lam\lam'}$ and $F(y)>1$ for small to moderate $y$.
The choice (II) of $\lam_e=\lam'_e=.4$, $P=P'=-1$ yields
a peaked spectrum with $\vev{\lam\lam'}>0.85$ at the peak.
Finally, the choice (III) of $\lam_e=\lam'_e=.4$, $P=1,P'=-1$
gives a broad spectrum, but never achieves large $\vev{\lam\lam'}$.
As earlier noted, large values of $\vev{\lam\lam'}$
are important for suppressing the $b\anti b$ continuum Higgs detection
background, with leading tree-level term $\propto 1-\vev{\lam\lam'}$.
Thus, the peaked spectrum choice (II) is most suited to Higgs studies.
In fact, because $\vev{\lam\lam'}$ increases rapidly as $y$ increases
just past the peak location, it is always possible to find a
value of $y$ for which $F(y)\sim 95\%$ of its peak value while
$\vev{\lam\lam'}\sim 0.9$.  A final important point is to note that
it is really very important for both $e$ beams to be polarized in
order to minimize the $1-\vev{\lam\lam'}$ component of the background
and that luminosity {\it and} polarization at the peak are very significantly
reduced if one beam is unpolarized.
Current technology only allows for large $e^-$ polarization at
high luminosity.  Unless techniques for achieving large $e^+$
polarization at high luminosity are developed \cite{epluspol}, Higgs studies
at a $\gam\gam$ collider demand $e^-e^-$ collisions.  Thus,
it may be very difficult to perform Higgs studies at a 2nd `parasitic'
interaction region during $e^+e^-$ operation.

Let us now turn to the relevance of the particular $x$ values illustrated
in Fig.~\ref{f:gamgamlum_plot}.
If the laser energy is adjustable, $x\sim 4.8$
is often deemed to be an optimal choice (yielding $y_{\rm max}\sim 0.82$)
in that it is the largest value consistent with being below
the pair creation threshold, while at the
same time it maximizes the peak structure (at $y\sim 0.8$) for
the case (II) spectrum.  More realistically,
however, the fundamental laser wavelength will be fixed;
the Livermore group has determined that a wavelength
of 1.054 microns is the most technologically feasible value ---
see Section 5 of Chapter 13, p. 359-366 of Ref.~\cite{Abe:2001nr}.
The subpulse energy of the Livermore design is 1 Joule.
This results in a probability of $\sim 65\%$ that a given electron
in one bunch will interact with a photon.
Higher values for the subpulse energy
are possible, but would result in more multiple interactions
and increased non-linear effects. The subpulse energy chosen is
felt to be a good compromise value for achieving good luminosity without
being overwhelmed by such effects.

For a fixed wavelength, $x$ will vary as the machine energy is varied.
For a wavelength of $\lam= 1.054~\mu$, representative
values are $x=1.86$ at a machine energy of $\rts=206\gev$, 
for which $P\lam_e<0,P'\lam_e'<0$ yields
a spectrum peaking at $E_{\gam\gam}\sim 120\gev$ (as appropriate for
a light Higgs boson), and $x=5.69$
at $\rts=630\gev$, for which $P\lam_e<0,P'\lam_e'<0$ yields
a spectrum peaking at $E_{\gam\gam}\sim 500\gev$ (as appropriate for
a heavy Higgs boson).
However, as illustrated in Fig.~\ref{f:gamgamlum_plot},
the peaking for $x=1.86$ is not very strong as compared to higher
$x$ values. 
Further, the value of $1-\vev{\lam\lam'}$ at the peak (to which backgrounds
for Higgs detection are proportional) for $x=1.86$
is somewhat larger than for large $x$ values.
Fortunately, the Livermore group has developed a technique by which
the laser frequency can be 
tripled.\footnote{In order to triple the laser photon
frequency, one must employ non-linear optics. The efficiency
with which the standard $1.054~\mu$ laser beam is converted to $0.351~\mu$
is 70\%. Thus, roughly 40\% more laser power is required 
in order to retain the subpulse power of 1 Joule as deemed
roughly optimal in the Livermore study.} 
In this way, the $x$ value can be tripled for a given $\rts$,
allowing for a much more peaked spectrum, and smaller
$1-\vev{\lam\lam'}$ at the peak, for the light Higgs case.
For $\lam\sim 1/3~\mu$, a spectrum peaked at $E_{\gam\gam}=120\gev$
is obtained by operating at $\rts=160\gev$, 
yielding $x=4.334$. The spectra for this case is also plotted in 
Fig.~\ref{f:gamgamlum_plot}. The much improved peaking for $x=4.334$
as compared to $x=1.86$ is apparent.  Regarding $x=5.69$,
it has been argued in the past that $x>4.8$ 
is undesirable in that it leads to pair creation. However, our studies,
which include these effects, indicate that the resulting backgrounds
are not a problem. 

We will return to the importance of including the full dependence
of the acceptance on $z$ and $z'$ shortly.  For now, let us continue
to neglect this dependence and review a few more of the `standard'
results.

\section{Realistic \boldmath$E_{\gam\gam}$ spectra}

\begin{figure}[htb]
\centerline{
\psfig{figure=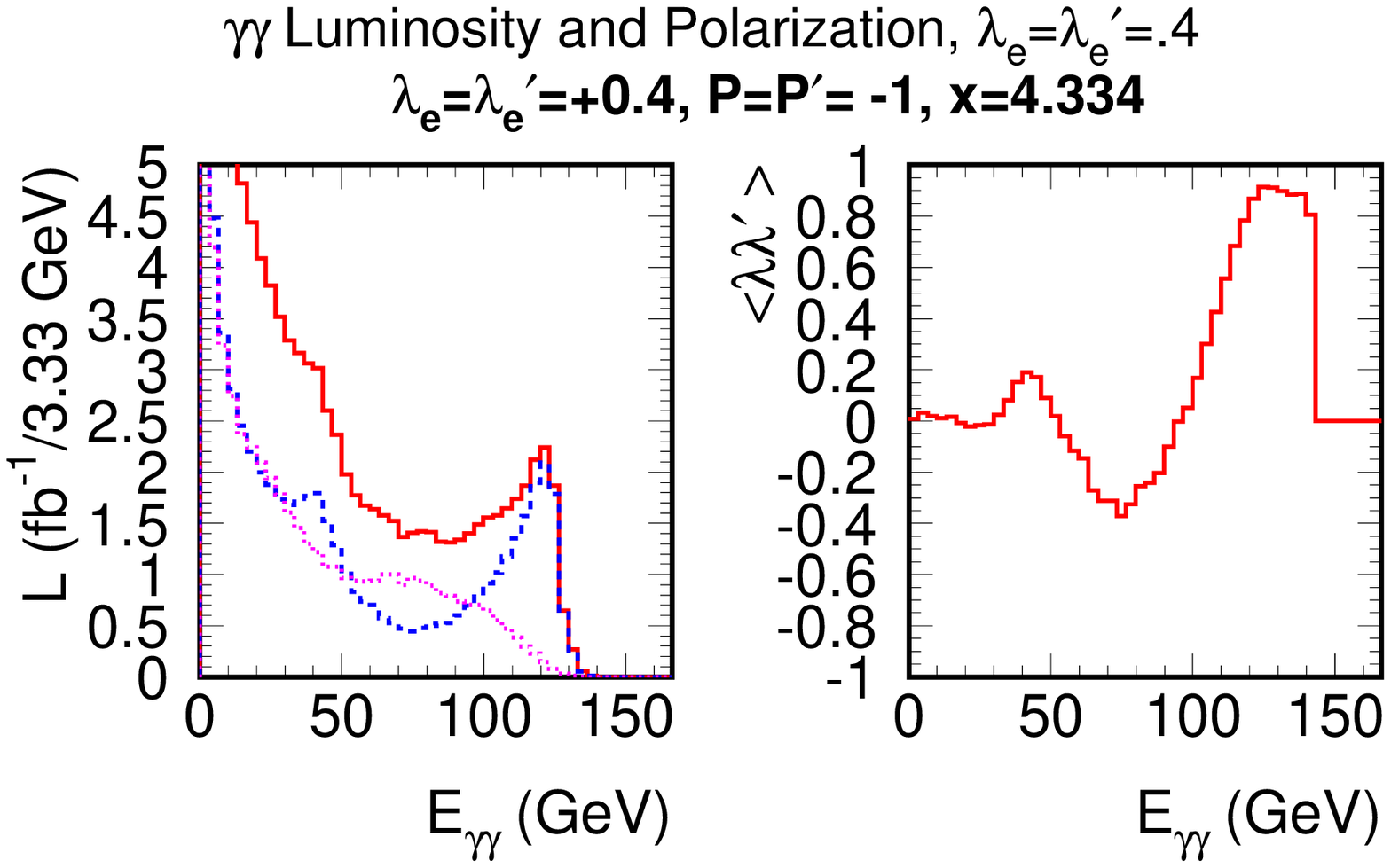,width=19cm,angle=0}}
\vspace*{-3.2in}
\caption{We plot the CAIN \cite{cainref} predictions for the $\gam\gam$ 
luminosity, $L=d\call/dE_{\gam\gam}$, in units of $\fbi/3.33 \gev$
(3.33 GeV being the bin size) for circularly polarized [case (II)] 
  photons assuming a $10^7$ sec year, $\rts=160\gev$, 
80\% electron beam polarization,
and a 1.054/3 micron laser wave length. Beamstrahlung
and other effects are included. The dashed (dotted) curve gives the component
of the total luminosity that derives from the $J_z=0$ ($J_z=2$) two-photon
configuration. 
Also plotted is the corresponding value of $\vev{\lam\lam'}$
[given by $\vev{\lam\lam'}=(L_{J_z=0}-L_{J_z=2})/(L_{J_z=0}+L_{J_z=2})$].
} 
\label{fig:higgsspec}
\end{figure}

There are important corrections to the naive luminosity distributions just
considered. First,
the luminosity at low $E_{\gam\gam}$ is affected by two conflicting 
corrections. Finite $\rho$ suppresses the
low-$E_{\gam\gam}$ luminosity.  However, this effect is more than
compensated by beamstrahlung, secondary collisions between scattered electrons
and photons from the laser beam and other non-linear effects.  
The result is a substantial enhancement of the luminosity in the
low-$E_{\gam\gam}$ region.  This is illustrated in \Fig{fig:higgsspec}
for case (II) polarization orientation choices
and for $\rts=160\gev$, which yields $x=4.33$
for a 1.054 micron laser source running with the 
`frequency tripler', and a (CP-IP) separation 
between the photon conversion point (CP) and photon-photon
interaction point (IP) of 1 mm. We also note that all the spectra
considered here were obtained for flat electron beams.
(For a given CP-IP separation, round electron beams would give 
a factor of roughly two larger luminosity. However, we chose
the flat beam configuration for consistency with the final-focus 
and collimation arrangements
that will be used in $\epem$ collisions.) 
As expected from \Fig{f:gamgamlum_plot}, the spectrum shows a peak at
$E_{\gam\gam}=120\gev$ (as might correspond to a light Higgs boson mass).
However, the low-$E_{\gam\gam}$ tail is now quite substantial.
This implies that it will be very important to achieve 
a small mass resolution, $\gamres$, for the final
state reconstruction. The luminosity
$\Delta\call_{\gam\gam}$ in the bin centered at $E_{\gam\gam}=120\gev$ 
is equivalent to  
$d\call/dE_{\gam\gam}\sim 0.66\fbi/\gev$ per $10^7$ sec year. The
corresponding luminosity at TESLA could be as much as a factor of 2
larger due to higher repetition rate and larger charge per bunch.
If one wishes to avoid a large low-$E_{\gam\gam}$ tail, then
it is necessary to have a significantly different configuration, including
much larger CP-IP separation and/or a high-field sweeping magnet.  
These options were considered (also using the CAIN program)
in the Asian Committee for Future Accelerators (ACFA) 
report~\cite{Abe:2001gc}, where
a CP-IP separation of 1~cm was adopted
and a 3 Tesla sweeping magnet was employed.~\footnote{For earlier NLC studies, 
a CP-IP separation of 0.5 cm was used
and sweeping magnets were not incorporated.} 
The disadvantage of this arrangement is a substantially lower value
for $d\call/dE_{\gam\gam}$ at the peak, at least for the 
corresponding bunch charge, repetition rate and spot size employed 
in \cite{Abe:2001gc}. As noted above, we have 
$d\call/dE_{\gam\gam}\sim 0.66 \fbi/\gev$ per year, 
which should be 
compared to $\sim 0.13\fbi/\gev$ per year for the ACFA report choices.
The latter leads to a much larger error for the precision studies of
a light SM-like Higgs boson (despite the assumption of 100\%
polarization for the $e$ beams).
In the TESLA 
Technical Design Report (TDR)~\cite{Aguilar-Saavedra:2001rg}, 
a CP-IP separation of 2.1 mm (2.7 mm)
is used for $\rts=500\gev$ ($\rts=800\gev$). 
A flat beam configuration is employed. Combining information from
Fig. 1.4.7 and Table 1.4.1 ($200\gev$ numbers) in Part VI (Appendices)
of the TESLA TDR \cite{Aguilar-Saavedra:2001rg}, we estimate that
the TESLA design will give $d\call/dE_{\gam\gam}\sim 1.8 \fbi/\gev$ per year,
more than a factor of 2 better~\footnote{The TESLA table and figure
are based upon assuming 85\% polarization for the two electron
beams.  For 80\% polarization, our estimate is that
the difference between the TESLA luminosity and ours would
be about a factor of 2, as quoted earlier.} than our $\sim 0.66\fbi/\gev$
that we shall employ for studying a Higgs with mass of $120\gev$. 

Turning to the important average $\vev{\lam\lam'}$, we note that
the naively predicted value for $\vev{\lam\lam'}$ at the luminosity
peak is about 0.86 (see \Fig{f:gamgamlum_plot}), rising
rapidly to higher values as $y$ increases. For instance,
$\vev{\lam\lam'}\sim 0.96$ at the point where the luminosity
has fallen only 25\% from its peak value.
From Fig.~\ref{fig:higgsspec} 
we see that the CAIN Monte Carlo predicts that 
this behavior of $\vev{\lam\lam'}$ is smoothed out somewhat
after including the beamstrahlung contribution,
but the value at the luminosity peak of $\vev{\lam\lam'}\sim 0.85$
is nearly the same as predicted in the naive case.~\footnote{For 
$\vev{\lam\lam'}\sim 0.85$, the heavy quark background to Higgs detection
will be dominated by its $J_z=\pm 2$ component
(proportional to $1-\vev{\lam\lam'}$); even after radiative
corrections, the $J_z=0$ component of the background is significantly
smaller once cuts isolating the 2-jet final states are imposed. See
Appendix B.}

The above results are still somewhat misleading due to the fact
that we have not yet incorporated the dependence of the acceptance
function $A(z,z',\zth)$.  For the Higgs signal that is
independent of $\zth$, it is useful to define
\bea
&&\half\sum_{\lam,\lam'}\int dzdz' \int d\zth {d\call_{\gam}^{\lam}(\lam_e,P,z)\over dz}
{d\call_{\gam}^{\lam'}(\lam_e^\prime,P',z') \over dz'}
A(z,z',\zth)[1,\lam\lam']\nonumber\\&&
\equiv\int dy 
{d\call_{\gam\gam}^{\rm eff}(\lam_e,\lam_e^\prime,P,P',y)\over dy}
[1,\vev{\lam\lam'}^{\rm eff}(y)]\,,
\label{dldyeff}
\eea
yielding
\bea
N(\gam\gam\to{\h}\to X)&=&{4\pi^2\Gamma({\h}\to\gam\gam)
\br({\h}\to X)(1+\vev{\lam\lam'}^{\rm eff}(y))\over \rts \mh^2}
\left.{d\call_{\gam\gam}^{\rm eff}\over dy}\right|_{y=\mh/\rts}\,.
\eea
The effective luminosity
and $\vev{\lam\lam'}$ depends on the cut $|\zth|<0.5$
and the standard LC detector acceptances, including, in particular,
the requirement that the jets pass fully through the vertex detector
and be fully reconstructed (with little energy in the uninstrumented
forward and backward regions).
For $E_{\gam\gam}$ substantially below the peak
region, the peak being 
in the vicinity of $E_{\gam\gam}\sim 120\gev$, the effective
luminosity for Higgs production is only  slightly suppressed
(beyond the obvious factor of 0.5 coming from the $|\zth|<0.5$ cuts).

\section{Studying a light SM-like Higgs boson}

Consider first a SM-like Higgs boson $\h$ of relatively light mass;
SM-like Higgs bosons arise in many models containing physics beyond
the SM.
The $\h\to\gam\gam$ coupling receives contributions from loops
containing any charged particle
whose mass, $M$, arises in whole or part from the vacuum expectation
value of the corresponding neutral Higgs field.
(Of course, in the strict context of the SM, the masses
of all elementary particles derive entirely from the Higgs
field vacuum expectation value.) 
When the mass, $M$, derives in whole or part from the vacuum expectation
value ($v$) of the neutral Higgs
field associated with the $\h$, 
then in the limit of $M\gg\mh$ for the particle in the loop, the
contribution asymptotes to a  value  that depends 
on the particle's spin (\ie\ the contribution does not decouple).
As a result, a measurement of $\Gamma(\h\to\gam\gam)$ 
provides the possibility of revealing the presence of 
heavy charged particles that acquire their mass via the Higgs mechanism.
Of course, since the mass deriving from the SM-like neutral Higgs
vev $v$ is basically proportional to
some coupling times $v$, if the coupling is perturbative
the mass of the heavy particle is unlikely to be much 
larger than $0.5-1$ TeV. 
In addition, we note that  $\br(\h\to X)$ 
is entirely determined by the spectrum
of particles with mass $<\mh/2$, and is not affected by heavy states
with $M>\mh/2$. Consequently,
measuring $N(\gam\gam\to \h\to X)$ provides
an excellent probe of new heavy particles with mass deriving
from the Higgs mechanism.  We emphasize that in models beyond the
SM, particles can acquire mass from mechanisms other than
the Higgs mechanism. If there is a SM-like Higgs boson 
in such an extended model the loop contributions from the charged particles that acquire
a large mass from some such alternative 
mechanism will decouple as $({\rm mass})^{-2}$
and $\gam\gam\to\h$ will not be sensitive to their presence.

\begin{figure}[htb]
\centerline{
\psfig{figure=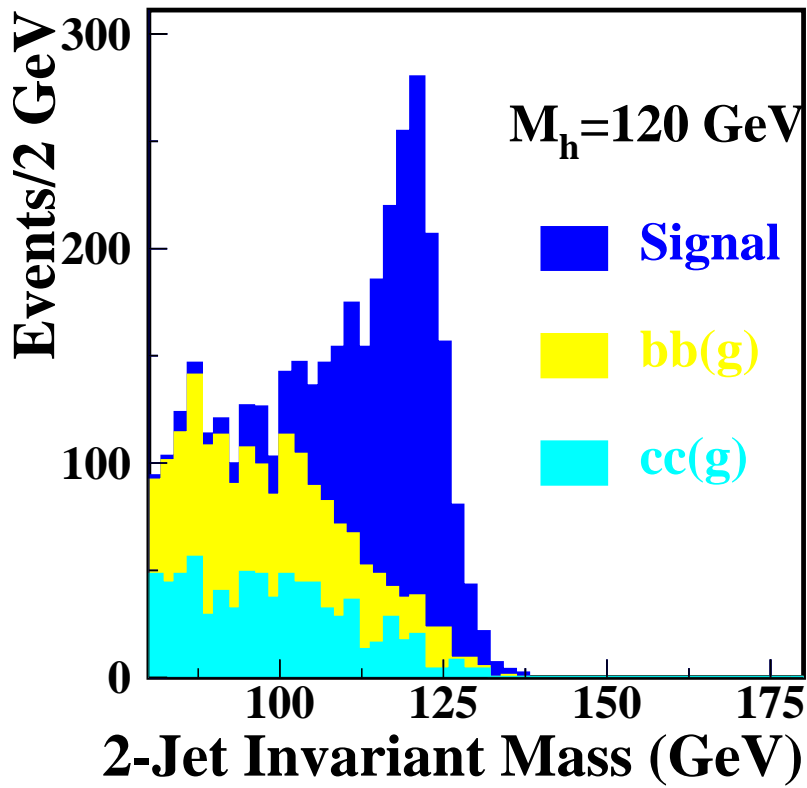,width=12cm,angle=0}}
\vspace*{0.1cm}
\caption{
Higgs signal and heavy quark backgrounds in units of events per 2 GeV
for a Higgs mass of 120~GeV and assuming a running year of $10^7$ sec.
We have employed the cuts as given in the text.
}
\label{fig:higgs}
\end{figure}

If there are no new particles that acquire mass via the Higgs
mechanism, a precision measurement of $\Gamma({\widehat h}\to\gam\gam)$
can allow one to distinguish between a ${\widehat h}$ that is part
of a larger Higgs sector and the SM $\hsm$. \Fig{fig:higgs}
shows the di-jet invariant
mass distributions for the $\mhsm=120\gev$ Higgs signal and
for the $b\anti b(g)$ and $c\anti c(g)$ backgrounds,
using the luminosity distribution
of \Fig{fig:higgsspec}, after
all cuts.  Our analysis is similar, but not identical, to
that of Ref.~\cite{jikia}. See also 
\cite{Melles:1999xd,Melles:1998rp,Muhlleitner:2001kw,Muhlleitner:2000jj}.
Both employ JETSET fragmentation using
the Durham algorithm choice of $y_{\rm cut}=0.02$ for defining the jets.
Further, we employ the event mixture predicted
by PYTHIA (passed through JETSET) \cite{pythiajetset}
and we use the LC Fast Monte Carlo detector simulation within ROOT \cite{Brun:1997pa},
which includes calorimeter smearing and detector configuration as
described in Section 4.1 of Chapter 15 of Ref.~\cite{Abe:2001nr}.  
The signal is generated using PANDORA plus PYTHIA/JETSET \cite{pandora}.
We have employed the following cuts.
\bit
\item Only tracks and showers with $|\cos\theta|<0.9$ in the laboratory
frame are accepted.
\item Tracks are required to have momentum greater than 200 MeV and showers
must have energy greater than 100 MeV.
\item We then focus on the two most energetic jets in the event (with jets
defined using $y_{\rm cut}=0.02$).
\item We require these two jets to be back-to-back in three dimensions
using the criteria $|p_i^1+p_i^2|<12\gev$ for $i=x,y,z$.
\item 
We require $|\cos\theta^*|<0.5$, where $\theta^*$ is the
angle of the two most energetic jets in their center of mass relative
to the beam direction.
The alternative of $|\cos\theta|<0.5$ results in very little
change for $E_{\gam\gam}>80\gev$ once the preceding back-to-back
cut has been applied.
\eit
We note that even though we do not explicitly require exactly two-jets
in the final state, the third and fourth cuts listed above, especially the
back-to-back requirement, results in 90\% of the retained
events containing exactly two jets.

We employ the two most energetic jets (after imposing the cuts given above)
to reconstruct the Higgs boson signal.
Our mass resolution for the narrow-width Higgs boson signal 
is  $4.76\pm 0.13 \gev$ (for a Gaussian 
fit from $-1\sigma$ to $+10\sigma$)\footnote{We
employ this range in order to avoid the rapidly rising background
at low masses and the mass distribution tail at masses
below the resonance peak coming from reconstruction.} 
which is similar to the $\sim 6\gev$ found in \cite{jikia}.
We believe that the difference in mass resolution
is due primarily to differences in the Monte Carlo's employed.
If we keep only events with 
$M_{2~\rm jet}\geq 110\gev$,
there are roughly 1450 signal events and about 335 background events, 
after all cuts. 
This would yield a measurement of
$\Gamma(\hsm\to\gam\gam)\br(\hsm\to b\anti b)$ with an accuracy of  
$\sqrt{S+B}/S\sim 2.9\%$.~\footnote{The more optimistic error of close
to $2\%$ quoted in \cite{jikia} for $\mhsm=120\gev$ is 
based upon a higher peak luminosity.
We estimate a factor $\lsim 2$ larger peak luminosity at TESLA
coming primarily from rep rate and bunch charge density.
The TESLA analyzes also assume a somewhat higher beam polarization.
The result is that TESLA errors will be about a factor
of $\sqrt 2$  smaller than errors we estimate, as is 
consistent with the $2\%$ vs. $2.9\%$ error at $\mhsm=120\gev$.
The error for the ACFA design of Ref.~\cite{Abe:2001gc} is about
7.6\% for (we believe) about 3 years of running, which is much larger
than the error we achieve after just one year of operation. 
This difference is largely due to the factor of nearly 5
smaller value of $d\call/dE_{\gam\gam}$ at the peak and would have been
even greater if more realistic $<100\%$ polarization for the $e$
beams had been employed.}
The error for this measurement increases to about $10\%$
for $\mhsm\sim 160\gev$ given the predicted signal rate,  $S:B\sim 1:1$ 
and $\vev{\lam\lam'}\sim 0.85$ at the peak.
These accuracies are those estimated for one $10^7$ sec year of operation.
Deviations due to ${\widehat h}\neq \hsm$ in an extended Higgs
sector model typically exceed 3\% if the other heavier Higgs bosons
have masses below about 500 GeV (so that there are significant
corrections to the decoupling limit). To obtain the above results,
excellent $b$ tagging is essential to eliminate backgrounds
from light quark states. 
We have not simulated $b$-tagging. Rather we have assumed 
(as in \cite{jikia}) 70\% 
efficiency for double-tagging $b\anti b$ events (after having already made
the necessary kinematic cuts), for which there is a 3.5\%
efficiency for tagging $c\anti c$ events as $b\anti b$, 
a rejection factor of 20. This rejection factor is very essential
since, crudely speaking, the $c\anti c$ background is a factor
of 16 ($=(q_c/q_b)^4$) larger without this rejection. After including
the tagging rejection, the $c\anti c$ and $b\anti b$ backgrounds
are roughly comparable.  

We should note explicitly that we have performed our background
and signal cross section calculations at tree-level.  Various studies
have appeared in the literature showing that under some circumstances
higher order corrections and other effects can be quite important.  
We have explicitly
chosen our cuts so that they are not. In particular,
we have employed cuts that primarily retain only events with two
jets. It is the processes with extra radiated
gluons (which are included as part
of the NLO radiative corrections) that can cause the largest
corrections since the associated cross sections are not
proportional to $1-\vev{\lam\lam'}$. As discussed
in more detail in Appendix B, NLO corrections to two-jet
events, while sizable, will not significantly impact our results.
The primary reason for employing tree-level
computations is the importance of being able to perform full
simulation analyzes, something that is only possible in the context
of PANDORA and JETSET for the signal and in the context of PYTHIA
and JETSET for the background.  
We estimate our errors are not more than 10\%-20\% 
as a result of ignoring higher order corrections.
Appendix B is devoted to a more detailed discussion
of the relevant issues.

\section{\boldmath The $\hh$ and $\ha$ of the MSSM}

\begin{figure}[th]
\centerline{\psfig{file=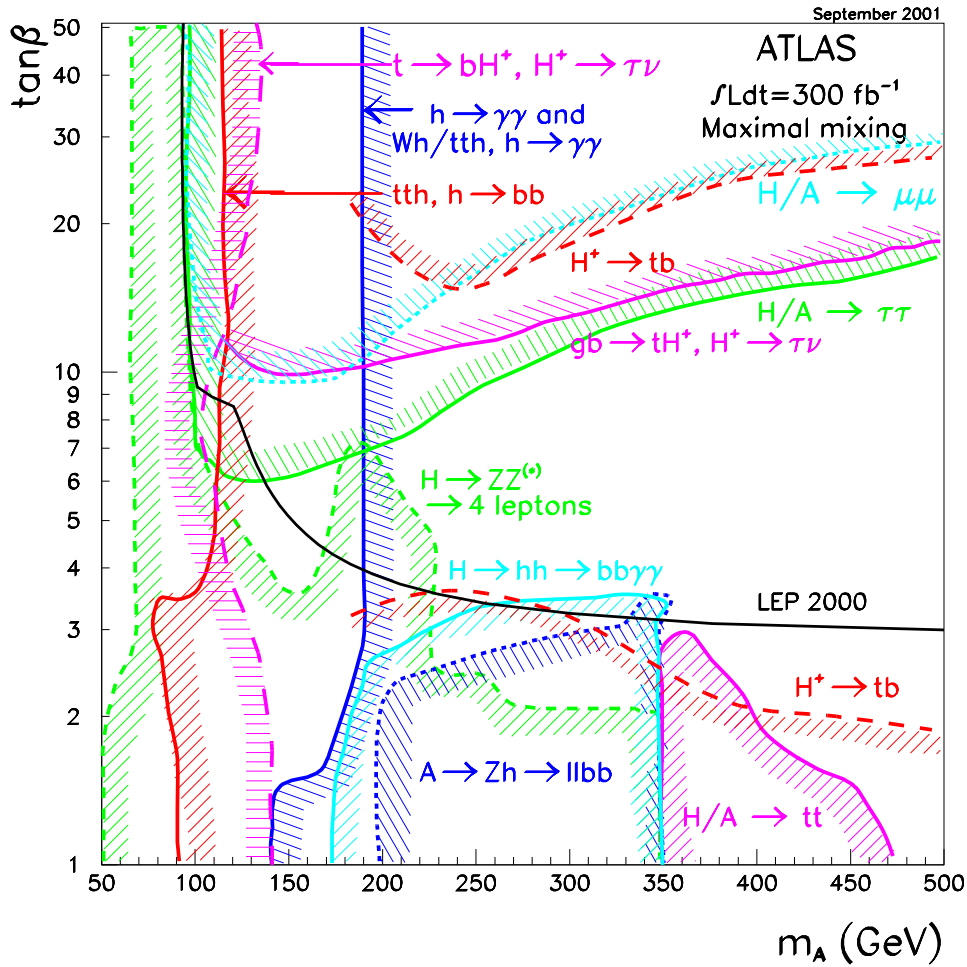,width=14cm}}
\caption[0]{$5\sigma$ discovery contours for MSSM Higgs boson detection
in various channels are shown in the $[\mha,\tanb]$ parameter plane, 
assuming maximal mixing and an integrated luminosity of $L=300\fbi$
for the ATLAS detector. This figure is preliminary \cite{atlasmaxmix}.}
  \label{f:atlasmssm}
\end{figure}

In many scenarios, it is very possible that by combining 
results from $\gamma\gamma \to\hl\to b\anti b$ with other types of
precision measurements for the SM-like Higgs boson, we will
observe small deviations and suspect the presence of heavy Higgs
bosons. Giga-$Z$~\footnote{The phrase ``Giga-$Z$'' refers
to operating the future LC at $\rts=\mz$. The high LC luminosity would
allow the accumulation of a few$\times 10^9$ $Z$ pole events
after just a few months of running. 
By combining such operation with a high-precision $WW$
threshold scan to determine $\mw$ to within $\pm 6$~MeV,
the standard $S,T$ parameters
could then be determined with much greater accuracy than
is currently possible using LEP data.}
precision measurements could provide additional indirect evidence
for extra Higgs bosons through a very precise determination
of the $S$ and $T$ parameters, which receive
corrections from loops involving the extra Higgs bosons.  However, 
to directly produce the heavier Higgs in $\epem$ collisions is likely to 
require large machine energy. For example, In the 2HDM $\epem\to \hh\ha$ 
pair production would be the most relevant process in the decoupling limit, but
requires $\rts>\mhh+\mha$, with $\mhh+\mha\sim 2\mha$ as the decoupling
limit sets in. The alternatives of $b\anti b \hh$ and $b\anti b \ha$ production
will only allow $\hh$ and $\ha$ detection if $\tanb$ is 
large~\cite{Grzadkowski:2000wj}. Either low or high $\tanb$ is
also required for LHC discovery of the $\hh,\ha$ if they have
mass $\gsim 250\gev$.  This is illustrated in
\Fig{f:atlasmssm}. After accumulation of $L=300\fbi$ at the LHC, 
the $\hh,\ha$ will be detected except in the wedge of parameter
space with $\mha\gsim 250\gev$ and moderate $\tanb$ (where only the
$\hl$ can be detected). If the LC is operated at $\rts=630\gev$,
then detection of $\epem\to\hh\ha$ will be possible for $\mha\sim\mhh$
up to nearly 300 GeV. In this case, the parameter region 
for which some other means of detecting the $\hh,\ha$ must be
found is the portion of the LHC wedge with $\mha\gsim 300\gev$.
We will explore the possibility of finding the $\hh$ and $\ha$
in $\gam\gam$ collisions.  Earlier work along this line appears
in \cite{Muhlleitner:2001kw,Muhlleitner:2000jj}.
Our results will incorporate CAIN predictions for the luminosity
and polarizations of the colliding back-scattered photons
using 80\%  polarization for the electron beams (which we
believe is more realistic than the 100\% polarization assumed in 
\cite{Muhlleitner:2001kw,Muhlleitner:2000jj}).

We will show that single $\hh,\ha$ production via $\gam\gam$
collisions will allow their discovery throughout a large fraction
of this wedge. The event rate, see \Eq{ngamgam},
can be substantial due to quark loop contributions (mainly $t$
and, at high $\tanb$, $b$) and loops containing other new
particles (\eg\ the charginos, $\ldots$ of supersymmetry).
In this study, we will also assume that the superparticle
masses (for the charginos, squarks, sleptons, \etc.) are sufficiently
heavy that (a) the Higgs bosons do not decay to superparticles 
and (b) the superparticle loop contributions to the $\gam\gam$
coupling are negligible.

Assuming no reliable preconstraints on $\mha,\mhh$, 
an important question is whether it is best to search for the $\hh,\ha$
by scanning in $\rts$ (and thereby in $E_{\gam\gam}$,
assuming type-II peaked spectrum configuration) or running at fixed $\rts$
using a broad $E_{\gam\gam}$ spectrum part of the time and a peaked spectrum
the rest of the time \cite{Gunion:1993ce}.
As we shall discuss, if covering the wedge
region is the goal, then running at a single
energy, part of the time with a peaked $E_{\gam\gam}$ luminosity distribution
and part of the time with a broad distribution (in ratio 2:1), 
would be a somewhat preferable approach.

\begin{figure}[htb]
\centerline{\psfig{file=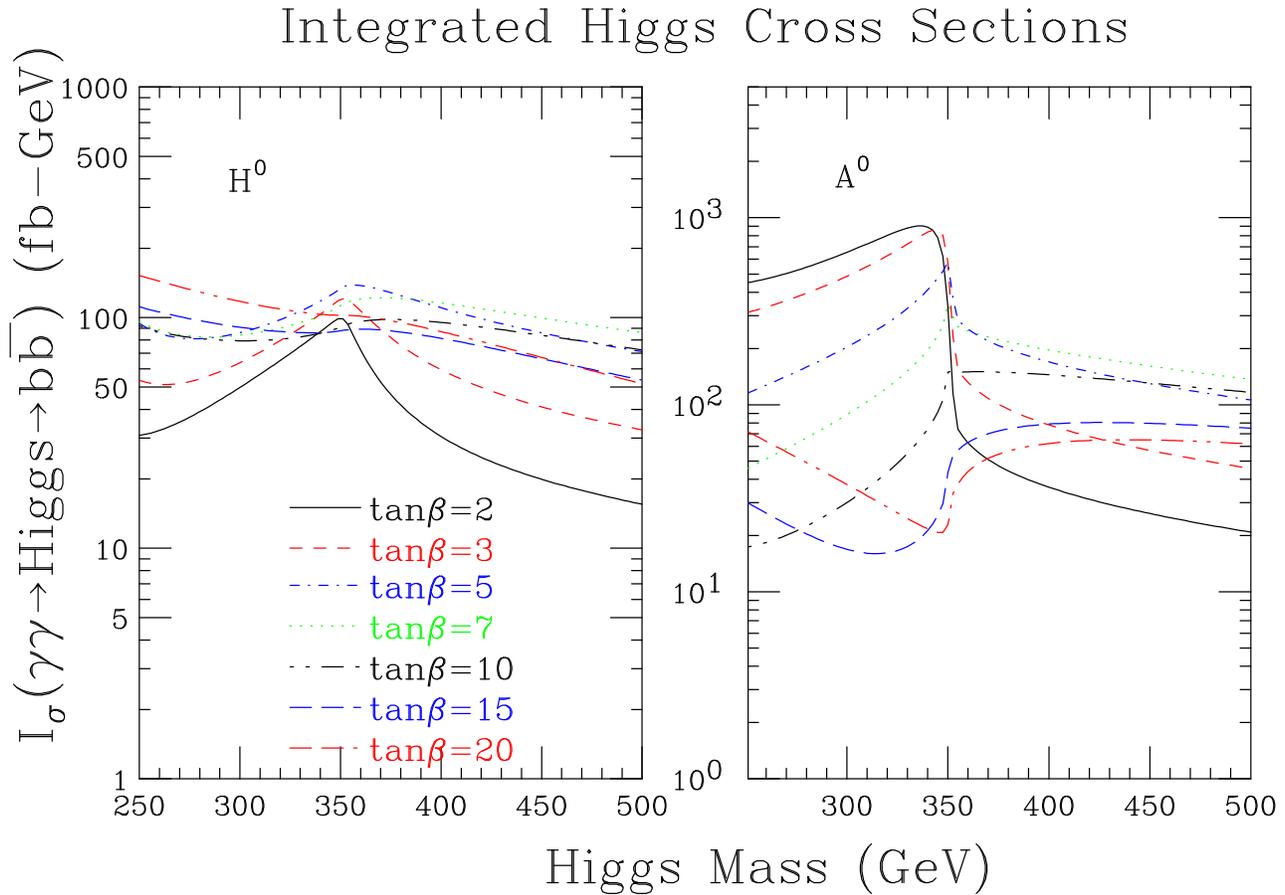,height=12cm}}
\caption[0]{We plot the integrated $\hh$ and $\ha$ Higgs cross sections
 $I_\sigma$,
as defined in Eq.~(\ref{ngamgam}), as a function of Higgs mass, for a variety
of $\tanb$ values. We employ the maximal-mixing scenario with
$\msusy=1\tev$. Supersymmetric particle loops are neglected.  
}
\label{f:sigeff}
\end{figure}

The first important input to the calculations is the effective integrated cross
section, $I_\sigma$, as defined in Eq.~(\ref{ngamgam}), for the $\hh$
and $\ha$. These cross sections are plotted as a function of Higgs mass
for a variety of $\tanb$ values in Fig.~\ref{f:sigeff}.
We have computed the cross sections using the $b\anti b$ branching ratios
and $\gam\gam$ widths obtained from HDECAY \cite{hdecayref}
using {\it input} masses of $\mha$ as plotted on the $x$-axes.
We have employed $\mt=175\gev$, exactly.  For 
Supersymmetry (SUSY) parameters,
we have chosen $\msusy=1\tev$ for all slepton and squark soft-SUSY-breaking
masses and $\mu=+1\tev$.  For $A_t$ we have assumed the maximal-mixing
choice of $A_t=\mu/\tanb+\sqrt 6 \msusy$. In addition, we have
taken $A_b=A_\tau=A_t$. Our plots have been restricted to $\mha\leq 500\gev$
due to the fact that if the LC is operated at 
$\rts=630\gev$ (corresponding to $x\sim 5.69$
for 1 micron laser wavelength) we can potentially probe Higgs masses as
high as $\sim 500\gev$. 

\begin{figure}[htb]
\centerline{\psfig{file=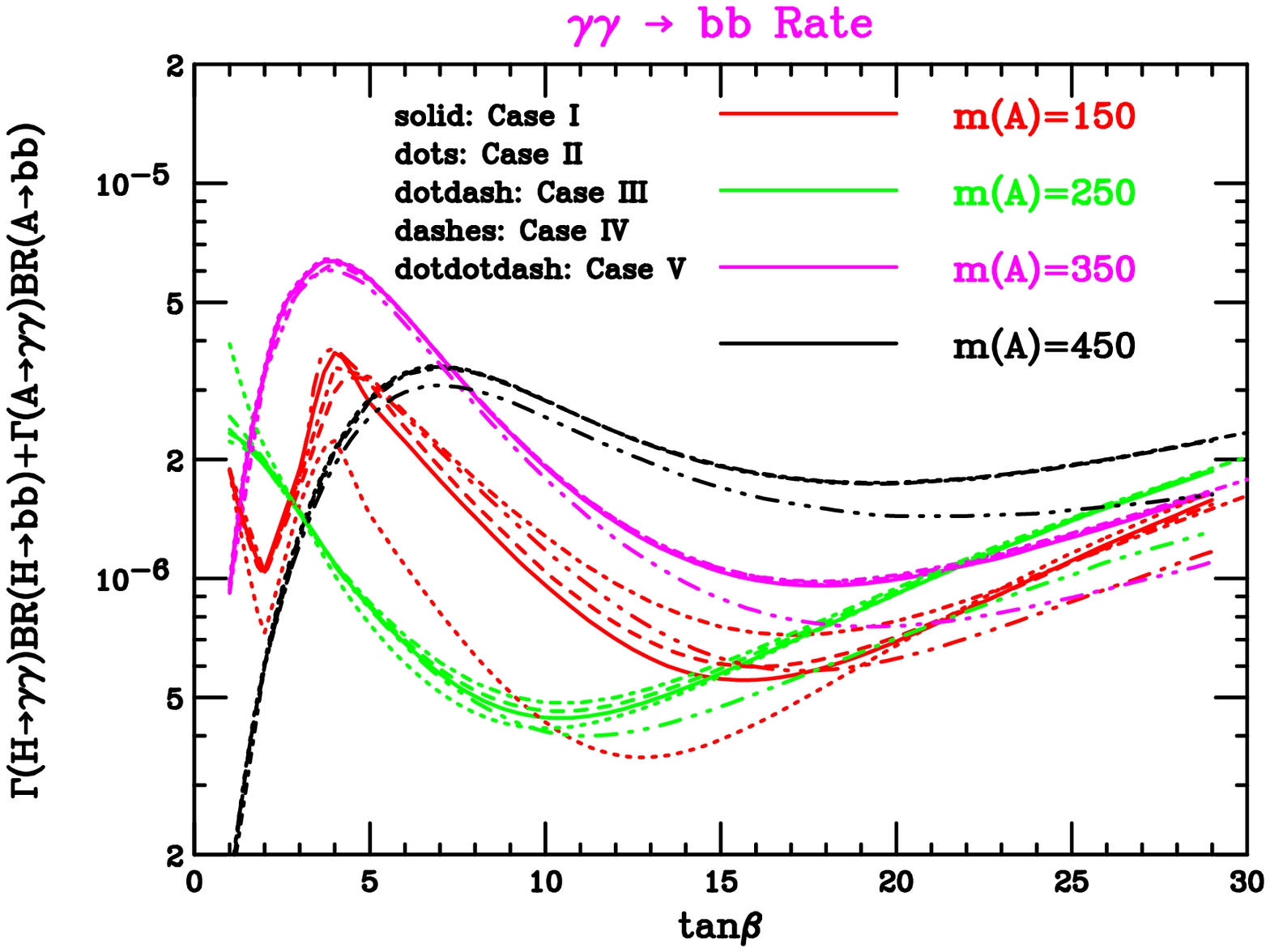,height=12cm}}
\caption[0]{We plot the sum $\Gamma(\hh\to\gam\gam)\br(\hh\to b\anti b)+
\Gamma(\ha\to\gam\gam)\br(\ha\to b\anti b)$ as a function
of $\tanb$ for several $\mha$ values. The signal rate 
$N(\gam\gam\to \hh,\ha\to b\anti b)$ is roughly proportional to this quantity.
Results for the five cases delineated in the text are shown.}
\label{modelcomps}
\end{figure}

An interesting question is the extent to which these inputs are model
dependent in that they are sensitive to
other parameters of the MSSM. Our study has been performed 
for the maximal-mixing scenario with $\mt=175\gev$ and $\msusy=1\tev$,
assuming that all SUSY particles are heavy enough to not
significantly influence the $\gam\gam\to\hh,\ha$ couplings and
heavy enough that $\hh,\ha\to $ SUSY decays are not significant.
(In the context of HDECAY, we have set IOFSUSY=1.)
If SUSY particles are moderately light, there will be some, but not
dramatic modifications to the couplings and some dilution of
the $\hh,\ha\to b\anti b$ branching ratios. These effects will
be minimal at the higher $\tanb$ values in the wedge region,
but could make discovery in the $b\anti b$ channel
difficult for some of the lower $\tanb$ points. One would undoubtedly
try to make use of the SUSY decay channels themselves to
enhance the net signal for $\gam\gam\to\hh,\ha$.
Even if SUSY particles are all heavy, 
there could be some variation as one moves from the maximal-mixing 
scenario to the  no-mixing scenario, and so forth.  Further,
there are certain non-decoupling
loop corrections to the relation between $m_b$ and the $\hh,\ha\to
b\anti b$ Yukawa couplings that could either enhance or diminish
the $\gam\gam\to \hh,\ha\to b\anti b$ rates~\cite{Carena:2001bg}.
(These are not currently incorporated into the standard version
of HDECAY.) 
We have performed a limited exploration by considering
five cases. Computations for cases I-IV are performed using version 
2.0 of HDECAY, \ie\ that available as of September 2001.
\bed
\item{I:} The maximal-mixing scenario defined above.
\item{II:} The maximal-mixing scenario as above, but with $\mu=-1\tev$.
\item{III:} The no-mixing scenario defined by $A_b=A_\tau=A_t=\mu/\tanb$,
with $\msusy=\mu=1\tev$.
\item{IV:} The maximal-mixing scenario, as in case I, but with $\mu=0$.
\item{V:} In this case, we employ the maximal mixing scenario with
$\msusy=\mu=1\tev$, but employ a modified version of HDECAY
(provided by the authors of Ref.~\cite{Carena:2001bg})
in which the $\Delta \lambda_b$ corrections to the Higgs $b\anti b$
vertices are included.  These arise from loop corrections involving
supersymmetric particles (neglected in cases I-IV), and are most
substantial when $\tanb$ is large.  These corrections do not
vanish (i.e. do not decouple) even when SUSY particle masses are large.
The corrections would have opposite sign to those plotted for $\mu=-1\tev$. 
\eed
The results in each of the above five cases for 
$\Gamma(\hh\to\gam\gam)\br(\hh\to b\anti b)+
\Gamma(\ha\to\gam\gam)\br(\ha\to b\anti b)$
(to which the signal rate $N(\gam\gam\to\hh,\ha\to b\anti b)$ is roughly
proportional) are plotted in Fig.~\ref{modelcomps} as a function
of $\tanb$ for several $\mha$ values. We observe that, although
there is considerable model dependence for the relatively low
mass of $\mha=150\gev$, this model dependence becomes
quite minimal when comparing cases I-IV
for $\mha\geq 250\gev$, \ie\ in the wedge region
of interest. However, results for case V show that SUSY loop corrections
can impact the predicted signal event rate once $\tanb$ is large enough,
but remains minimal for $\mha\leq 500\gev$ and $\tanb$ values
in the wedge region.

\begin{figure}[p!]
\vspace*{-.3in}
\centerline{\psfig{file=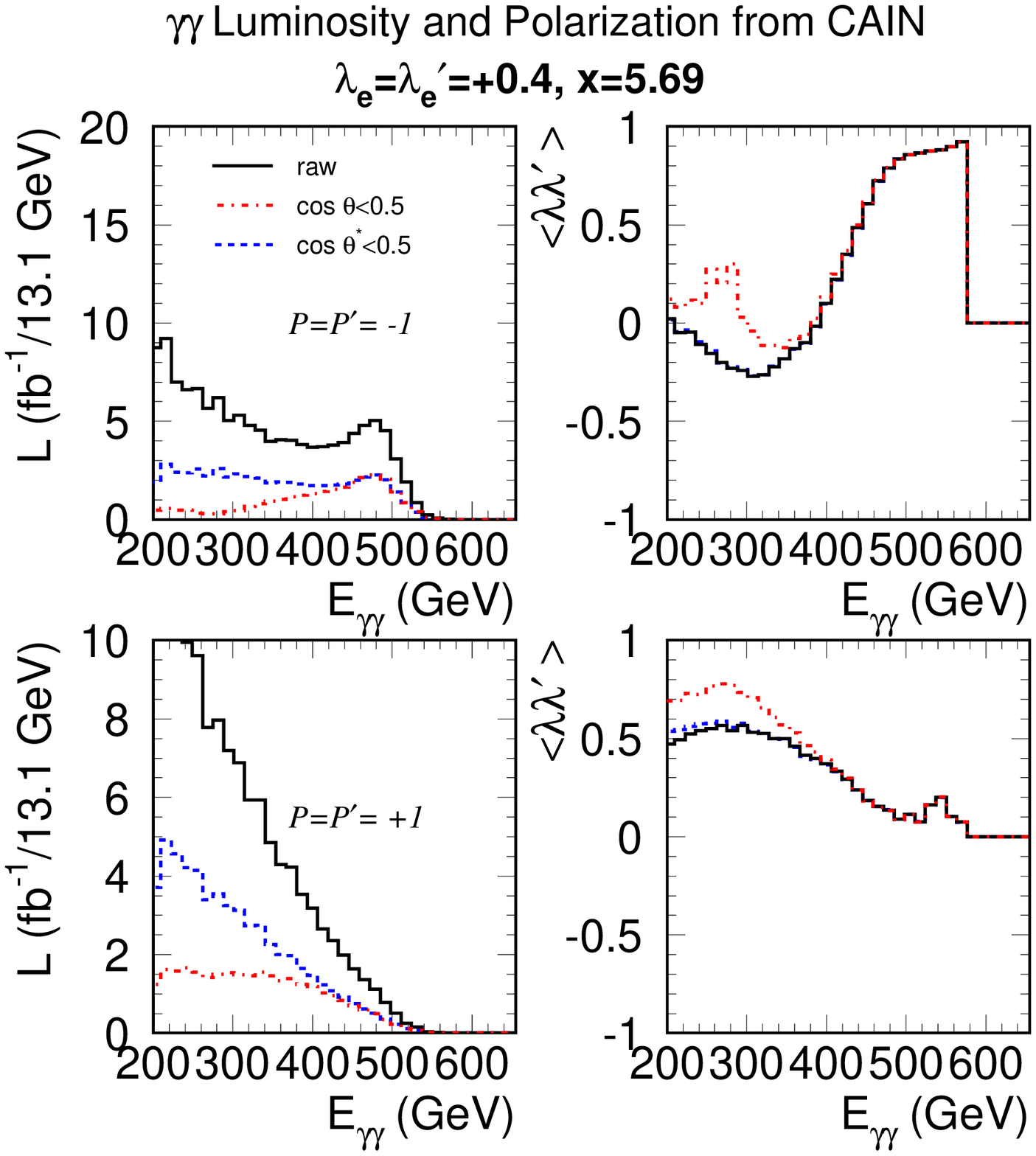,height=18cm}}
\caption[0]{Luminosity, in units $\fbi/13.1\gev$, for a $10^7$ sec year and
associated $\vev{\lam\lam'}$ are plotted for $\rts=630\gev$
($x=5.69$ for $1.054~\mu$m laser wavelength),
assuming 80\% electron beam polarizations, for polarization
orientation cases (I) and (II). Results are plotted for 3 different
cases. The solid lines show the results before any cuts or reconstruction
efficiencies are incorporated. The dashed and dash-dot
lines assume that the two most energetic jets are produced uniformly
(as for a spin-0 boson decaying to two jets)
in $\cos\theta^*$, where $\theta^*$ is the two-jet axis
angle relative to the beam direction in the two-jet rest frame. 
The dashed lines show the results after requiring $|\cos\theta^*|<0.5$.
The dash-dot lines show the results after requiring $|\cos\theta|<0.5$
for the $\theta$'s of the two most energetic jets in the laboratory rest frame.
}
\label{f:cainlums}
\end{figure}

The next important inputs are values of ${d\call \over dE_{\gam\gam}}$
and $\vev{\lam\lam'}$ for the peaked spectrum (type-II) and broad
spectrum (type-I) electron-laser-photon polarization configurations. 
The luminosity and polarization results from the CAIN \cite{cainref}
Monte Carlo program are plotted as the solid curves in \Fig{f:cainlums}. 
Note again the luminosity enhancement at low $E_{\gam\gam}$
relative to naive expectations. In the case of the type-II spectrum,
the luminosity remains quite large even below the $E_{\gam\gam}$
peak at $E_{\gam\gam}=500\gev$, and that $\vev{\lam\lam'}$
is large for $E_{\gam\gam}>450\gev$. In the case of the
type-I spectrum, the luminosity grows is substantial for $E_{\gam\gam}=400\gev$
and rises rapidly with decreasing $E_{\gam\gam}$. In addition,
reasonably large $\vev{\lam\lam'}$
is retained for $250<E_{\gam\gam}<400\gev$.
However, in both cases, the values
of $\vev{\lam\lam'}$ are always small enough
that the $J_z=2$ part of the $b\anti b$ background to Higgs detection
will be only partially suppressed by the $1-\vev{\lam\lam'}$ factor,
and will be dominant.

The final ingredient is to assess the impact of the cuts
required to reduce the $b\anti b(g)$ and $c\anti c (g)$ backgrounds
to an acceptable level.
In order to access the Higgs bosons with mass substantially below
the machine energy of 630 GeV, we must employ cuts that remove
as little luminosity for $E_{\gam\gam}$ substantially below $\rts$ 
as possible while still eliminating most of the background.
For this purpose, a cut on $|\cos\theta^*|<0.5$ (where $\theta^*$
is the angle of the two most energetic jets
relative to the beam direction in the two-jet rest frame) is far more
optimal than is a cut of $|\cos\theta|<0.5$ (where $\theta$
is the angle of a jet in the laboratory frame).  This is illustrated
in Fig.~\ref{f:cainlums} where it is seen that the former cut
on $\theta^*$ leads to much higher luminosity than the latter cut
on $\theta$. Thus, even though slightly larger $\vev{\lam\lam'}$
is obtained using the $\theta$ cuts, much better signals 
(relative to background) are achieved using the $\theta^*$ cut.
A second cut is that imposed upon the two-jet mass distribution.
The optimal value for this cut depends upon the Higgs widths,
the degree of degeneracy of the $\hh$ and $\ha$ masses,
and the detector resolutions and reconstruction techniques. 

\begin{figure}[htb]
\centerline{\psfig{file=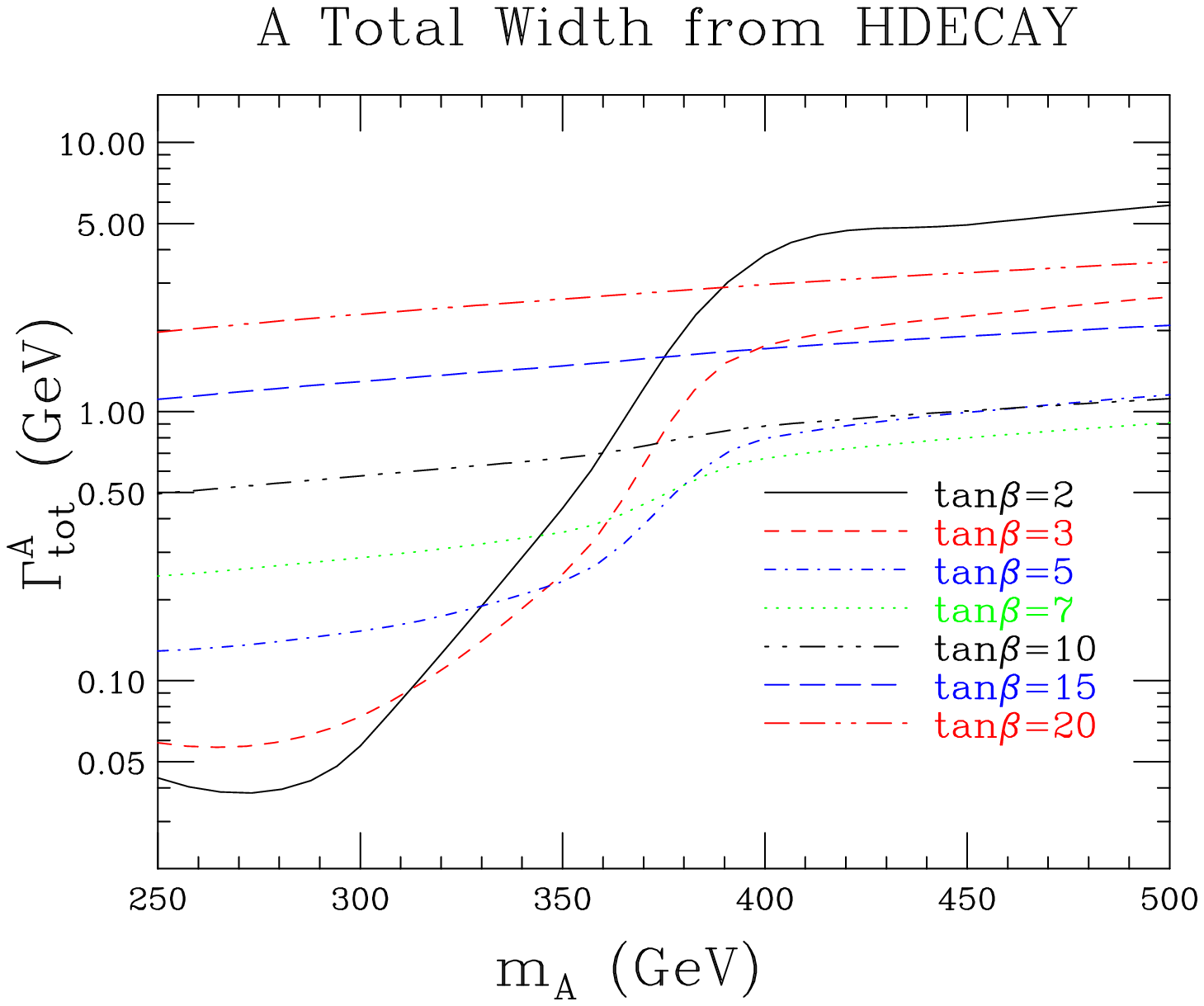,height=12cm}}
\caption[0]{We plot the total width of the $\ha$ as a function of $\mha$
for our standard set of $\tanb$ values. Results are those from HDECAY
for the earlier defined maximal mixing scenario with $\msusy=\mu=1\tev$.
Supersymmetric particle loops are neglected.  
}
\label{f:totwidth}
\end{figure}

Figure~\ref{f:totwidth} shows the total $\ha$ width as a function
for $\mha$ for our standard set of $\tanb$ values.
For the $\tanb$ range inside the problematical wedge ($15>\tanb>3$), the
$\ha$ (and also the $\hh$) is still relatively narrow, 
with widths below $\sim 3$ GeV.
In fact, the width of the two-jet mass distribution will
probably derive mostly
from detector resolutions and reconstruction procedures.
A full Monte Carlo analysis for heavy 
Higgs bosons with relatively small widths is not yet available.
However, there are many claims in the literature that the resulting
mass resolution will almost certainly be better than 
$\Delta m_{\rm 2-jet}=30\% \sqrt{m_{\rm 2-jet}}$ 
(the result obtained assuming $\Delta E_{\rm jet}=18\%\sqrt{E_{\rm jet}}$ 
for each of the back-to-back
jets)~\cite{Abe:2001nr,Abe:2001gc,Abe:2001wn,Aguilar-Saavedra:2001rg}.
Very roughly this corresponds to a full-width at half maximum of
about $6\gev$ in the mass range from $250-500\gev$ of interest.

\begin{figure}[htb]
\centerline{\psfig{file=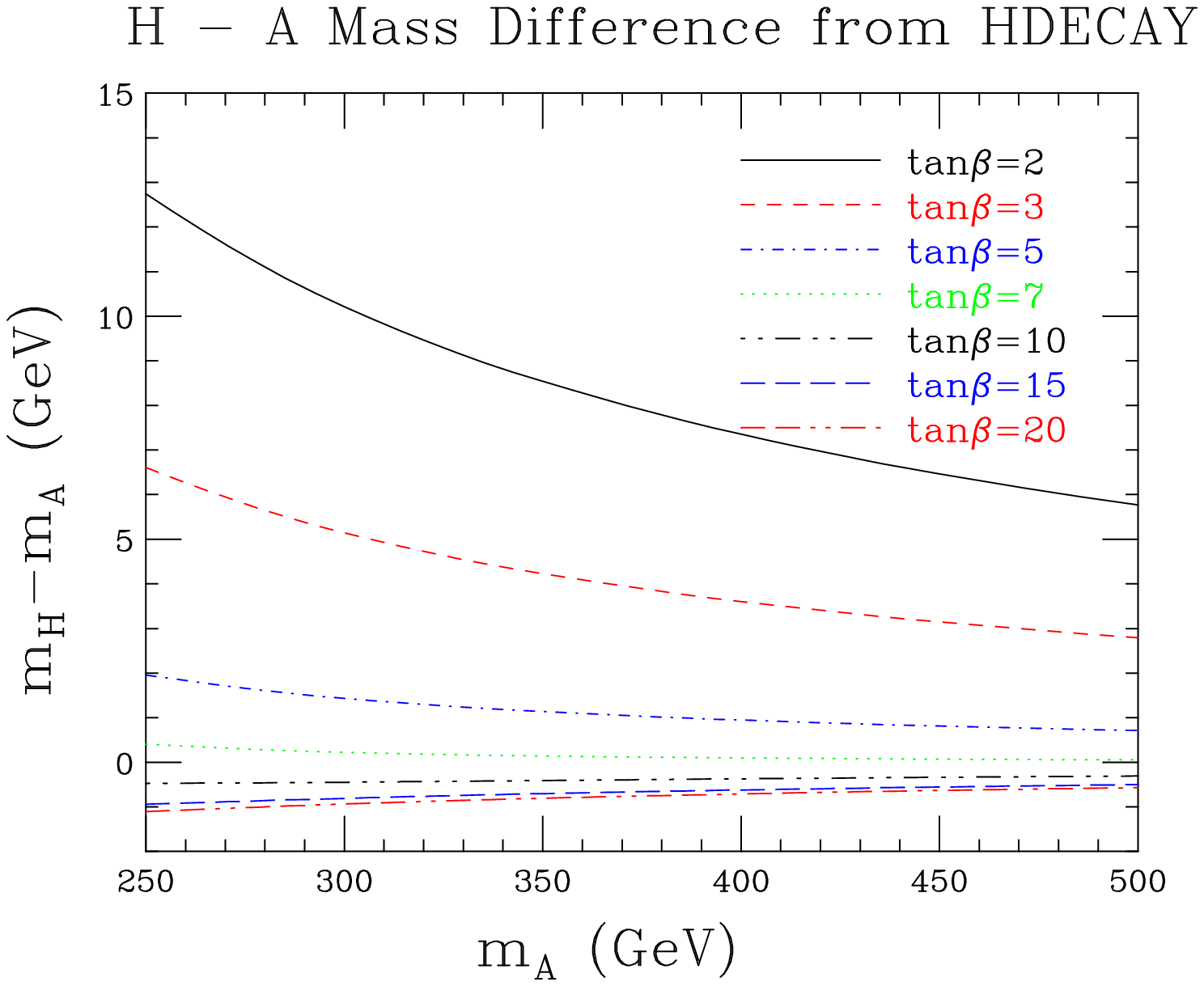,height=12cm}}
\caption[0]{We plot difference $\mhh-\mha$ as a function of $\mha$
for our standard set of $\tanb$ values. Results are those from HDECAY
for the earlier defined maximal mixing scenario with $\msusy=\mu=1\tev$.
Supersymmetric particle loops are neglected.  
}
\label{f:massdiff}
\end{figure}

The second important ingredient in understanding the nature of
$\hh,\ha$ signal is the degree to which they are degenerate in mass.
The degree of non-degeneracy is plotted
in Fig.~\ref{f:massdiff}. For $\tanb=2$ and 3, the mass differences
at lower $\mha$ are such that the $\hh$ and $\ha$ peaks would
remain substantially separated even after including $\sim 6\gev$
experimental mass resolution. However, starting
with $\tanb=5$, and for larger $\mha>2m_t$ in the $\tanb=2,3$ cases, the
mass difference is sufficiently small and their total widths
sufficiently large that 
after including experimental mass resolution there will be considerable
overlap between the $\hh$ and $\ha$ peaks. A centrally located 10 GeV bin
would pick up a large fraction of the $\hh$ and $\ha$ events..
The assumption that 50\% of the total number of $\hh$ and $\ha$
events fall into one 10 GeV bin centered on $\mha$
is thus an approximate way of taking
into account both the $6\gev$ experimental mass resolution, the 
few GeV total widths and
the non-degeneracy.  While 50\% is probably an overestimate
for $\tanb=2,3$ and lower $\mha$, it is not much of an overestimate
because, for these parameter cases, the $\ha$ signal is
much stronger than the $\hh$ signal in any case -- see
Fig.~\ref{f:sigeff}. The 50\% assumption is probably a conservative
approximation for $\tanb=5$ and above, and is probably only
a bit of an overestimate for $\tanb=3$ and $\mha>350\gev$.
A full simulation of both 
the $\hh$ and the $\ha$ peaks as a function of $\tanb$
and $\mha$ is required to do the job properly.
However, we have found that the existing Monte Carlo's seem to give
too large an experimental mass resolution. Further refinement
of the Monte Carlo's will be required before a complete simulation
will be possible.

Our full list of cuts is then as follows.
\bit
\item Only tracks and showers with $|\cos\theta|<0.9$ in the laboratory
frame are accepted.
\item Tracks have to have momentum greater than 200 MeV and showers
must have energy greater than 100 MeV.
\item We focus on the two most energetic jets in the event
(with jets defined using the Durham algorithm with $y_{\rm cut}<0.02$).
\item We require these two jets to be back-to-back in two dimensions
using the criteria $|p_i^1+p_i^2|<50\gev$ for $i=x,y$ (transverse
to the beam).
\item 
We require $|\cos\theta^*|<0.5$ where $\theta^*$ is the
angle of the jets relative to the beam direction 
in the two-jet center of mass.
As discussed, the alternative of $|\cos\theta|<0.5$ is not desirable
for retaining large luminosity at lower $E_{\gam\gam}$ in 
the broad band spectrum. It also does not significantly alter
the statistical significances for the peaked spectrum case. 
\eit
After the back-to-back and $\cos\theta^*$ cuts, about 95\%
of the events retained contain exactly two jets.

 Finally, we estimate the number of events 
with $\mha-5\gev\leq m_{2-{\rm jet}}\leq \mha+5\gev$ as follows.
As in the light Higgs study, we assume an efficiency of $70\%$
for double-tagging the two jets as $b\anti b$.
In addition to the reconstruction efficiency, which we find to be nearly
constant at 35\%, and the $b$-tagging efficiency of 70\%, 
we assume that only 50\% of the Higgs events fall within this 10 GeV bin.
In effect, these reconstruction, $b$-tagging and mass acceptance efficiencies
result in a net efficiency of 12.25\% for retaining Higgs events.
The efficiency for the $b\anti b$ background is much smaller due primarily to
the fact that the  reconstruction efficiency is far smaller than the
35\% that is applicable for the Higgs events.  This is due
primarily to the very forward/backward nature of the background
events as compared to the uniform distribution in $\cos\theta^*$
of the Higgs events. The $c\anti c$
background before $b$-tagging is substantially larger than
the $b\anti b$ background.  However, after double-tagging 
(we employ a probability of 3.5\% for double-tagging a
$c\anti c$ event as a $b\anti b$ event), the $b\anti b$ and $c\anti c$
backgrounds are comparable.

Higher order (NLO) corrections to the $J_z=0$ $c\anti c$ and $b\anti b$
backgrounds can be substantial.  However, the $J_z=\pm 2$ backgrounds
are so much larger (after our cuts, in particular the two-jet cuts)
that even if the $J_z=0$ background is increased by a factor
of 5 to 10 by the NLO corrections, the total background would increase by
only 5\% to 10\%.  For a more detailed discussion, see Appendix B. 

\begin{table}[ht]
\renewcommand{\arraystretch}{0.7}
  \begin{center}
\begin{tabular}[c]{|c|c|c|c|c|c|c|}

\hline $\mha(\gev)$ & 250 & 300 & 350 & 400 & 450 & 500 \\
\hline
$\tanb=2$ 
 & 121
 & 141
 & 54.6
 & 5.11
 & 1.60
 & 0.465
  \\
$\tanb=3$ 
 & 91.0
 & 110
 & 92.1
 & 10.8
 & 3.44
 & 0.790
 \\
$\tanb=5$ 
 & 52.0
 & 57.5
 & 94.2
 & 22.2
 & 7.59
 & 1.80
 \\
$\tanb=7$ 
 & 35.4
 & 34.1
 & 60.3
 & 24.8
 & 9.19
 & 2.26
 \\
$\tanb=10$ 
 & 27.6
 & 21.6
 & 31.6
 & 19.1
 & 7.57
 & 1.92
 \\
$\tanb=15$ 
&  35.8
&  21.3
&  17.2
&  12.7
&  5.15
&  1.30
 \\
$\tanb=20$ 
& 56.9
& 30.8
& 16.5
& 11.9
&  4.67
&  1.15
 \\
\hline
$B(b\anti b+c\anti c)$ 
 & 272
 & 90
 & 70
 & 13
 & 5
 & 1
  \\
\hline
    \end{tabular}
    \caption{We give net signal ($\hh\to b\anti b$ plus $\ha\to b\anti b$) 
and net background ($b\anti b+c\anti c$) rates
after cuts, assuming one $10^7$ sec year  of operation in polarization
configuration I. Background rates are those for a 10 GeV bin centered
on the given value of $\mha$. Signal rates are total rates before
restricting to the 10 GeV bin, but after tagging and acceptance efficiencies.}
    \label{sbratesi}
\vspace*{.2in}
\begin{tabular}[c]{|c|c|c|c|c|c|c|}
\hline $\mha(\gev)$ & 250 & 300 & 350 & 400 & 450 & 500 \\
\hline
$\tanb=2$ 
 & 38.8
 & 44.1
 & 24.2
 & 4.78
 & 5.79
 & 3.72
  \\
$\tanb=3$
 & 29.2
 & 34.3
 & 40.8
 & 10.1
 & 12.4
 & 8.05
  \\
$\tanb=5$ 
 & 16.7
 & 18.0
 & 41.7
 & 20.8
 & 27.5
 & 18.3
 \\
$\tanb=7$
 & 11.4
 & 10.7
 & 26.7
 & 23.2
 & 33.3
 & 23.0
  \\
$\tanb=10$  
 & 8.85
 & 6.78
 & 14.0
 & 17.9
 & 27.4
 & 19.5
 \\
$\tanb=15$ 
 & 11.5
 & 6.69
 & 7.61
 & 11.9
 & 18.7
 & 13.3
 \\
$\tanb=20$ 
& 18.2
& 9.65
& 7.30
& 11.1
& 16.9
& 11.7
 \\
\hline
$B(b\anti b+c\anti c)$
 & 555
 & 271
 & 130
 & 86
 & 8
 & 2
  \\
\hline
    \end{tabular}
    \caption{We give net signal ($\hh\to b\anti b$ plus $\ha\to b\anti b$) 
and net background ($b\anti b+c\anti c$) rates
after cuts, assuming one $10^7$ sec year of operation in polarization
configuration II. Background rates are those for a 10 GeV bin centered
on the given value of $\mha$. Signal rates are total rates before
restricting to the 10 GeV bin, but after tagging and acceptance efficiencies.
}
    \label{sbratesii}
  \end{center}
\end{table}

\begin{figure}[p!]
\leavevmode
\begin{center}
\epsfxsize=7in\epsffile{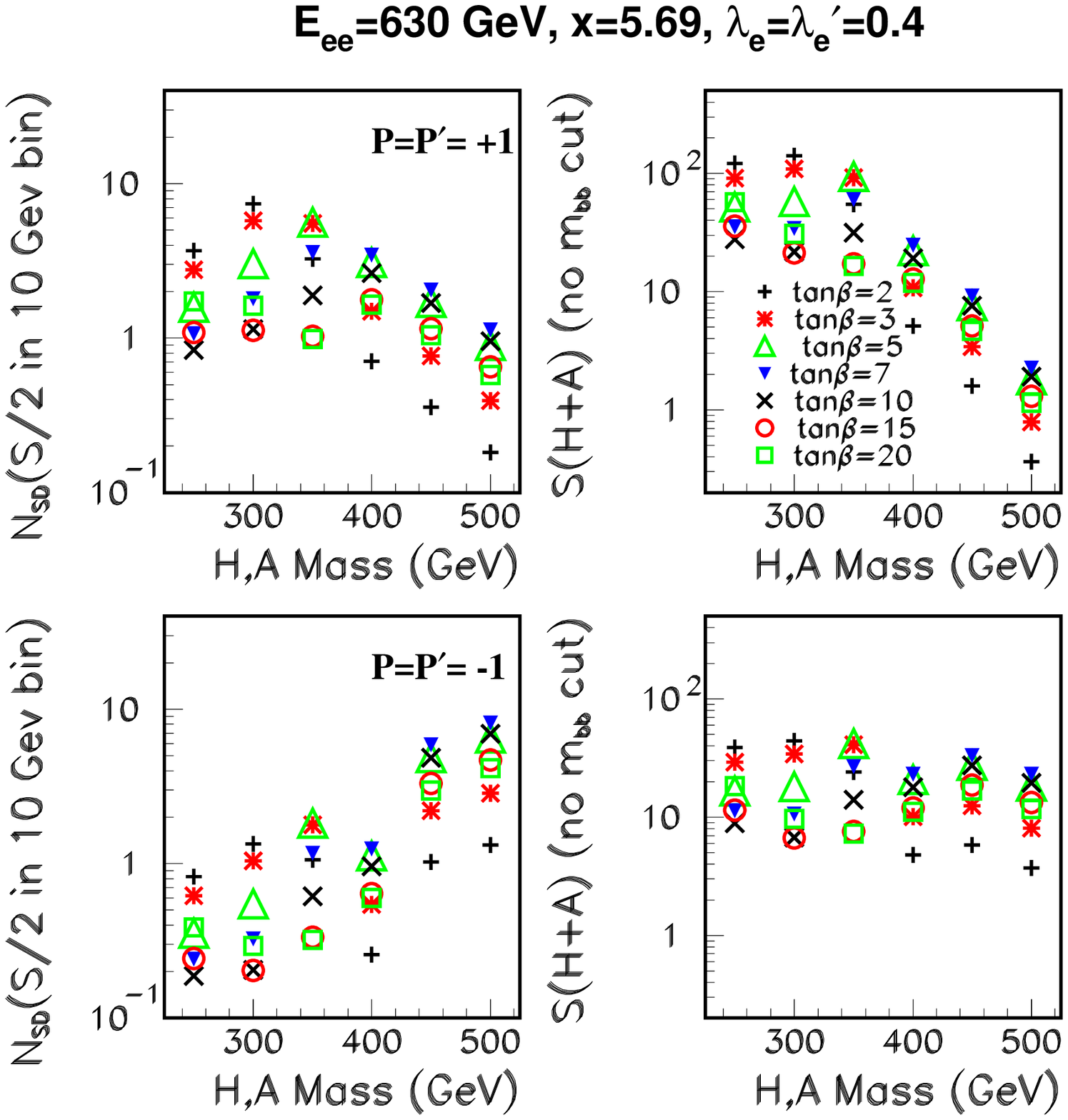}
\end{center}
\caption[0]{
For the luminosity spectra and  $\vev{\lam\lam'}$'s of \Fig{f:cainlums},
we plot in the upper (lower) right-hand windows the 
signal rates (without any $m_{\rm 2-jet}$ cut) for the various 
$[\mha,\tanb]$ cases considered assuming $\rts=630\gev$ operation
for one $10^7$ sec year (each) in the 
broad spectrum type-I (peaked spectrum type-II) configurations. 
In the upper (lower) left-hand windows we present
the corresponding statistical significances.
These are computed using the background rates 
obtained from our simulation (after cuts and tagging) for a 10 GeV
bin centered on the given $\mha$ assuming that 50\% of
the total number of Higgs events fall into that bin.  
}
\label{cainanalysis}
\end{figure}

\begin{figure}[h!]
\leavevmode
\begin{center}
\epsfxsize=7in\epsffile{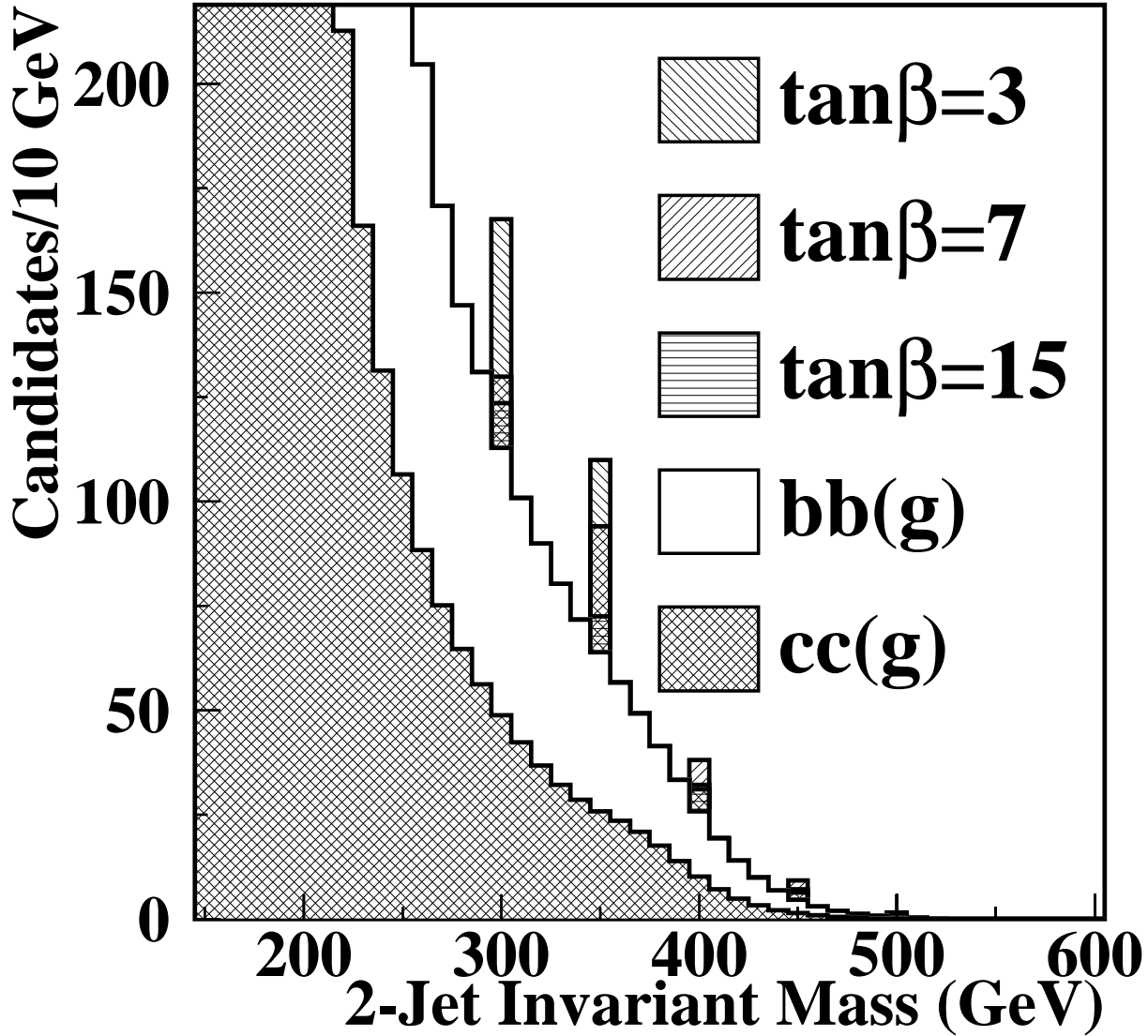}
\end{center}
\vspace*{-.4in}
\caption[0]{
For the luminosity spectra and  $\vev{\lam\lam'}$'s of \Fig{f:cainlums},
we illustrate the signal and background rates
for the various 
$[\mha,\tanb]$ cases considered assuming $\rts=630\gev$ and 
broad spectrum type-I  operation for one $10^7$ sec year. 
The signals shown assume that
50\% of the total number of signal events fall into the single 10 GeV bin
shown. Signals in the side bins are not shown.
Note that overlapping signal hatching types occur when a smaller signal
rate for one $\tanb$ value is drawn on top of a larger signal rate
for another $\tanb$ value. Such overlaps should not be confused
with the $c\anti c (g)$ background cross-hatching.
}
\label{signalploti}
\end{figure}
\begin{figure}[h!]
\leavevmode
\begin{center}
\epsfxsize=7in\epsffile{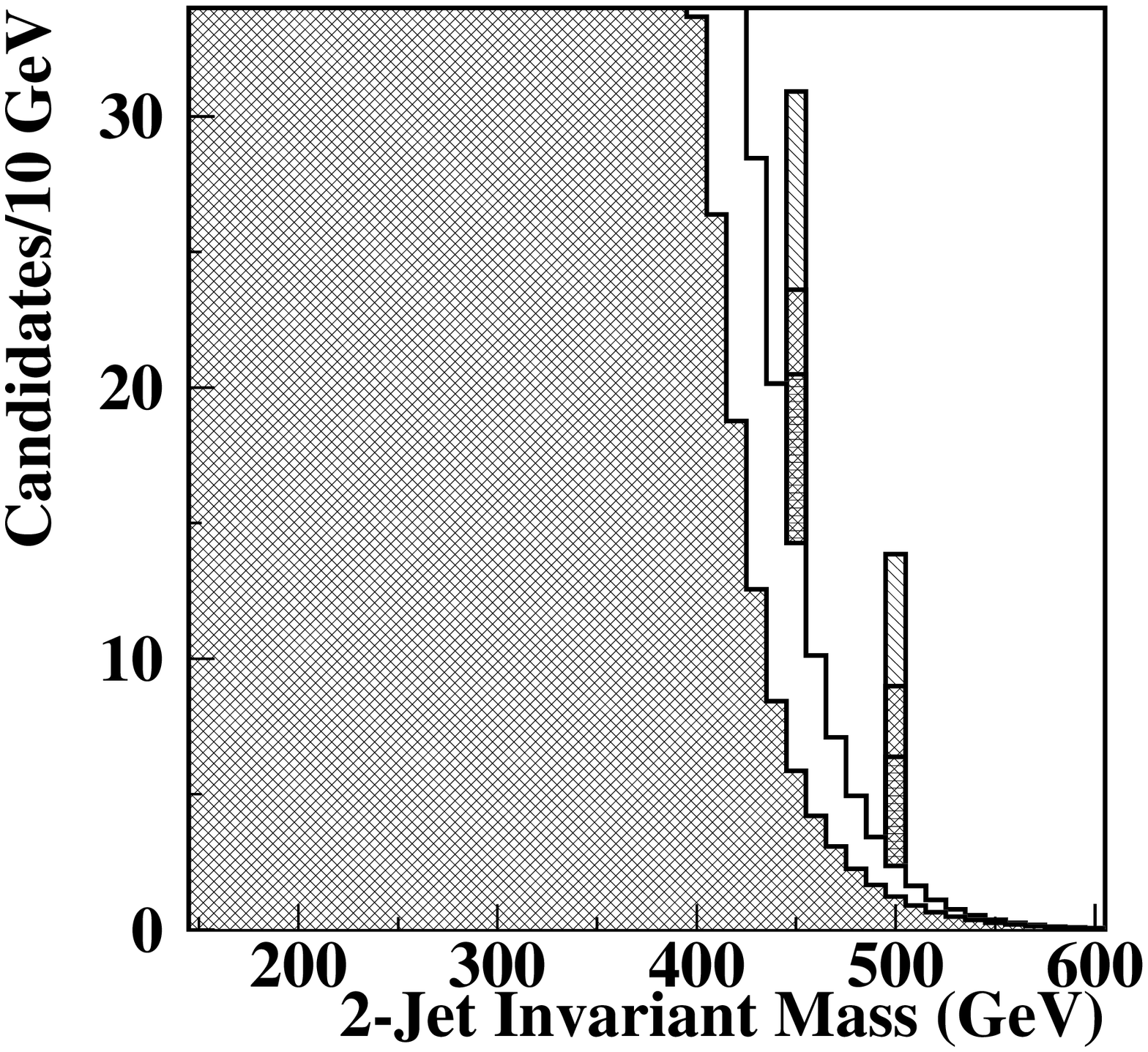}
\end{center}
\vspace*{-.4in}
\caption[0]{As in Fig.~\ref{signalploti} except for peaked
spectrum type-II operation.
}
\label{signalplotii}
\end{figure}

In Tables ~\ref{sbratesi} and \ref{sbratesii}, 
we tabulate signal and background rates
for the 42 $[\mha,\tanb]$ cases considered for polarization configurations
I and II, respectively.
These same net signal rates are also plotted in
the right-hand windows of Fig.~\ref{cainanalysis}. In the
left-hand windows of Fig.~\ref{cainanalysis} we plot 
the corresponding statistical significances
assuming that 50\% of the signal events 
fall into a 10 GeV bin centered on the given value of $\mha$.
As noted earlier, this width is meant to approximate the correct
result after allowing for the slight non-degeneracy between
$\mha$ and $\mhh$ (in the $\mha\gsim 250\gev$ region of interest)
and the expected experimental resolution of $\lsim 6\gev$
in the $250-500\gev$ mass region. There is an
important point as regards the rates
and results we give for $\mha=350\gev$.  
As can be seen from Fig.~\ref{f:sigeff}, the rates (especially
that for $\gam\gam\to \ha\to b\anti b$) will be very
sensitive to where exactly one is located relative to the $\mha=2\mt$
threshold for $t\anti t$ decay. We have deliberately run HDECAY
in such a way that our $\mha=350\gev$ point is actually slightly
above this threshold. This is because we are especially interested
in results starting with the $350\gev$ mass.  For points
just below our plotted points, the $\ha$ and, hence, net signal
is much stronger.

To illustrate the nature of these signals
relative to background, we show in Figs.~\ref{signalploti} and 
\ref{signalplotii}
the backgrounds as a function of 2-jet invariant mass with the
signals (including the 50\% factor and plotting only
the central 10 GeV bin) superimposed.  Results for the different
$\tanb$ cases and different spectra are shown. 
For all these computations, we have employed
the luminosities and polarizations plotted in
Fig.~\ref{f:cainlums}. We observe that many of the $[\mha,\tanb]$
cases considered will yield an observable $4\sigma$ signal.
Of course, we are most interested in our ability to cover the LHC wedge
in which the neutral $\hh,\ha$ Higgs bosons cannot be detected.

\begin{figure}[t!]
\leavevmode
\begin{center}
\epsfxsize=7in\epsffile{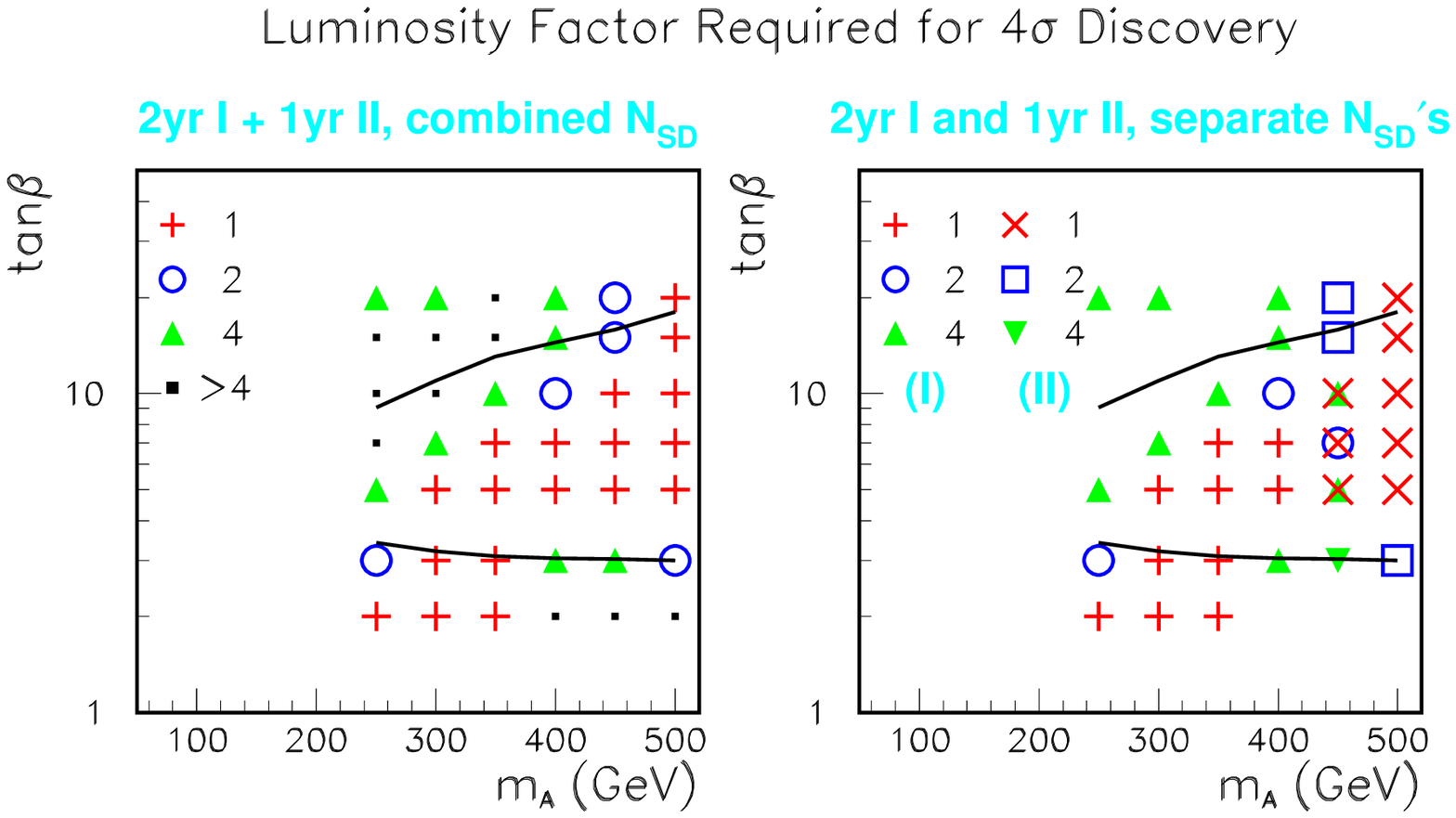}
\end{center}
\vspace*{-3.7in}
\caption[0]{Assuming a machine energy of $\rts=630\gev$,
we show the $[\mha,\tanb]$ points for which two $10^7$ sec year
of operation 
using the type-I $P\lam_e,P'\lam_e'>0$ polarization configuration
and one $10^7$ sec year of operation using the type-II $P\lam_e,P'\lam_e'<0$
configuration will yield $S/\sqrt{B}\geq4$.  In the left-hand window we
have combined results from the type-I and type-II running using
$S/\sqrt B=\sqrt{ S_I^2/B_I+S_{II}^2/B_{II}}$. In the right-hand
window, we show the separate results for $S_I/\sqrt{B_I}$ and 
$S_{II}/\sqrt{B_{II}}$. 
The solid curves indicate the wedge region from the LHC plot
of Fig.~\ref{f:atlasmssm} --- the lower black curve is that from the
LEP (maximal-mixing) limits, but is
somewhat higher than that currently claimed by the 
LEP Electroweak Working Group,
while the upper solid curve is that above which $\hh,\ha\to\tauptaum$
can be directly detected at the LHC.  Also shown are the additional points for
which a $4\sigma$ signal level is achieved if the total
luminosity is doubled or quadrupled (the `2' and `4' symbol cases)
relative to the one-year luminosities we are employing. 
(The small black squares in the left-hand window indicate the additional
points sampled for which even a luminosity increase of a factor
of 4 for both types of running does not yield a $4\sigma$ signal.)
Such luminosity
increases could be achieved for some combination of longer running time and/or
improved technical designs. For example, the factor of `2' 
results probably roughly apply to TESLA.   
Cuts and procedures are as described in the text.
}
\label{wedgeplot}
\end{figure}

Our ability to `cover' the wedge is illustrated in Fig.~\ref{wedgeplot}.
At $\mha=250\gev$, cases with $\tanb=3,5,7$ fall into the LHC wedge.
At $\mha=300,350\gev$, cases with $\tanb=3,5,7,10$ fall into the LHC wedge.  
At $\mha=400,450,500\gev$, cases with $\tanb=3,5,7,10,15$ fall into
the LHC wedge.  Altogether we have considered 26 points that
are in the LHC wedge.
Very roughly, after running
for two $10^7$ sec years using the broad type-I spectrum 
it will be possible to detect a $4\sigma$ signal for about 7 of
the 13 $[\mha,\tanb]$ cases with $\mha=300,350,400\gev$ in the LHC wedge.
(We do not include $\mha=250\gev$ in our counting since
$\hh\ha$ pair production would certainly be observable for
$\mha=250\gev$ for $\rts=630\gev$.)
These are cases with low to moderate $\tanb$.
After running for one $10^7$ sec year using the type-II peaked spectrum,
we predict a $4\sigma$ signal for  7 of the 10 cases with
$\mha=450,500\gev$ in the LHC wedge.
These are cases with higher $\tanb$.
If results for these 2+1 years of operation are combined,
the statistical significance at a given parameter space point
is only slightly improved (broad/I and peaked/II signals do not
overlap much). In all, we would be able to detect
a $\geq 4\sigma$ Higgs signal for $15/23 \gsim 65\%$ 
of the wedge cases considered.
Obviously, further improvements in luminosity or mass resolution would
be helpful for guaranteeing complete coverage of the wedge region.
If both type-I and type-II luminosities are doubled, the 15/23 becomes 
18/23. Further, for $\rts=630\gev$ it is very probable that
one could see $\hh\ha$ pair production  for $\mha=300\gev$, in which case 
$\gam\gam$ collision operation with factor `2'
type accumulated luminosity would allow detection of $\gam\gam\to\hh,\ha$ 
throughout most of the remaining portion of the wedge in which they cannot
be seen by other means.
Finally, we note that other channels than $b\anti b$ are available.
At low $\tanb$, we expect that 
the $\hl\hl$ channel for the $\hh$ and the $Z\hl$ channel for
$\ha$ will provide observable signals for the remaining wedge points
with $\mha\leq 2\mt=350\gev$. The $t\anti t$ channels might provide
further confirmation for $b\anti b$ signals for wedge points with 
$\mha>450\gev$. The single most difficult wedge point is $\mha=400,\tanb=15$,
which is at the edge of the LHC wedge region.

It is important to realize that
if the LHC was able to detect the $\hpm$ Higgs bosons
in some portion of the wedge region, for example  
using the $\hpm\to \tau^\pm\nu_\tau$ decay mode, 
a reasonably accurate $\sim \pm 25\gev$ determination of $\mhpm$
would emerge. If studies of the SUSY particles
indicate that the MSSM is the correct theory, then 
we would employ the model prediction that $\mha\sim\mhh\sim \mhpm$
and run the $\gam\gam$ collider with type-II peaked 
spectrum at the $\rts$ value yielding $E_{\rm peak}\sim \mhpm$.
Unfortunately, the latest simulation results, as represented
in Fig.~\ref{f:atlasmssm}, indicate that the $\hpm$
can only be detected if $\tanb$ is larger than the upper boundary
of the wedge region.  However, these studies are being continually
refined. Ultimately, the actual situation will only
be known once the LHC starts operation.

We conclude that a $\gam\gam$ collider can
provide Higgs signals for the $\hh$ and $\ha$
over a possibly crucial portion of  parameter space in which
the LHC and direct $\epem$ collisions at a LC will not be able
to detect these Higgs bosons or their $\hpm$ partners. 
Indeed, the $\gam\gam$ collider
is very complementary to the LHC and $\epem$ LC operation
as regards the portion of $[\mha,\tanb]$ parameter space 
over which a signal for the heavy MSSM Higgs bosons can be detected.

\begin{table}[ht]
\large
\renewcommand{\arraystretch}{0.7}
  \begin{center}
\begin{tabular}[c]{|c|c|c|c|c|c|c|}
\hline $\mha(\gev)$ & 250 & 300 & 350 & 400 & 450 & 500 \\
\hline
$\tanb=2$ 
 & 0.51
 & 0.34
 & 0.20
 & 0.66
 & 0.46
 & 0.48
  \\
$\tanb=3$ 
 & 0.51
 & 0.27
 & $-$
 & 0.45
 & 0.30
 & 0.32
 \\
$\tanb=5$ 
 & 0.71
 & 0.34
 & 0.19
 & $-$
 & 0.56
 & 0.55
 \\
$\tanb=7$ 
 & $-$
 & 0.66
 & 0.23
 & 0.62
 & 0.67
 & 0.87
 \\
$\tanb=10$ 
 & $-$
 & $-$
 & 0.50
 & 0.64
 & 0.46
 & 0.53
 \\ 
$\tanb=15$ 
&  0.46
&  0.67
&  $-$
&  $-$
&  $-$
&  $-$
\\
\hline
    \end{tabular}
    \caption{We give the rough error for $\tanb$ based on
measuring a certain $\gam\gam\to \hh,\ha\to b\anti b$ rate
associated with Higgs discovery in the wedge region. These 
errors assume two years of operation in broad spectrum mode  
and one year of operation in peaked spectrum mode at $\rts=630\gev$.
The $-$'s indicate $[\mha,\tanb]$ cases for which the error
exceeds 100\%.
The errors are computed as described in the text.
Because of the finite difference approach, results are not
presented for $\tanb=20$, but errors there would be large.
}
    \label{tanberrors}
  \end{center}
\end{table}

If a $\hh,\ha$ signal is detected in the wedge region, one will
of course, reset the machine energy so that $E_{\rm peak}=\mha$
and proceed to obtain a highly accurate determination of the
$\gam\gam\to \hh,\ha,\to b\anti b$ rates and rates in other
channels.  These rates will provide valuable information about
SUSY parameters, including $\tanb$.  In fact, even before performing
this very targeted study, a rough determination of $\tanb$ is likely
to be possible just from the data associated with the initial discovery.
in Table~\ref{tanberrors}, we give those $[\mha,\tanb]$ points
and the approximate fractional error 
for $\tanb$ for those points at which this error would be below 100\%.
The finite difference approximation we employ is the following:
\bit
\item
We first compute the error $\delta[\half(S_I+S_{II})]=\sqrt{\half(S_I+S_{II})+(B_I+B_{II})}$, where $\half$ comes from the fact that we assume
that one-half of the signal events will fall into a 10 GeV bin
in the reconstructed 2-jet invariant mass and the $I$ and $II$
subscripts refer to the $S$ and $B$ rates for type-I and type-II
spectra, respectively.
\item
We estimated the sensitivity of $\half(S_I+S_{II})$ to $\tanb$ by computing
\beq
{\Delta \half(S_I+S_{II})(\tanb)\over \Delta\tanb}= \half{(S_I+S_{II})(\tanb+\Delta\tanb)-(S_I+S_{II})(\tanb)\over \Delta\tanb}
\eeq
using the $\tanb$ values of $2,3,5,7,10,15$ and corresponding $\Delta\tanb$
values of $1,2,2,3,5,5$.
\item The fractional error on $\tanb$ is then approximated as 
\beq
{\delta\tanb\over \tanb}\sim {\delta[\half(S_I+S_{II})]\over 
{\Delta \half(S_I+S_{II})(\tanb)\over \Delta\tanb}\tanb}\,.
\eeq
\eit
While the resulting ($1\sigma$) errors are not exactly small, this
determination can be fruitfully combined with other $\tanb$ determinations,
especially for the higher $\tanb$ cases where the other techniques
for determining $\tanb$ also have rather substantial errors.
More importantly, these results show clearly that a dedicated
measurement of the $\gam\gam\to \hh,\ha\to b\anti b$ rate
and the rates in other channels 
($\hh\to\hl\hl$, $\ha\to Z\hl$, $\hh,\ha\to t\anti t$)
are likely to yield a rather high precision determination of $\tanb$
after several years of optimized operation.

\begin{table}[ht]
\renewcommand{\arraystretch}{0.7}
  \begin{center}
\begin{tabular}[c]{|c|c|c|c|c|c|c|}
\hline $\mha(\gev)$ & 250 & 300 & 350 & 400 & 450 & 500 \\
\hline
$\tanb=2$ 
 & 30.1
 & 38.2
 & 164
 & 7.33
 & 0.987
 & 0
  \\
$\tanb=3$
 & 20.7
 & 26.7
 & 122
 & 14.8
 & 2.05
 & 0
  \\
$\tanb=5$ 
 & 11.0
 & 13.4
 & 58.9
 & 29.3
 & 4.45
 & 0
 \\
$\tanb=7$
 & 7.24
 & 7.84
 & 31.6
 & 31.6
 & 5.32
 & 0
  \\
$\tanb=10$  
 & 5.45
 & 4.75
 & 15.6
 & 23.4
 & 4.30
 & 0
 \\
$\tanb=15$ 
 & 6.67
 & 4.23
 & 7.87
 & 14.8
 & 2.80
 & 0
 \\
$\tanb=20$ 
& 10.4
& 5.79
& 6.85
& 12.9
& 2.40
& 0
 \\
\hline
$B(b\anti b+c\anti c)$
 & 620
 & 234
 & 94.0
 & 6.18
 & 0.46
 & 0.04
  \\
\hline
    \end{tabular}
    \caption{We give net signal ($\hh\to b\anti b$ plus $\ha\to b\anti b$) 
and net background ($b\anti b+c\anti c$) rates
after cuts, assuming one $10^7$ sec year of operation at $\rts=535\gev$
in polarization
configuration II. Background rates are those for a 10 GeV bin centered
on the given value of $\mha$. Signal rates are total rates before
restricting to the 10 GeV bin, but after tagging and acceptance efficiencies.
}
    \label{sbrates400ii}
  \end{center}
\end{table}

\begin{figure}[h!]
\leavevmode
\begin{center}
\epsfxsize=7in\epsffile{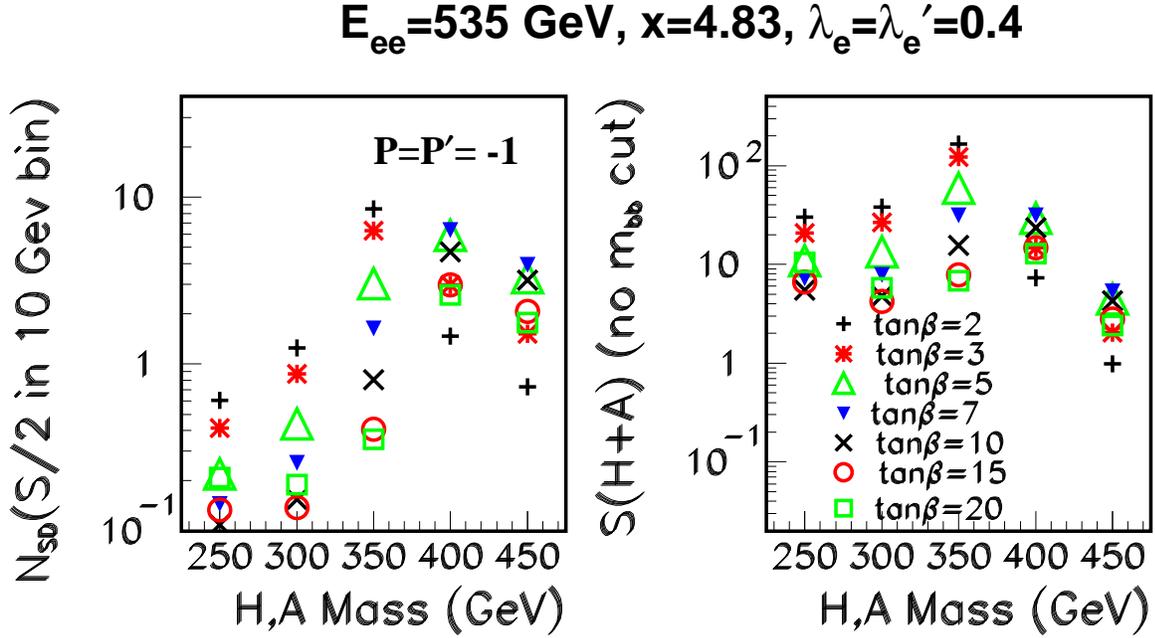}
\end{center}
\vskip -3.5in
\caption[0]{
For the luminosity spectra and  $\vev{\lam\lam'}$'s obtained
from CAIN for $\rts=535\gev$, $x=4.83$, and type-II (peaked)
spectrum, we plot in the right-hand window the 
signal rates (without any $m_{\rm 2-jet}$ cut) for the various 
$[\mha,\tanb]$ cases considered, assuming one $10^7$ sec year
of operation. In the left-hand window we present
the corresponding statistical significances.
These are computed using the background rates 
obtained from our simulation (after cuts and tagging) for a 10 GeV
bin centered on the given $\mha$ assuming that 50\% of
the total number of Higgs events fall into that bin.  
}
\label{cainanalysis400}
\end{figure}

\begin{figure}[h!]
\leavevmode
\begin{center}
\epsfxsize=7in\epsffile{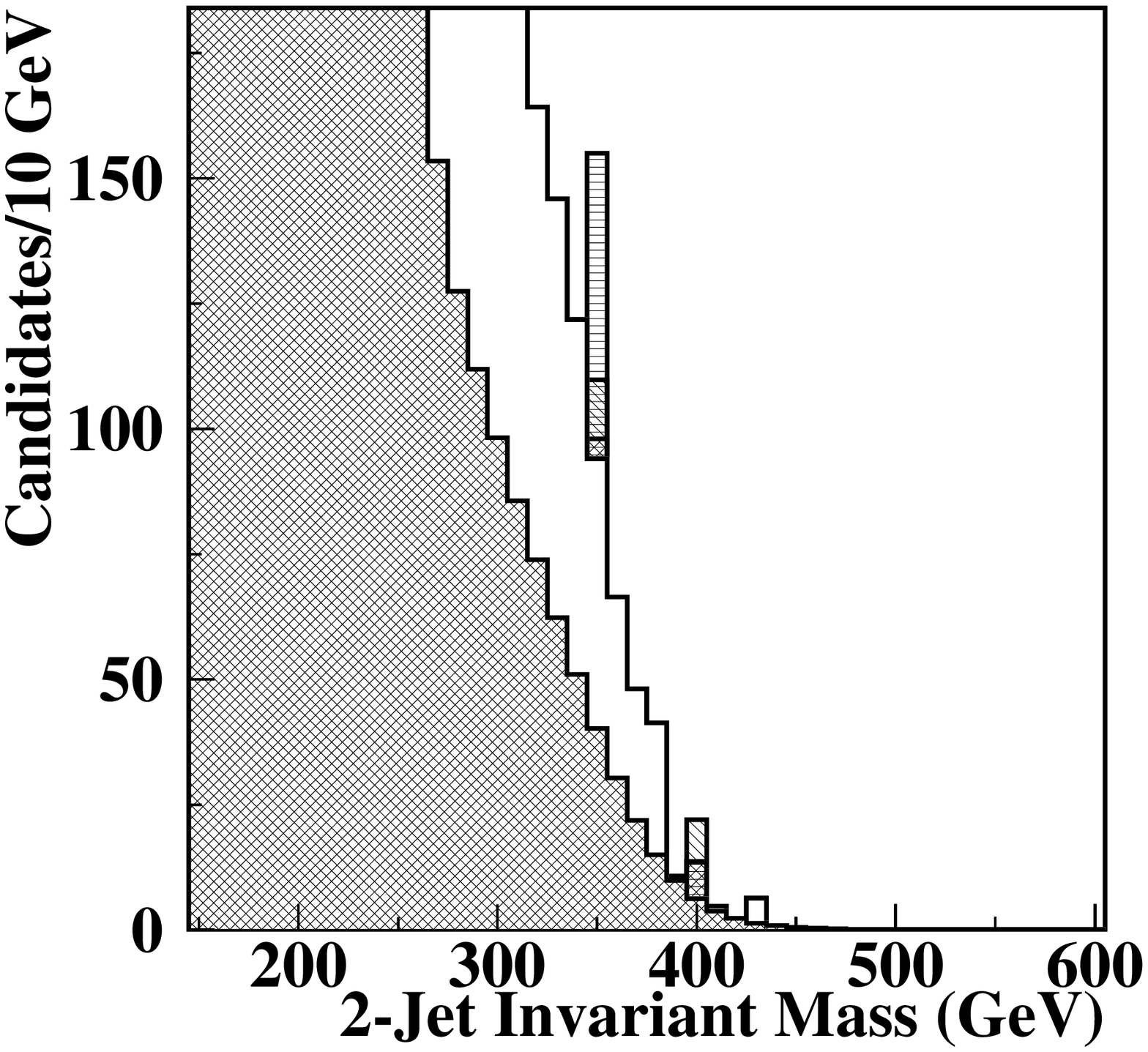}
\end{center}
\vspace*{-.4in}
\caption[0]{Signals for $\tanb=3,7,15$ and $\mha=350,400\gev$
as in Fig.~\ref{signalploti} except for peaked
spectrum type-II operation at $\rts=535\gev$, $x=4.83$.
}
\label{signalplot400ii}
\end{figure}

We now turn to a discussion of how the above running scenario
(2 years with broad spectrum and 1 year with peaked spectrum)
compares to running part of the time with a (type-II) spectrum peaked
at $E_{\gam\gam}=500\gev$ and part of the time with a spectrum
peaked at $E_{\gam\gam}=400\gev$ ($\rts=630\gev$, $x=5.69$ and $\rts=535\gev$,
$x=4.83$, respectively, for laser wavelength $\lambda=1.054~\mu$m).
We denote these two cases by `500' and `400', respectively. 
In the 400 case, we have followed exactly the same procedures
as in the 500 case, using CAIN to generate the luminosity spectra
and corresponding $\vev{\lam\lam'}$ 
and then using these to compute signal and background rates
in the $b\anti b$ final state, assuming running for one $10^7$ sec
year. These rates are tabulated in Table~\ref{sbrates400ii}.
The signal rates are also plotted in Fig.~\ref{cainanalysis400}
along with the corresponding statistical significances,
assuming that 50\% of the signal events fall into one 10 GeV
bin centered on $\mha$. Typical signals relative to background
for $\mha=350\gev$ and $400\gev$ and $\tanb=3$, $7,$ and $15$
are illustrated in Fig.~\ref{signalplot400ii}.
We should note that the $S/\sqrt B$ values are not very good
indicators of discovery potential at $\mha=450\gev$ because
of the very small numbers of $S$ and $B$ events.

\begin{figure}[p!]
\leavevmode
\begin{center}
\epsfxsize=7in\epsffile{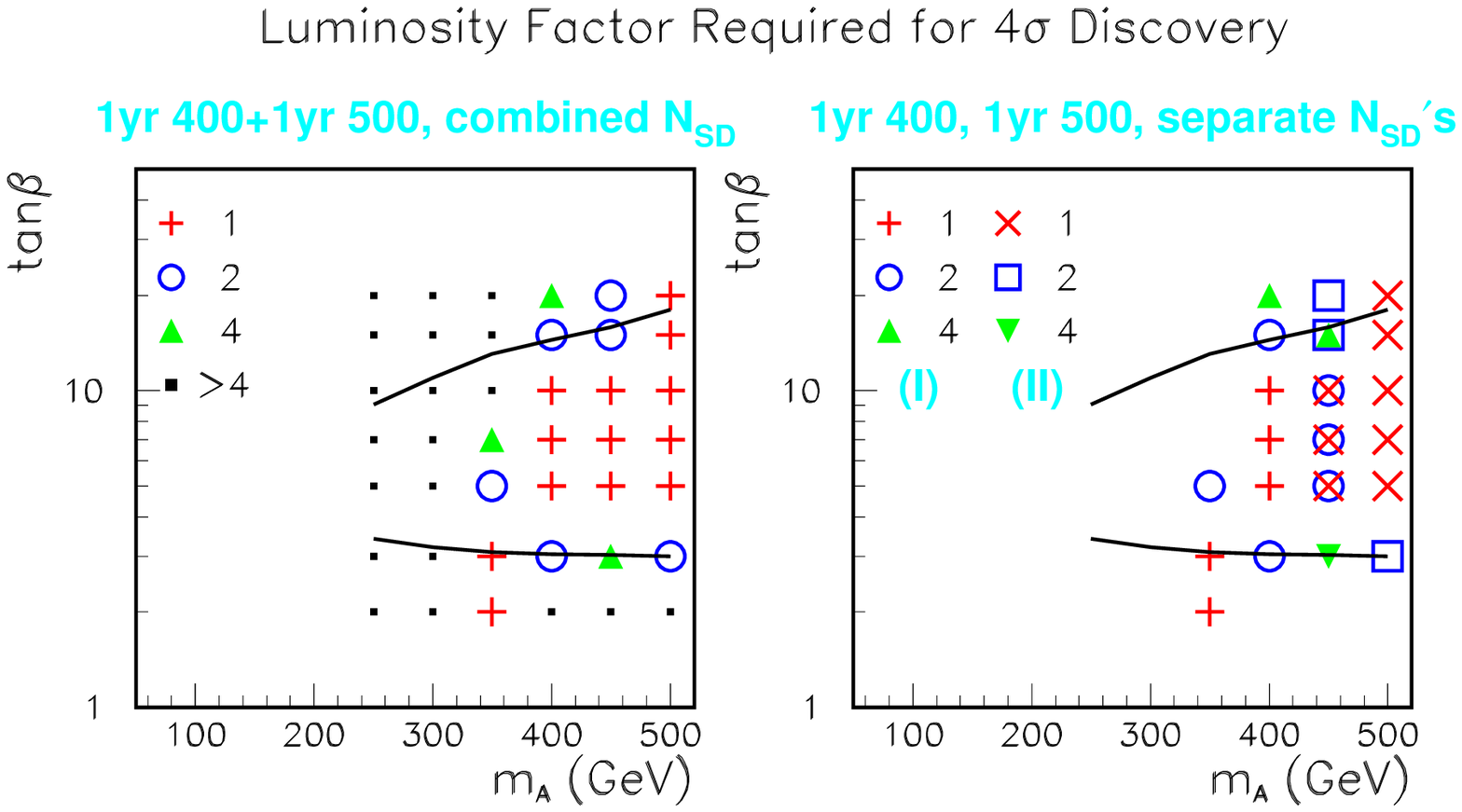}
\vskip -3.5in
\epsfxsize=7in\epsffile{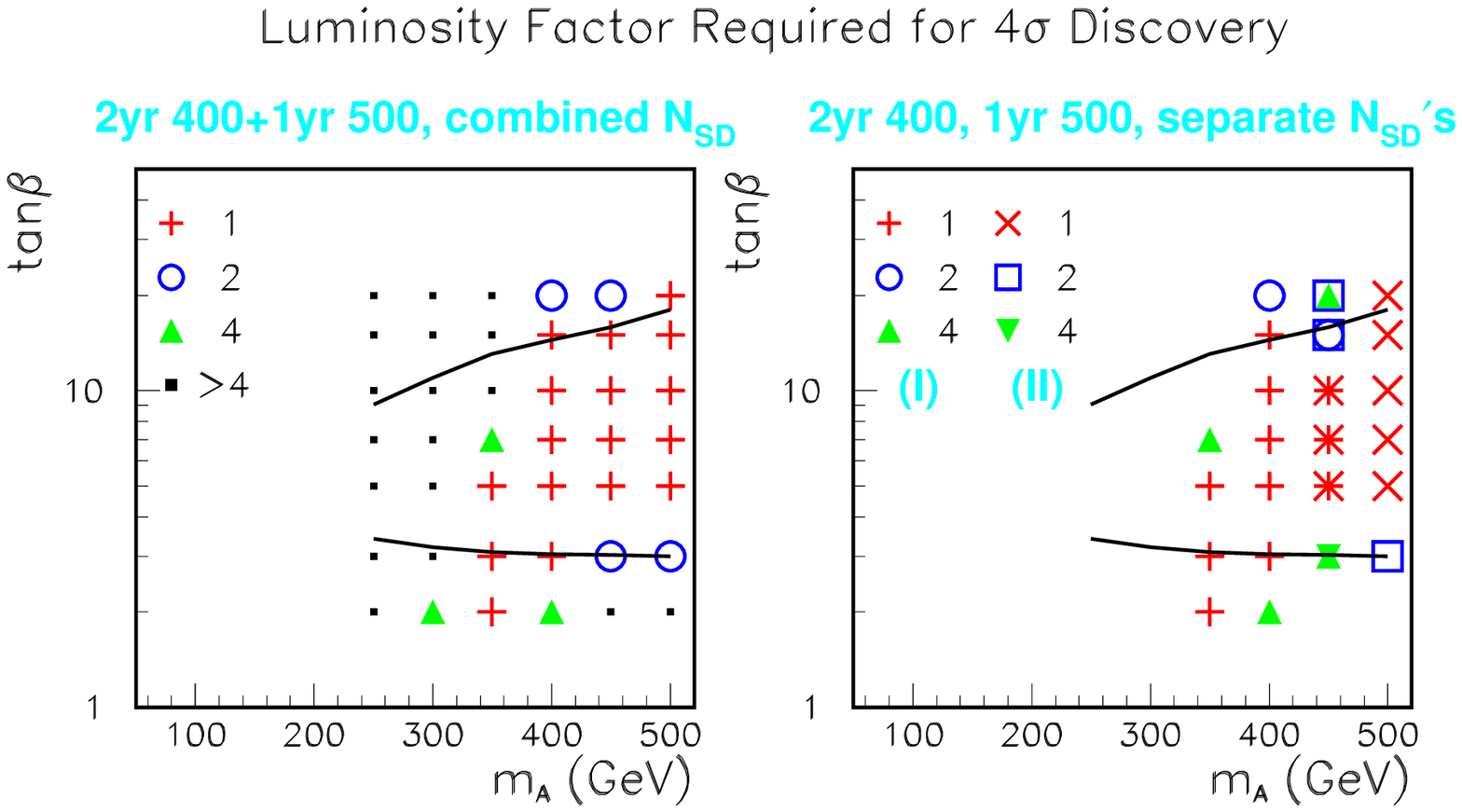}
\end{center}
\vspace*{-3.7in}
\caption[0]{In the upper windows, 
we plot the points in $[\mha,\tanb]$ parameter space
for which a $>4\sigma$ signal is obtained after 
one $10^7$ sec year
of operation at $\rts=535\gev$ and one year of operation at $\rts=630\gev$,
using type-II peaked spectrum in both cases.
In the left-hand window we
have combined results from the 400 and 500 running using
$S/\sqrt B=\sqrt{ S_{400}^2/B_{400}+S_{500}^2/B_{500}}$. In the right-hand
window, we show the separate results for $S_{400}/\sqrt{B_{400}}$ and 
$S_{500}/\sqrt{B_{500}}$. The notation and the solid curves  
outlining the LHC wedge are as specified in the caption for
Fig.~\ref{wedgeplot}. Exactly the same plots are presented
in the lower windows assuming two years of operation
at $\rts=535\gev$ and one year of operation at $\rts=630\gev$.
}
\label{wedgeplots400and500}
\end{figure}

The above results show that the 1-year 400 (type-II) 
plus 1-year 500 (type-II) option
gives better signals at $\mha=400\gev$  than does
the 2-year (type-I) 500 plus 1-year (type-II) 500 option,
but much worse signals at $\mha=300\gev$ and $350\gev$.  Going
to 2-year 400 (type-II) plus 1-year 500 (type-II) still does not provide
as good coverage of the wedge in an overall sense as the
2-year (type-I) 500 plus 1-year (type-II) 500 option.
We also expect, but have not explicitly performed the necessary study,
that 1-year 350 (type-II)~\footnote{As before, the `350' label
means operation at a $\rts$ such that the type-II spectrum
peaks at $E_{\gam\gam}=350\gev$.} 
plus 1-year 400 (type-II) plus 1-year 500 (type-II) operation,
would do a better job for $\mha\gsim 350\gev$ than the
2-year (type-I) 500 plus 1-year (type-II) 500 option, but would not
provide reliable signals in the wedge region for $\mha\lsim 325\gev$.

The ability to obtain a $>4\sigma$ signal in
nearly all of the $\mha\gsim 350\gev$ wedge 
using the 2-year (type-I) 500 plus 1-year (type-II) 500 option
is important since it is likely that the $\gam\gam$ collider will be run
at maximum energy for other physics reasons. 
Thus, if no signals for the $\hh$, $\ha$, {\it and} $\hpm$ are detected
at the LHC, we believe the optimal procedure 
at the $\gam\gam$ collider for the combined purposes
of discovering the $\hh,\ha$  Higgs bosons and pursuing other physics
studies (supersymmetric particle production in particular) will be
operation part time with type-I and  part time with type-II $\gam\gam$
luminosity spectra (roughly in the ratio 2:1).

Finally, we make a few remarks regarding the ability to detect
the $\hh,\ha$ for $\tanb$ values for which the LHC {\it would}
already have detected a signal.  Precision studies of 
the $\gam\gam\to\hh,\ha\to b\anti b$ rate (and rates in other channels
as well) would be an important source of information and cross
checks because of the many different types of particles in the MSSM that
potentially contribute to the $\gam\gam\to \hh,\ha$ couplings.
Fig.~\ref{modelcomps} shows that the minimum rate
in the $b\anti b$ final state occurs at $\tanb\sim 15$ when
$\mha\sim 250\gev$ (and also, though not plotted, $\mha\sim 300\gev$)
and at $\tanb\sim 20$ when $\mha\geq 350\gev$.  Thus, the $\gam\gam$
signals are actually weakest precisely in the upper part of
the wedge region and somewhat beyond.
Starting with $\tanb$ values sufficiently far above the wedge region, 
the signals become
stronger and stronger as $\tanb$ increases, asymptotically rising
as $\tan^2\beta$, but rising more like $\tanb$ in the $\tanb=30-50$ range.
Thus, if other physics studies force $\gam\gam$ running at maximal $\rts$,
it is quite possible to nonetheless have a strong signal for the $\hh,\ha$
if $\tanb$ is large enough that they are seen at the LHC.
 
\section{A decoupled light \boldmath $\ha$ of a general 2HDM}

As noted earlier, it is possible to construct a general two-Higgs-doublet
model that is completely consistent with precision electroweak constraints
in which the only Higgs boson that is light has no $WW/ZZ$ couplings
\cite{Chankowski:2000an}. 
(The particular models considered here are those constructed
in the context of a CP-conserving type-II 2HDM.)
This light Higgs could be either the $\ha$
or the $\hl$ (but with 2HDM parameters
chosen so that there is no $\hl\to WW,ZZ$ couplings). 
Here, we will study the case of a light $\ha$, since it (and
not a light $\hl$) could play a role in explaining the possible
discrepancy of the anomalous magnetic moment of the muon with the SM prediction
\cite{Cheung:2001hz}.~\footnote{In order for a light $\ha$ to
be the entire source of the originally published deviation in
$a_\mu$ large $\tanb$ is required~\cite{Cheung:2001hz}, sufficiently
large that LHC and/or LC detection would be probable. However,
recent improvements in the theoretical predictions for $a_\mu$
suggest that the $a_\mu$ deviation could be smaller than
originally thought. In this case, or if other mechanisms
contribute, the scenario we focus on of a moderately light $\ha$
and moderate $\tanb$ could be very relevant.}
As discussed in \cite{Chankowski:2000an},
the precision electroweak constraints imply that if the $\ha$
is light and the other Higgs bosons are heavy, then 
the couplings of the $\hl$ must be SM-like. Further, perturbativity
implies that the $\hl$ should not be heavier than about $1\tev$.
We would then be faced with a very unexpected scenario. The
LHC would detect the heavy SM-like $\hl$ and 
no supersymmetric particles would be discovered. The
precision electroweak
constraints (which naively require a very light $\hsm$
in the absence of additional physics) would demand
the existence of additional
contributions to $\Delta T$ (as could be verified by Giga-$Z$
operation of the LC). The general 2HDM  provides the additional
$\Delta T$ contribution via a mass splitting between the $\hpm$
and the $\hh$ (both of which would have masses of order a TeV). 
Detection of the light $\ha$ possibly needed
to explain the  $a_\mu$ deviation would be crucial
in order to learn of the existence of the extended Higgs sector. 

\begin{figure}[h!]
\leavevmode
\begin{center}
\epsfxsize=7in\epsffile{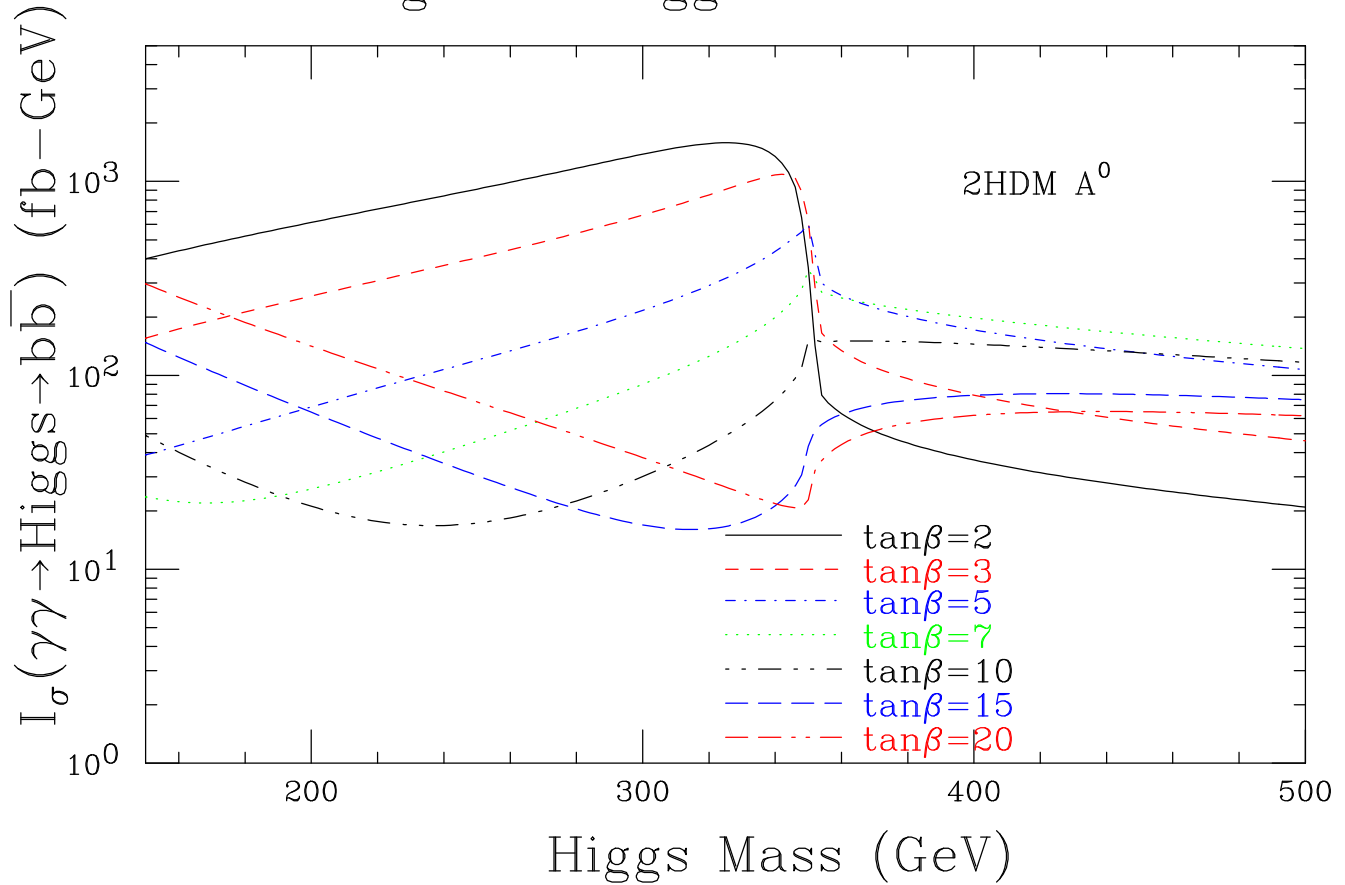}
\end{center}
\caption[0]{We consider a general 2HDM and 
plot the integrated $\ha$ Higgs cross section
 $I_\sigma$, 
as defined in Eq.~(\ref{ngamgam}), as a function of $\mha$, for a variety
of $\tanb$ values. We have assumed that all other Higgs bosons of
the 2HDM have masses of order $1\tev$.
}
\label{sigA2HDMbb}
\end{figure}

As for the $\hh$ and $\ha$ of the MSSM,
discovery of an $\ha$ with mass above 200 to 250 GeV could be difficult. 
If $\tanb$ is chosen in the moderate range, the $\ha$
will not be seen in $\epem\to \ha b\anti b$ or $\epem\to \ha t\anti t$
production \cite{Grzadkowski:2000wj,Chankowski:2000an}. 
Discovery of the $\ha$ would also be impossible
at the LHC in a wedge of parameter space 
very similar to (but somewhat more extended
in $\tanb$, assuming no overlapping resonance
with the opposite CP) than that found in the MSSM case.
Finally, such an $\ha$ can
only be seen in $\epem\to Z\ha\ha$ or $\epem\to \nu\anti\nu\ha\ha$ production
(through its quartic $WW\ha\ha,ZZ\ha\ha$ couplings) if
$\mha< 150\gev$ ($250\gev$) for $\rts=500\gev$ ($800\gev$).
 Thus, the ability to detect
the $\ha$ in a moderate $\tanb$ wedge beginning at
$\mha\gsim 250\gev$ using $\gam\gam$ collisions
might turn out to be of great importance.
In exploring this ability, we follow procedures closely analogous
to the MSSM study.

First, we need the integrated cross
section, $I_\sigma$ --- see Eq.~(\ref{ngamgam}). Results  
are presented in \Fig{sigA2HDMbb}. In computing $I_\sigma$ for
the 2HDM $\ha$, we assume that all the other 2HDM Higgs bosons
have mass of $1\tev$. The main difference with the MSSM is that
since we take the $\hl$ and $\hh$ to be heavy, there are no overlapping  
signal events from a 2nd Higgs boson. However, 
for $\mha<2\mt$ this loss of overlapping signal
is somewhat compensated by increased $\ha\to b\anti b$ branching ratio
due to the absence of $\ha\to Z\hl$ decays in the 
large-$\mhl$ scenario being envisioned. 

\begin{figure}[p!]
\leavevmode
\begin{center}
\epsfxsize=7in\epsffile{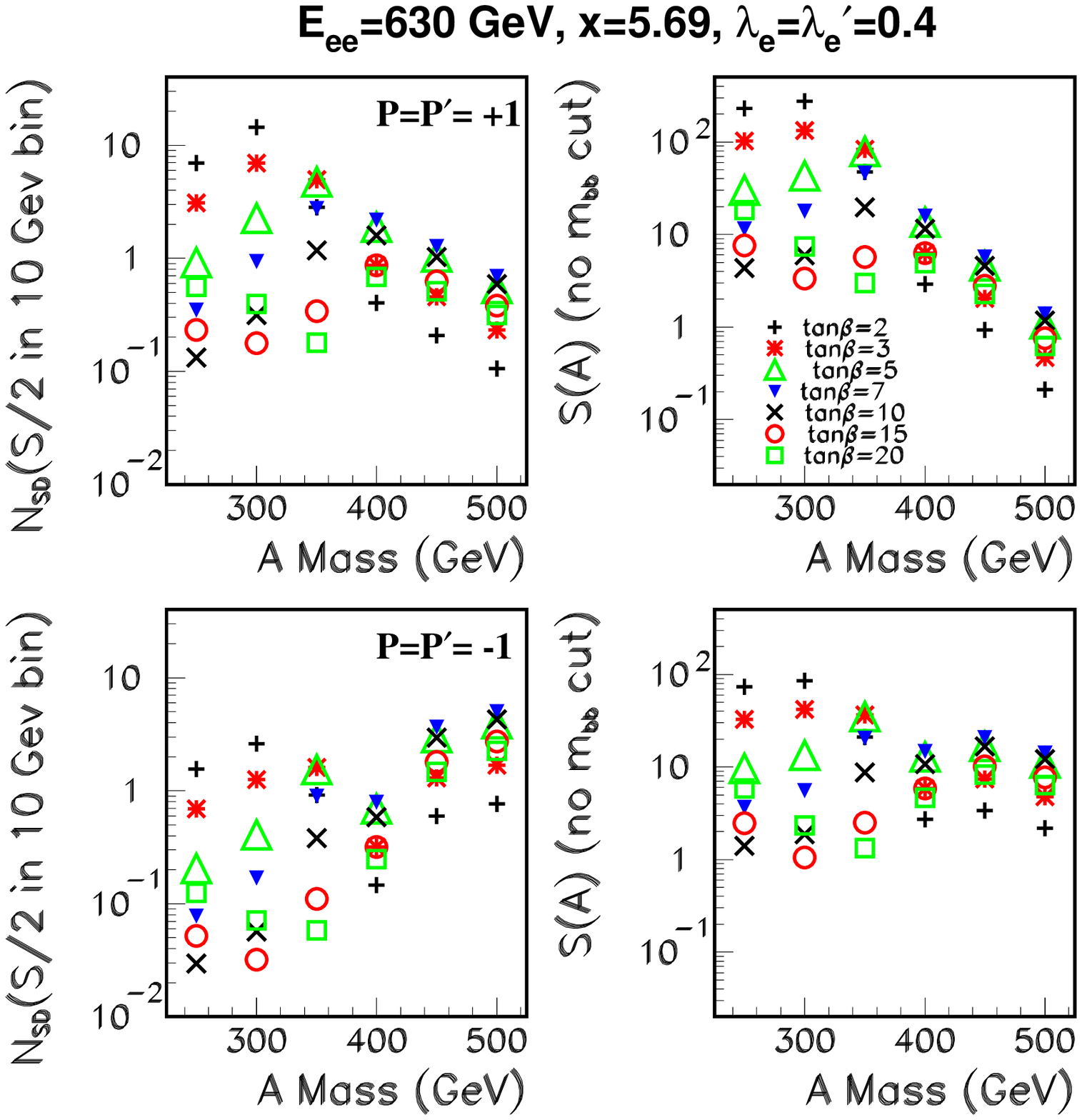}
\end{center}
\vspace*{-.2in}
\caption[0]{We consider a general 2HDM in which only the $\ha$
is light enough to be produced (all other Higgs bosons are taken
to have mass = 1 TeV).  In the right-hand window we plot 
the total $\ha$ signal rate after all cuts and efficiencies
for a variety of $\tanb$ and $\mha$ values, assuming $\rts=630\gev$
for the $\epem$ (or $\emem$) collisions and after accumulating
luminosities equivalent to one $10^7$ sec year of operation (each) using
the type-I broad $E_{\gam\gam}$ spectrum  
and the type-II peaked spectrum operation.
In the left-hand window we give the corresponding statistical significance 
of the signal $N_{SD}$ ($N_{SD}$ stands for 
the number of standard deviations)
for each of the sample $[\tanb,\mha]$ values
assuming that 50\% of the total signal rate falls within a 10 GeV
bin centered on the given $\mha$.
}
\label{a2hdm_analysis}
\end{figure}

\begin{figure}[t!]
\leavevmode
\begin{center}
\epsfxsize=7in\epsffile{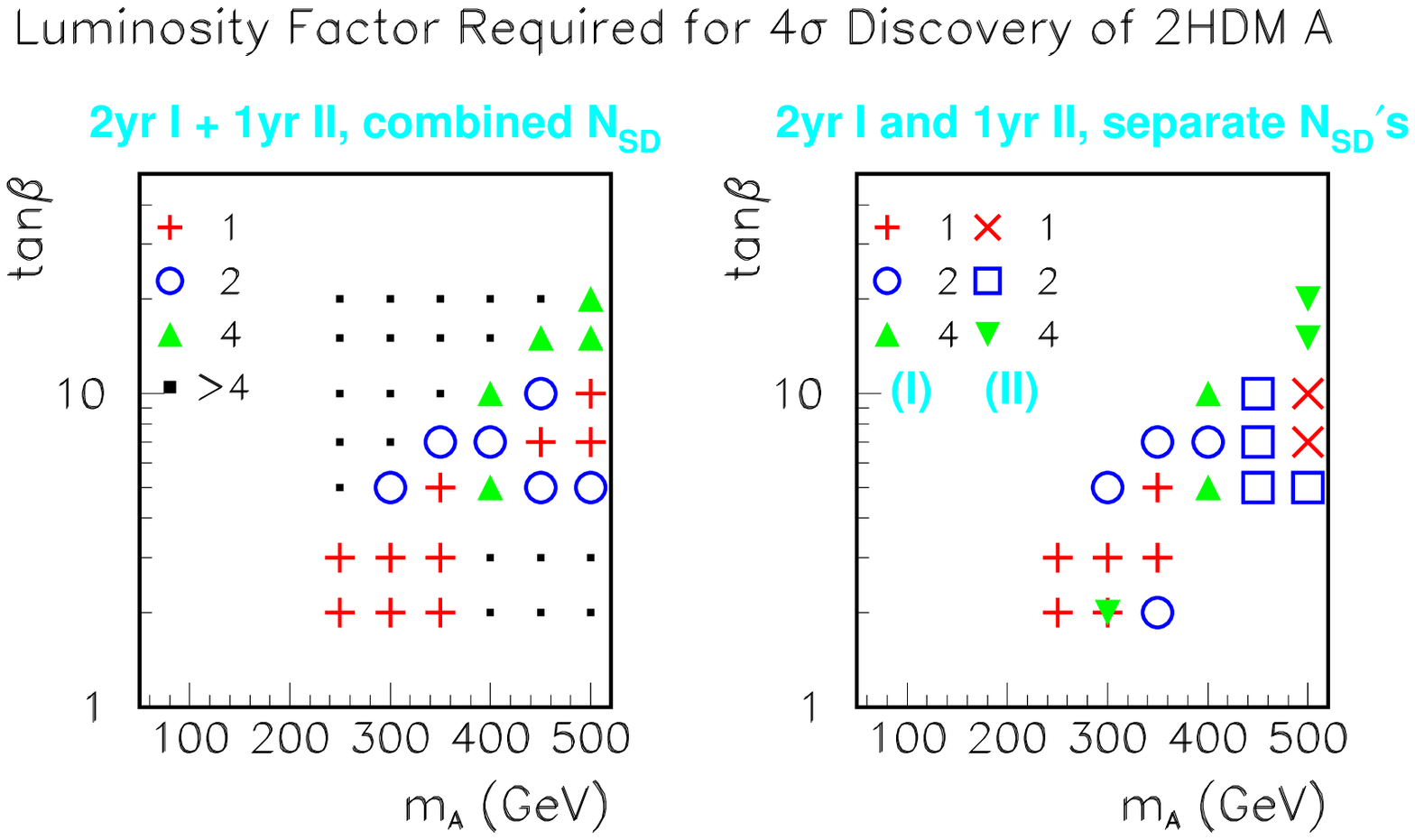}
\end{center}
\vspace*{-3.7in}
\caption[0]{Assuming a machine energy of $\rts=630\gev$,
we show the $[\mha,\tanb]$ points for which two $10^7$ sec years
of operation 
using the type-I $P\lam_e,P'\lam_e'>0$ polarization configuration
and one $10^7$ sec year of operation using the type-II $P\lam_e,P'\lam_e'<0$
configuration will yield $S/\sqrt{B}\geq4$ for
the $\ha$ of a general 2HDM, assuming all other 2HDM
Higgs bosons have mass of $1\tev$.  In the left-hand window we
have combined results from the type-I and type-II running using
$S/\sqrt B=\sqrt{ S_I^2/B_I+S_{II}^2/B_{II}}$. In the right-hand
window we show the separate results for $S_I/\sqrt{B_I}$ and 
$S_{II}/\sqrt{B_{II}}$. Also shown are the additional points for
which a $4\sigma$ signal level is achieved if the total
luminosity is doubled or quadrupled (the `2' and `4' symbol cases)
relative to the 2+1-year luminosities we are employing.
(In the left-hand window, the small black squares indicate the additional
points sampled for which even a luminosity increase of a factor
of 4 does not yield a $4\sigma$ signal.) 
Such luminosity
increases could be achieved for some combination of longer running time and/or
improved technical designs. For example, the factor of `2' 
results probably roughly apply to TESLA.   
Cuts and procedures are as described in the text.
}
\label{wedgeplota2hdm}
\end{figure}

Next, as in the MSSM case, we consider $\rts=630\gev$ and employ the CAIN
luminosity spectrum. Efficiencies and cuts are the same as in
the MSSM study. Assuming one year of $10^7$ sec operation (each) using
type-I (broad spectrum) and type-II (peaked spectrum), we give   
results for the total signal rate after all cuts and efficiencies
in Fig.~\ref{a2hdm_analysis}.  The corresponding statistical
significances, $N_{SD}=S/\sqrt B$, are also shown. 
In Fig.~\ref{wedgeplota2hdm}, we display those $[\mha,\tanb]$ points
for which two years of operation in type-I mode and one year
of operation in type-II mode would allow $4\sigma$ level
discovery of the $\ha$. (The additional points for which a $4\sigma$
signal would be achieved for 2 and 4 times as much luminosity
for both type-I and type-II operation are also displayed.)
We find that a reasonable fraction of the points in the wedge would
allow $\ha$ detection after 3 years of $\gam\gam$ collisions.
A $4\sigma$ signal is found for 10/42 of the
42 sampled points that might fall into the wedge in which
the $\ha$ would not be discovered by other means. For a factor
two higher integrated luminosity (e.g. after 6 years of operation
at the nominal luminosity predicted by CAIN for the current
design), this fraction would increase to 16/42.

Of course, one could also consider the 1-year 350 (type-II) plus
1-year 400 (type-II) plus 1-year 500 (type-II) running option,
which would provide somewhat improved signals for $\mha=350\gev$
and $\mha=400\gev$ than does the 2-year 500 (type-I)
plus 1-year 500 (type-II) option considered above.
However, the LHC/LC wedge in which the $\ha$ cannot be discovered
is quite large and certainly extends to $\mha$ values as low
as $200-250\gev$ to which only the latter option provides some
sensitivity (at lower $\tanb$).  Regardless of the running
option chosen, $\gam\gam$
collisions provide an important addition to our ability to
detect the $\ha$ of a general 2HDM in the scenario where
the other Higgs bosons are substantially heavier.

\section{Determining the CP nature of a Higgs boson}

Once one or several Higgs bosons have been detected, 
precision studies using the
peaked spectrum (II) with $\rts=m_{\rm Higgs}/y_{\rm peak}$ can be
performed. These include: determination of CP properties; a detailed scan to
separate the $\hh$ and $\ha$ when in the decoupling limit of a 2HDM; and
branching ratios, those for supersymmetric final states
being especially important in the MSSM 
context \cite{Gunion:1995bh,Gunion:1997cc,Gunion:1996qd,Feng:1997xv,Muhlleitner:2001kw,Muhlleitner:2000jj}.
By combining the $\gam\gam$ production cross
sections with the branching ratios, important information about
$\tanb$ and the masses of supersymmetric particles and their 
Higgs couplings could be extracted and be used to determine
much about the nature of soft supersymmetry breaking.

Determination of the CP properties of
any spin-0 Higgs $\widehat h$ produced in $\gam\gam$
collisions is possible since $\gam\gam\to {\widehat h}$ must proceed at one
loop, whether ${\widehat h}$ is CP-even, CP-odd or a mixture.
As a result, the CP-even and CP-odd parts of ${\widehat h}$ have
$\gam\gam$ couplings
of similar size.   However, the structure of the couplings is very different:
\beq
\cala_{CP=+}\propto \vec \eps_1\cdot\vec \eps_2\,,\quad
\cala_{CP=-}\propto (\vec\eps_1\times\vec \eps_2)\cdot \hat p_{\rm beam}\,.
\eeq
By adjusting the orientation of the photon polarization vectors with
respect to one another, it is possible to determine the relative
amounts of CP-even and CP-odd content in the resonance ${\widehat h}$
\cite{Grzadkowski:1992sa}. 
If ${\widehat h}$ is a mixture, one can use helicity asymmetries for this purpose
 \cite{Grzadkowski:1992sa,Kramer:1994jn}.
However, if ${\widehat h}$ is either purely CP-even
or purely CP-odd, then one must employ transverse linear polarizations
\cite{Gunion:1994wy,Kramer:1994jn}. 

For a Higgs boson of pure CP, one finds that the Higgs cross section 
is proportional to
\beq
{d\call\over dE_{\gam\gam}}\left(1+\vev{\lam\lam'}+{\cal CP} \vev{\lam_T\lam_T'}
\cos2\delta \right)
\label{tranxsec}
\eeq
where ${\cal CP}=+1$ (${\cal CP}=-1$) 
for a pure CP-even (CP-odd) Higgs boson and
and $\delta$ is the angle between the transverse polarizations of
the laser photons. Thus, one measure of  
the CP nature of a Higgs is the asymmetry
for parallel vs. perpendicular orientation 
of the transverse linear polarizations of the initial laser beams.
In the absence of background, this would take the form
\beq
\cala\equiv{N_{\parallel}-N_{\perp}\over N_{\parallel}+N_{\perp}}
={{\cal CP}\vev{\lam_T\lam_T'}\over 1+\vev{\lam\lam'}} \,,
\label{asymzerob}
\eeq
which is positive (negative) for a CP-even (odd) state. 
The $b\anti b(g)$ and $c\anti c(g)$ backgrounds result 
in additional contributions 
to the  $N_{\parallel}+N_{\perp}$ denominator, which dilutes the
asymmetry. The backgrounds do not contribute to the numerator
for CP invariant cuts. Since, as described below, total
linear polarization for the laser beams translates into
only partial polarization for the back-scattered photons
which collide to form the Higgs boson, both $N_{\parallel}$
and $N_{\perp}$ will be non-zero for the signal.
The expected value of $\cala$ must be carefully computed for a given model
and given cuts.

\begin{figure}[p]
\centerline{\psfig{file=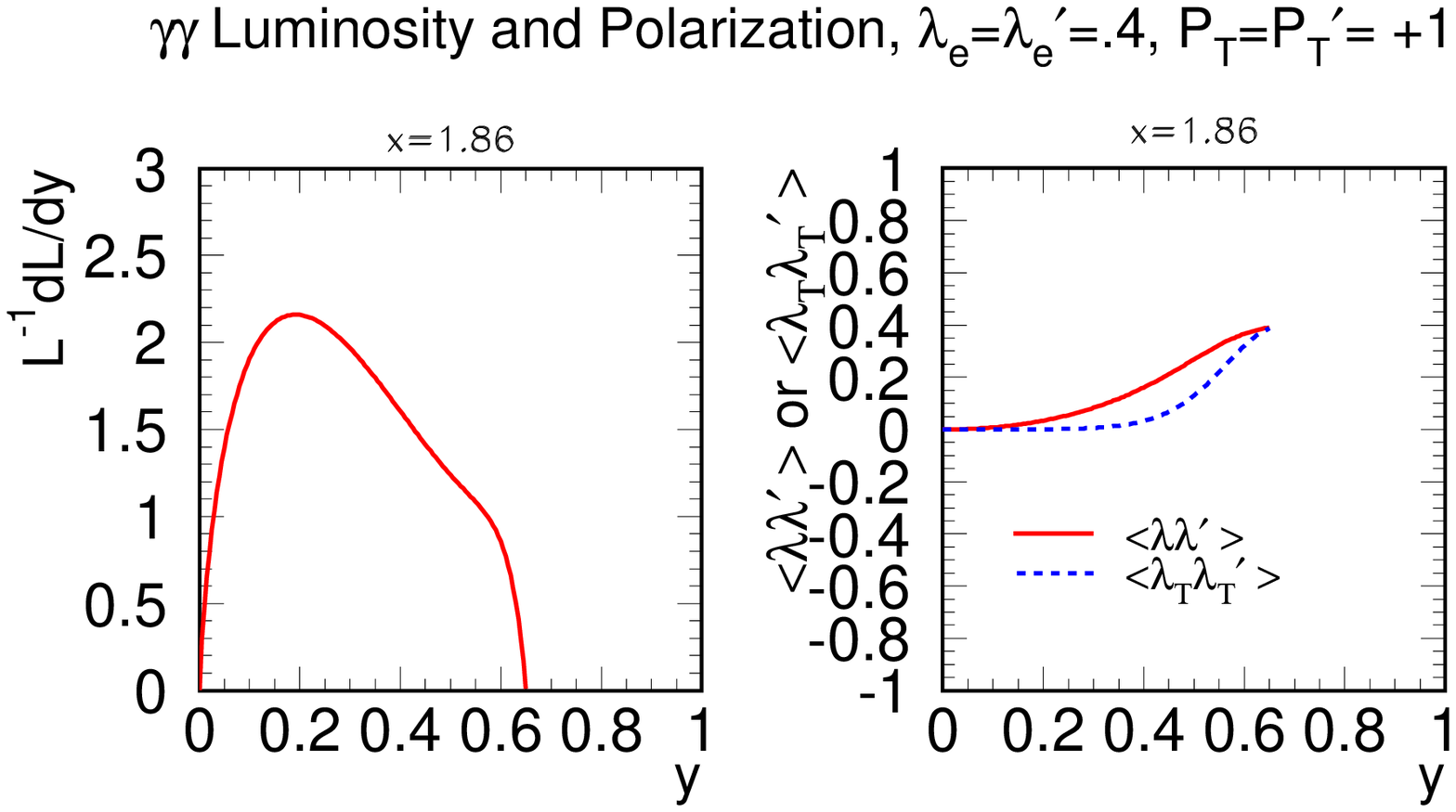,width=14cm}}
\vspace*{-2.7in}
\caption[0]{We plot the luminosities and corresponding $\vev{\lam\lam'}$
and $\vev{\lam_T\lam_T^\prime}$
for operation at $\rts=206\gev$ and $x=1.86$,
assuming 100\% transverse polarization for the laser photons
and $\lam_e=\lam_e^\prime=0.4$. These plots are for the naive
non-CAIN distributions.
}
  \label{linlumnaive}
\centerline{\psfig{file=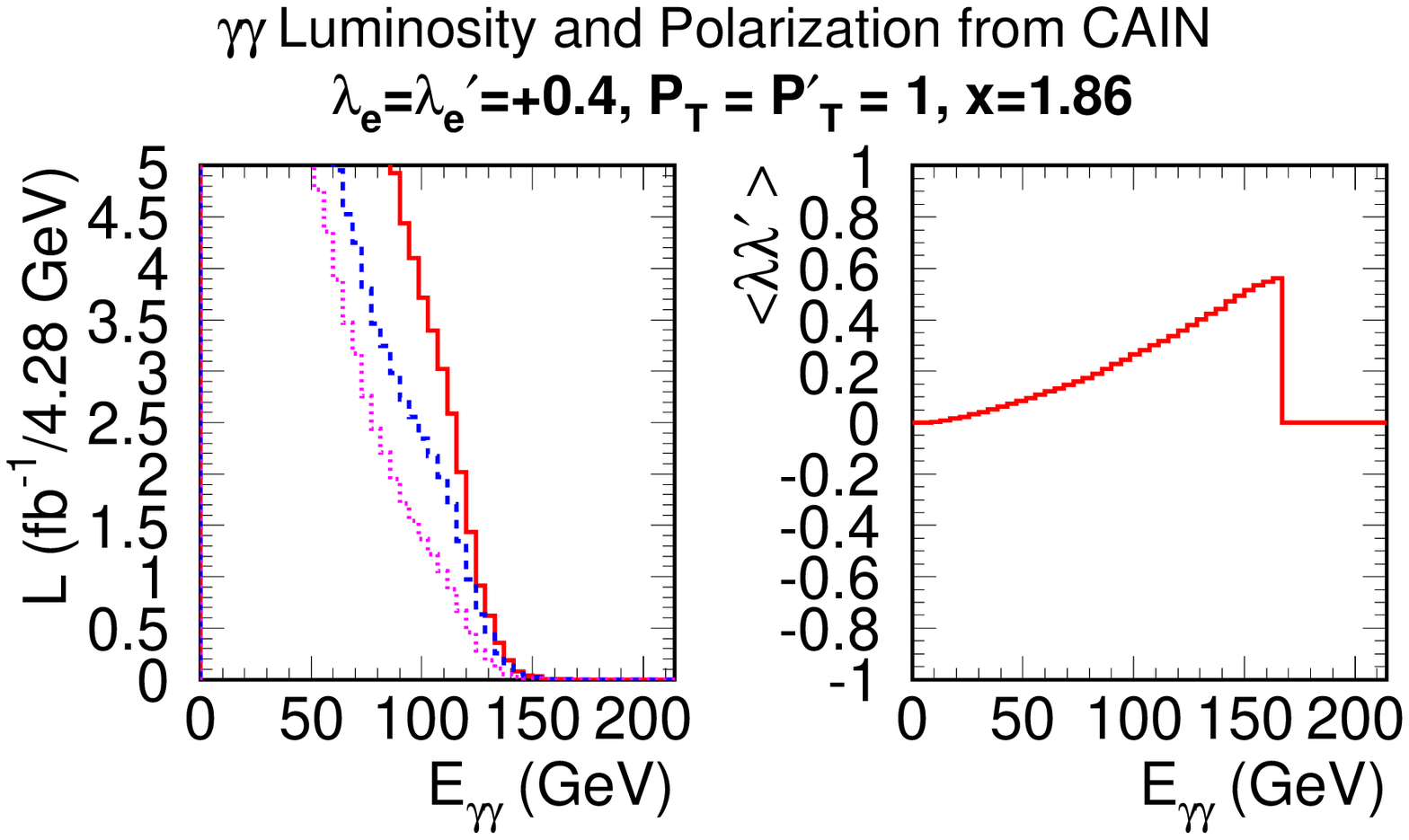,width=14cm}}
\vspace*{-2.7in}
\caption[0]{We plot the luminosity,
$L=d\call/dE_{\gam\gam}$, in units of $\fbi/4.28 \gev$ 
and corresponding $\vev{\lam\lam'}$ predicted by CAIN
for operation at $\rts=206\gev$ and $x=1.86$,
assuming 100\% transverse polarization for the laser photons
and $\lam_e=\lam_e^\prime=0.4$.
 The dashed (dotted) curve gives the component
of the total luminosity that derives from the $J_z=0$ ($J_z=2$) two-photon
configuration. The solid luminosity curve is the sum of these two
components and $\vev{\lam\lam'}=(L_{J_z=0}-L_{J_z=2})/(L_{J_z=0}+L_{J_z=2})$.
}
  \label{linlum}
\end{figure}

Using the naive analytic forms for back-scattered
photon luminosities and polarizations, one
finds that for 100\% transverse polarization of
the laser photon, the transverse polarization of the back-scattered photon
\footnote{Our $\lam_T$ is the same as $\xi_3$ --- see \cite{Gunion:1994wy} ---
for laser photon orientation such that $\xi_1=0$.
Recall that the longitudinal polarization in this same notation is
given by the Stoke's parameter $\xi_2$.}
is given by the electron-polarization-independent form
\beq
\lam_T={2r^2\over (1-z)^{-1}+(1-z)-4r(1-r)}\,,
\label{ptform}
\eeq
where $\lam_T$ is the appropriate Stoke's parameter and $r=zx^{-1}/(1-z)$ 
with $z=E_{\gam}/E_{e^-}$. The maximum of $\lam_T$, 
\beq
\lam_T^{\rm max}=2(1+x)/[1+(1+x)^2],
\eeq 
occurs at the kinematic limit,
$z_{\rm max}=x/(1+x)$ (i.e. $r=1$). 
This can be compared to the analytic form for the longitudinal polarization:
\beq
\lam={2\lam_e r x[1+(1-z)(2r-1)^2] \over (1-z)^{-1}+(1-z)-4r(1-r)}\,.
\eeq
At the kinematic limit, $z=z_{\rm max}=x/(1+x)$, 
the ratio of $\lam$ to $\lam_T$ is given by
\beq
{\lam\over \lam_T}= \lam_e x{2+x\over 1+x}\sim 1
\eeq
for $\lam_e=0.4$ and $x=1.86$.
 Substantial luminosity and values of $\lam_T$ close to the
maximum are achieved for moderately smaller $z$. From (\ref{ptform}),
operation at $x=1.86$ (corresponding to $\rts=206\gev$ and laser
wave length of $\lam\sim 1~\mu$) would allow 
$\lam_T^{\rm max}\sim\lam^{\rm max}\sim 0.6$.
Making these choices for both beams
is very nearly optimal for the CP study for the
following reasons. First, these choices will maximize 
${d\call\over dE_{\gam\gam}}\vev{\lam_T\lam_T'}$ 
at ${E_{\gam\gam}=120\gev}$. As seen
from earlier equations, it is
the square root of the former quantity that
essentially determines the accuracy with which
the CP determination can be made. 
Second, $\lam_e=\lam_e'=0.4$ results in $\vev{\lam\lam'}>0$. This is 
desirable for suppressing the background. (If there
were no background, Eq.~(\ref{asymzerob}) implies
that the optimal choice would be to employ
$\lam_e$ and $\lam_e'$ such that $\vev{\lam\lam'}<0$.
However, in practice the background is very substantial and it
is very important to have $\vev{\lam\lam'}>0$ to suppress
it as much as possible.) 
In \Fig{linlumnaive}, we plot  
the naive luminosity distribution  
and associated values of $\vev{\lam\lam'}$ and $\vev{\lam_T\lam_T'}$
obtained for $\lam_e=\lam_e'=0.4$ and 100\% transverse polarization
for the laser beams.

As discussed in \cite{Gunion:1994wy}, the
asymmetry studies discussed below are not very sensitive to
the polarization of the colliding $e$ beams.
Thus, the studies could be performed in parasitic fashion during
$e^-e^+$ operation if the $e^+$ polarization is small. (As emphasized
earlier, substantial $e^+$ polarization would be needed for precision
studies of other $\hsm$ properties.)

\begin{figure}[h!]
\centerline{\psfig{file=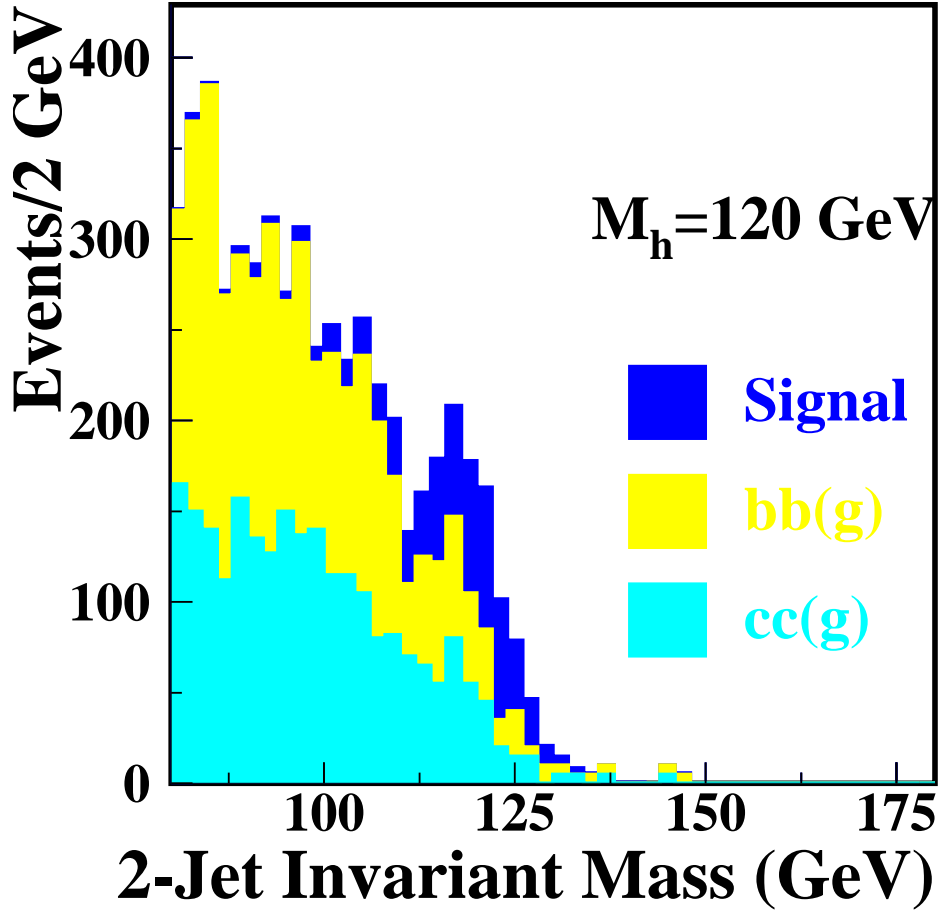,width=14cm}}
\caption[0]{We plot the signal and $b\anti b$ and $c\anti c$
backgrounds for a SM Higgs boson with $\mhsm=120\gev$
assuming $\gam\gam$ operation at $\rts=206\gev$ and $x=1.86$,
based on the luminosity and polarization distributions of Fig.~\ref{linlum}
for the case of linearly polarized laser photons.
The cross sections presented are those for $\delta=\pi/4$, \ie\
in the absence of any contribution from the transverse polarization
term in Eq.~(\ref{tranxsec}).
}
  \label{linsignal}
\end{figure}

The luminosity distribution predicted by the CAIN
Monte Carlo for transversely polarized laser photons
and the corresponding result for $\vev{\lam\lam'}$ are plotted in
\Fig{linlum}. We note that
even though the luminosity spectrum is not peaked, it is very nearly the same 
at $E_{\gam\gam}=120\gev$ as in the circular polarization case.
As expected from our earlier discussion
of the naive luminosity distribution,
at $E_{\gam\gam}=120\gev$ we find  $\vev{\lam\lam'}\sim 
\vev{\lam_T\lam_T'}\sim 0.36$. Since CAIN includes multiple interactions
and non-linear Compton processes, the luminosity is actually
non-zero for $E_{\gam\gam}$ values above the naive kinematic limit of
$\sim 132\gev$.  Both $\vev{\lam\lam'}$ and $\vev{\lam_T\lam_T'}$
continue to increase as one enters this region.  However, the luminosity
becomes so small that we cannot make effective use of this region
for this study. 
We employ these luminosity and polarization results 
in the vicinity of $E_{\gam\gam}=120\gev$
in a full Monte Carlo for Higgs production and decay as outlined
earlier in the circular polarization case. All the same cuts
and procedures are employed.  

The resulting
signal and background rates for $\delta=\pi/4$
are presented in \Fig{linsignal}.
The width of the Higgs resonance peak is $5.0\pm 0.3\gev$ (using a Gaussian
fit), only slightly larger than in the circularly polarized case.
However, because of the shape of the luminosity distribution,
the backgrounds rise more rapidly for $m_{b\anti b}$
values below $120\gev$ than in the case of circularly polarized laser beams. 
Thus, it is best to use a slightly higher cut on the $m_{b\anti b}$
values in order to obtain the best statistical significance for the signal.
We find $\sim 360$
reconstructed two-jet signal events with $m_{b\anti b}\geq 114\gev$
in one year of operation, with roughly 440 background events
in this same region. This corresponds to a precision
of $\sqrt{S+B}/S\sim 0.079$ for the measurement
of $\Gamma(\hsm\to\gam\gam)\br(\hsm\to b\anti b)$.
Not surprisingly, this is not as good as for the circularly polarized
setup, but it is still indicative of a very strong Higgs signal.
Turning to the CP determination, let us assume that 
we run 1/2 year in the parallel
polarization configuration and 1/2 year in the perpendicular
polarization configuration.
Then, because we have only 60\% linear
polarization for the colliding photons for $E_{\gam\gam} \sim 120\gev$, 
$N_{\parallel}\sim 180[1+(0.6)^2]+273\sim 518$ and
$N_{\perp}\sim 180[1-(0.6)^2]+273=388$.
For these numbers, $\cala=130/906\sim 0.14$.
The error in $\cala$
is $\delta\cala=\sqrt{N_{\parallel}N_{\perp}/N^3}\sim 0.016$
($N\equiv N_\parallel+N_\perp$), yielding 
${\delta\cala\over\cala}={\delta {\cal CP}\over {\cal CP}}\sim 0.11$.
This measurement would thus provide a 
fairly strong confirmation of the CP=+ nature of the $\hsm$
after one $10^7$ sec year devoted to this study.

\section{Conclusions}

In this paper, we have explored the various ways in which
a $\gam\gam$ collider could contribute to our understanding
of Higgs physics. We have confined our study to the $b\anti b$
final state. We have shown the following.
\bit
\item For a SM-like Higgs boson, it will
be possible to determine $\Gamma(\gam\gam\to \h)\br(\h\to b\anti b)$
with excellent precision, \eg\ $\sim 2.9\%$ accuracy for $\mh\sim 120\gev$.
This accuracy will be achieved after just one $10^7$ sec year
of operation, using the frequency tripler technology
and a peaked $E_{\gam\gam}$ spectrum is the most optimal.

By using the excellent $\sim 1\%-2\%$ measurement of $\br(\h\to b\anti b)$,
one can extract a $\sim 2.9\%$ measurement for $\Gamma(\h\to\gam\gam)$.
As discussed in the introduction, deviations of this width
from its SM expectations could be very revealing. In particular,
at this level of accuracy, deviations that might be present
as the result of the SM-like Higgs boson being part of a larger Higgs
sector, such as that of the MSSM, would typically be visible if some
of the other Higgs bosons were not too much heavier than $500\gev$ or so.
In the MSSM context, the precise magnitude of the deviations 
might thus allow extraction of the crucial mass scale $\mha$.
If $\mha$ is known with sufficient accuracy, one would
know more or less exactly what $\rts$ to employ so that 
detection of $\gam\gam\to\ha,\hh$ at the $\gam\gam$ collider would
be straightforward and would become a high priority.

\item Even if there is no predetermination of $\mha$,
detection of $\gam\gam\to\hh,\ha$ is still likely to be possible
for a large fraction of the problematical `wedge' of moderate-$\tanb$
parameter space, described earlier, for which the $\hh,\ha$
will not be observable either at the LHC or at a LC.
For instance, for a LC of $\rts=630\gev$, the wedge begins 
at $\mha\gsim 300\gev$ (the approximate upper reach of the $\epem\to\hh\ha$
pair production process) whereas the $\gam\gam$ collider can
potentially allow detection of the $\hh,\ha$ up to 
the $E_{\gam\gam}$ spectrum limit of about 500 GeV. Indeed,
using just $b\anti b$ final states, our results show that
$\hh,\ha$ detection will be possible in 
somewhat more than 65\% of the wedge
after two ($10^7$ sec) years of operation using a broad spectrum
and one year of operation using a peaked spectrum.
By also considering
$\hh\to \hl\hl$, $\ha\to Z\hl$ and $\hh,\ha\to t\anti t$ final states,
we estimate that somewhat more than 85\% of the wedge parameter
region with $\mha\lsim 500\gev$ 
would provide a detectable signal after a total of 
two to three years of operation.

Thus, by combining $\rts=630\gev$ $\gam\gam$ collider operation
with LC studies of $\epem$ collisions and LHC searches for the MSSM
Higgs bosons, we would have an excellent chance of finding
all the neutral Higgs
bosons of the MSSM Higgs sector (if they have mass $\lsim 500\gev$), 
whereas without the $\gam\gam$
collider one would detect only the $\hl$ (at both the LC and LHC)
in the problematical parameter space wedge. In short, if we detect
supersymmetric particles at the LHC and LC consistent with the MSSM structure
and find only the $\hl$ at the LC and LHC, $\gam\gam$ operation
focusing on Higgs discovery will be a high priority.

\item
The one caveat to this very optimistic set of conclusions 
regarding the $\hh,\ha$ is that
if SUSY particles are light (masses $\lsim \mha/2$), they will alter the
predictions for the $\hh,\ha\to\gam\gam$ couplings and diminish
the $\hh,\ha\to b\anti b$ branching ratios. If these effects
are very strong, as possible at lower $\tanb$, 
detection of the $\hh,\ha$ in the $b\anti b$
channel could become 
significantly more difficult, both in $\gam\gam$ collisions
and at the LHC --- SUSY decay channels would need
to be employed.  However, at the larger $\tanb$ values
in the wedge region under consideration,  
the $b\anti b$ coupling is strongly enhanced and it is unlikely
that these effects would be sufficiently large to significantly
alter our conclusions.

\item
It is important to note that 
the $\gam\gam\to\hh,\ha\to b\anti b$ rate has a minimum
at $\tanb\sim 15$ ($\tanb\sim 20$) for $\mha\lsim 300\gev$ 
($\mha\geq 350\gev$), \ie\ $\tanb$ values that are just large enough
to be above the wedge region at higher $\mha$.
Thus, the  $\gam\gam\to\hh,\ha\to b\anti b$ 
rate increases for still higher $\tanb$
(roughly linearly for $\tanb$ in the $30-50$ range).
Consequently, if the $\hh,\ha$
are discovered at the LHC because $\tanb$ is large,
and yet other physics considerations
force $\gam\gam$ operation at maximal $\rts$ (rather than 
at the $\rts$ such that $E_{\rm peak}\sim \mha$) there
is a good possibility that the 
$\gam\gam\to \hh,\ha\to b\anti b$ signal will be quite substantial
(if one chooses the appropriate, type-I or type-II, spectrum for the $\mha$
value found at the LHC).  This would then
provide an opportunity for a relatively precise measurement
of the very interesting $\gam\gam\to\hh,\ha$ couplings
that will not be accessible by any other means. This in turn
could lead to significant information about other SUSY parameters.
In particular, as illustrated in the main part of the paper, 
$\tanb$ can be determined with reasonable accuracy
from the $\gam\gam\to\hh,\ha\to b\anti b$ rate
if the masses and properties
of the SUSY particles are known from LHC and/or LC data.
Most notably, the larger $\tanb$ is, the more
accurate will be this determination. In contrast, most
other techniques for determining $\tanb$ (\eg\ from neutralino, chargino, gluino, \etc\ cross sections and branching ratios) become increasingly
insensitive to $\tanb$ as $\tanb$ increases.

\item After three ($10^7$ sec) years of operation
(2 with type-I spectrum and 1 with type-II spectrum), 
it will be possible to detect the $\ha$ of 
a general two-Higgs-double model (in particular, one with parameters
such that all other Higgs bosons are heavy, including the SM-like
neutral Higgs) over a substantial portion of the parameter
space in which it cannot be detected in any other LC or LHC modes.
\item Determination of  the CP nature of any Higgs boson that can be
observed will be possible in $\gam\gam$ collisions by employing
transversely (linearly) polarized laser beam photons.  In particular,
we studied the case of a light SM-like Higgs boson with $\mh=120\gev$,
and showed that the error in determination of its ${\cal CP}=+1$
would be $\delta {\cal CP}/{\cal CP}\sim 0.11$.
\eit
For these various purposes, there is no question that maximizing
the luminosity will be very important.  
In the case of the NLC design we consider, the results stated above
would require 1 $10^7$ sec year of operation at low $\rts$ for the light
Higgs precision study, 1 year of operation at low $\rts$
in the linearly polarized mode for the ${\cal CP}$ study, and
3 years of operation for the $\hh,\ha$ search (one 
in the peaked spectrum mode and two in the broad spectrum mode
if one is constrained to run at the maximal $\rts=630\gev$
assumed in our study).
The extra factor of 2 in luminosity that might be achievable
at TESLA would prove an advantage. Further optimization of the NLC
design might also be possible and is strongly encouraged.
For instance, going to a round beam configuration keeping the CP-IP
separation at 1 mm might yield as much as a factor of two increase
in luminosity.

We should note that our studies have only included hadronic backgrounds
due to direct (QED) processes and have not yet incorporated backgrounds
resulting from the hadronic structure of the photon. The photon can
``resolve'' into quarks or gluons plus spectator jets. Hadronic production
could then occur through $\gam\gam^*$ (1-resolved) or $\gam^*\gam^*$ 
(2-resolved) processes. Resolved photon backgrounds have two contributions
to the background to Higgs production.
The first is that 
in which a quark or gluon `constituent' of one of the back-scattered
photons is responsible for initiating a two-body scattering
process that creates a pair of high-$p_T$ $b$ or $c$ jets.
(As discussed, for example, in Ref.~\cite{Baillargeon:1994af}
good $b$-tagging efficiency and purity, as employed here,
is required in order
to eliminate other resolved photon two-jet backgrounds, such as $gb$
or $gc$ final states.)
However, it is generally the case that
such contributions to the background are 
numerically unimportant unless the Higgs mass is far below the
maximum $E_{\gam\gam}$. This was first concluded in 
Ref.~\cite{Baillargeon:1994af} and more recently confirmed
in Ref.~\cite{jikia}. 
In the $\mh=120\gev$ cases we study, the Higgs mass is quite close to
the maximum $\gam\gam$ energy, and in the $\hh,\ha$ studies
the Higgs mass is at least 50\% of the maximum $\gam\gam$ energy.
For such choices, this kind of resolved photon background is not important.
In addition, any residual resolved photon background of this type could
be further reduced by vetoing events
in which there is an extra ``remnant'' jet in the forward 
and/or backward region --- such jets would tend to have transverse
and longitudinal 
momentum of order $\third\mh$ to $\half\mh$ for the configurations
we employ and would, therefore, be readily visible in the detector.

The second type of background from resolved photon processes arises 
when a resolved photon scattering process underlies
the primary Higgs production reaction.  These events arise when
back-scattered photons other than those involved in the Higgs
production reaction also interact.  
This can happen either using back scattered photons arising in the
same bunch crossing or photons from two different bunch crossings
within the same detector readout interval. Cross sections (before cuts)
for producing relatively soft jets deriving from 
these resolved photon processes are several orders of magnitude larger than
the corresponding direct $\gam\gam \to X$ cross section.
Thus, such additional scatterings primarily yield additional low-$p_T$ jets
that would underlie the $b\anti b$ jets arising from Higgs production.
They would thus make it less efficient to isolate the true 2-$b$-jet signal
using cuts that require exactly two reconstructed jets which are rather
precisely back-to-back. Mass resolution could also deteriorate,
as might the efficiency for $b$-tagging.
The level of this background
is determined by the number of back-scattered
photons created in each bunch crossing as well
as the number of bunch crossings over
which the detector integrates.  At TESLA, the bunch spacing
is 337 ns and it might be possible to design
the detector so that there would be only one crossing
per detector readout.  In this case,
only the underlying $\gam\gam$ interactions from this
single crossing would need to be considered.
For the NLC parameters considered here, 
the bunch spacing is only 2.8 ns (as desirable for $\gam\gam$ operation
in order to maximize the bunch charge for the same total current).
In general, the detector will integrate over a number of bunch crossings
and it will therefore be desirable to minimize this number. 
This may turn out to be an important factor in determining the
NLC detector design. On the other hand, 
although it may only be necessary to integrate
over one bunch crossing at TESLA, the bunch charge
will be roughly 30\% higher and there will be more back-scattered
photons (that can give rise to underlying $\gam\gam$
interactions) per crossing than for the NLC design.
Thus, a detailed examination of this background is required
in both the TESLA and NLC cases.
In particular, the performance of the $b$-tagging and energy flow
algorithms will be critical and will depend upon the 
occupancies in the vertex detector and calorimeter, respectively.   
Overall, our ability to reconstruct the (two-jet) component
of the Higgs resonance in the presence of underlying
soft jet structure from resolved photon interactions
is critically dependent upon detector
design features. Absent the required studies in the context
of a detailed detector design, we cannot currently determine
whether the resulting resolved photon backgrounds will be a problem
at either machine or which machine will yield the smallest resolved
photon background. 

We should note that our results have assumed 80\% polarization
for both the $e$ beams used to back-scatter the laser photons.
Only the ${\cal CP}$ studies would remain little altered if
one of the beams does not have substantial polarization.
Because of substantially increased background levels,
comparable results for the other studies/searches would require
significantly more integrated luminosity if only one beam
has large polarization. As a result, if one is to be able
to perform these $\gam\gam$ studies parasitically during normal
$\epem$ operation of the LC, 
substantial $e^+$ polarization will be very important.
Another issue related to simultaneously studying $\epem$
collisions and $\gam\gam$ interactions is the bunch spacing. If the design
$1.4$~ns bunch spacing for $\epem$ is employed, then our luminosities
will be decreased by about 40\%.

\bigskip
\centerline{\bf Acknowledgments}
\bigskip
We would like to thank M. Battaglia, T. Hill, M. Spira,
V. Telnov, M. Velasco and P. Zerwas
for useful discussions.

\clearpage
\section{Appendix A}

In this Appendix, we give the machine and beam parameters that we have
assumed in computing $\gam\gam$ luminosities (using 
the CAIN Monte Carlo) for the various running options
considered in this paper.  These parameters are presented
in Table~\ref{params}.
\begin{table}[h]
\begin{center}
\caption{Parameters for the various beam energy and polarization
options considered in this paper.}
\bigskip
\begin{tabular}[c]{|c|c|c|c|c|}
\hline
Energy (GeV) & 80 & 103 & 267.5 & 315\\
\hline
$\beta_x/\beta_y$~(mm) & 1.4/0.08  & 1.5/0.08 & 4/0.065  & 4/0.08 \\
$\epsilon_x/\epsilon_y~(\times 10^{-8})$ & 360/7.1 & 360/7.1 & 360/7.1 & 360/7.1 \\
$\sigma_x/\sigma_y ~({\rm nm})$ & 179/6.0 & 0 164/5.3 & 166/3.0 & 153/3.0 \\
$\sigma_z$~(microns) & 156 &  156 & 156 & 156 \\
$ N~(\times 10^{10})$ & 1.5 & 1.5 & 1.5 & 1.5 \\
$ e^-~{\rm Polarization}~ (\%)$ & 80 & 80 & 80 & 80\\
repetition rate (Hz) & 120$\times$95 & 120$\times$95 & 120$\times$95 & 120$\times$95 \\
Laser $\lambda$ (microns) & 0.351 & 1.054 & 1.054 & 1.054 \\
 CP-IP distance (mm) & 1 & 1 & 1 & 1 \\
\hline
\end{tabular}
\label{params}
\end{center}
\end{table}

We note that our designs have emphasized fairly flat beams 
which would be most appropriate if the $\gam\gam$ collider
interaction region is running parasitically at the same time
as the main interaction region is exploring
$e^-e^-$ collisions.

\section{Appendix B: Higher-order corrections}

Papers that have considered higher order QCD corrections 
to Higgs production and the background cross sections, and that have examined
implications for Higgs detection include 
\cite{Borden:1994fv,Jikia:1996bi,Melles:1999xd,Melles:1998rp,Muhlleitner:2001kw,Muhlleitner:2000jj,Jikia:2000rk}.
Some of the corrections found in these papers are large under certain
circumstance.  The purpose of this appendix is to explain why
these corrections are relatively small for the cuts 
and the colliding photon luminosities and polarizations (predicted
by CAIN) employed and
to demonstrate that it is much more important to have as accurate
a simulation as possible in a realistic experimental approach.

Let us discuss the background cross sections first.  
The tree-level $J_z=0$ and $J_z=\pm 2$ cross sections are given
in Eqs. (\ref{jz0bkgnd}) and (\ref{jz2bkgnd}). The $m_q^2/s$ suppression
of the $J_z=0$ background implies that radiative corrections to
this component of the cross section can be large.  
The exact magnitude of these corrections depends critically
upon the laser beam configuration (in particular, circular or linear)
and the precise cuts employed. The radiative corrections
are dramatically reduced by employing cuts that isolate
two-jet final states.  In the case of circularly
polarized laser beams, if two-jet final states
are isolated by employing $y_{\rm cut}=0.02$ (the first of the
two-jet cuts we use) 
the $J_z=0$ background can still be increased relative to the tree-level
expectation by up to a factor of 10~\cite{srprivcom}.
This factor will be reduced by the additional back-to-back
cut that we employ (which also discriminates against the radiation
of an additional gluon at the partonic level), but still might
be large. In contrast, radiative corrections to the $J_z=\pm2$
background cross sections are relatively small 
($\lsim 10\%$ typically \cite{srprivcom}).
For linearly polarized laser beams, the higher-order corrections
to the two-jet final states are quite modest in size~\cite{Jikia:2000rk}.  
Thus, the most important
question is whether or not we need to worry about the large
corrections to the $J_z=0$ background in the case of circularly
polarized laser beams. 

 \begin{figure}[th]
\centerline{\psfig{file=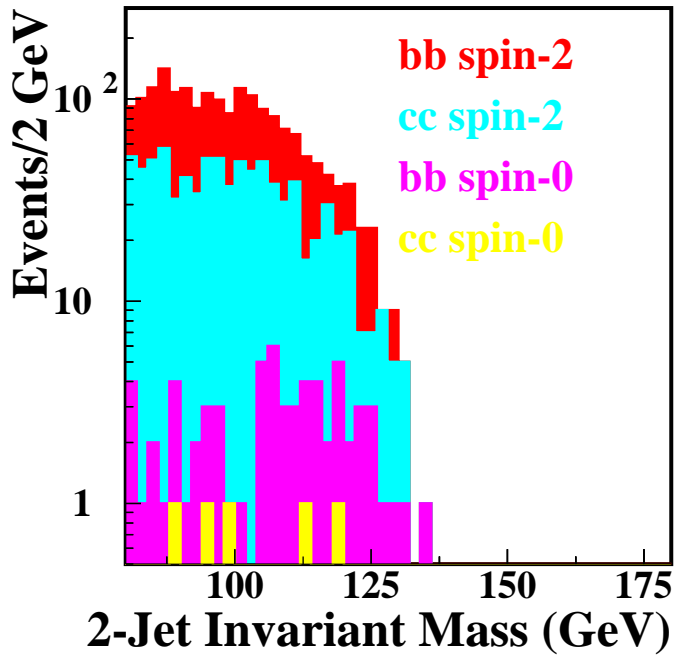,width=10cm}}
\caption[0]{For $\rts=160\gev$ and frequency tripler
operation, we plot, as a function of 2-jet mass (in GeV),
the $J_z=0$ and $J_z=2$ backgrounds (events per $10^7$ sec
year per bin) for $c\anti c (g)$
and $b\anti b (g)$ production as obtained from PYTHIA (modified
to incorporate correct tree-level $J_z=0$ and $J_z=2$
cross sections) with processing via JETSET and ROOT 
after all cuts, most especially including
cuts imposed to isolate only the 2 jet final state. Colliding photon 
luminosities and polarizations employed are those
predicted by the CAIN Monte Carlo assuming 100\% circular
polarization ($P_c=P_c'=-1$) for the laser beams and 
80\% polarization ($\lam_e=\lam_e'=0.4$) for the electron beams.}
  \label{nlc80bkgd}
\end{figure}

In Fig.~\ref{nlc80bkgd},
we plot the tree-level predictions for the $J_z=0$ and $J_z=\pm 2$
$c\anti c (g)$ and
$b\anti b(g)$ backgrounds obtained by running PYTHIA/JETSET and processing
using ROOT to impose the 2-jet final state cuts delineated
in the main body of the paper. The sum of the $J_z=0$ and $J_z=\pm2$
backgrounds plotted in this figure gives the net backgrounds displayed
in Fig.~\ref{fig:higgs} for the SM Higgs boson with mass of 120 GeV. 
These background levels include the
expected luminosity from CAIN in the $J_z=0$ and $J_z=\pm 2$
initial two-photon configurations for 80\% electron
beam polarization. What is immediately apparent
is that the background is overwhelmingly dominated by the $J_z=\pm 2$
backgrounds.  From Fig.~\ref{nlc80bkgd} we
see that even if the QCD corrections to the $J_z=0$ backgrounds 
were as large as a factor of 10, this would affect the total
background at a level no greater than 20\%.
This conclusion differs from that of previous works largely due to
to the fact that the value of $\vev{\lam\lam'}$ obtained
in CAIN and assuming the fairly realistic
80\% polarizations ($\lam_e=\lam_e'=0.4$) is not sufficiently close
to unity to more than partially suppress the $J_z=\pm 2$ background.

 \begin{figure}[th]
\centerline{\psfig{file=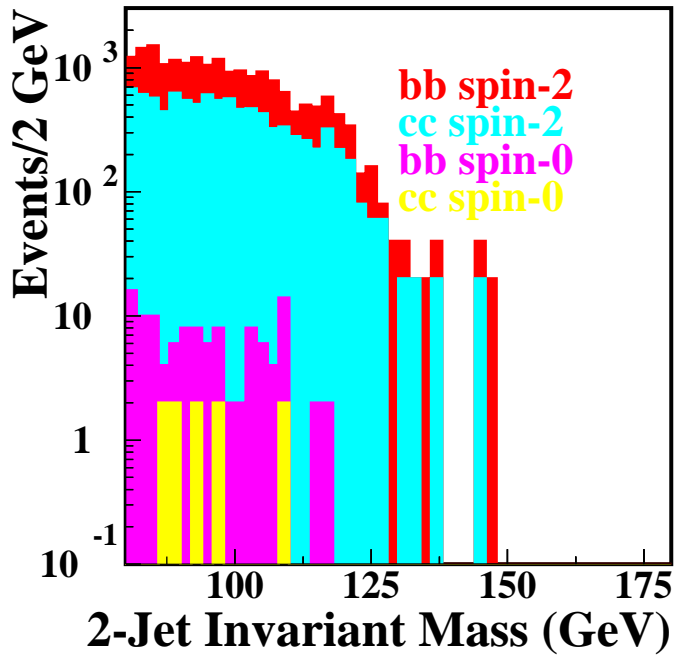,width=10cm}}
\caption[0]{For $\rts=206\gev$ and without frequency tripler
operation, we plot, as a function of 2-jet mass (in GeV),
the $J_z=0$ and $J_z=2$ backgrounds (events per $10^7$ sec
year per bin) for $c\anti c (g)$
and $b\anti b (g)$ production as obtained from PYTHIA (modified
to incorporate correct tree-level $J_z=0$ and $J_z=2$
cross sections) with processing via JETSET and ROOT 
after all cuts, most especially including
cuts imposed to isolate only the 2 jet final state. Colliding photon 
luminosities and polarizations employed are those
predicted by the CAIN Monte Carlo assuming 100\% linear
polarization for the laser beams and 
80\% polarization ($\lam_e=\lam_e'=0.4$) for the electron beams.}
  \label{nlc103bkgd}
\end{figure}

In Fig.~\ref{nlc103bkgd}, 
we plot the tree-level predictions for the $J_z=0$ and $J_z=\pm 2$
$c\anti c (g)$ and
$b\anti b(g)$ backgrounds in the case of linearly polarized
laser beams as employed in constructing Fig.~\ref{linsignal}.
As above, these were obtained by running PYTHIA/JETSET and processing
using ROOT to impose the 2-jet final state cuts delineated
in the main body of the paper. The sum of the $J_z=0$ and $J_z=\pm2$
backgrounds plotted in this figure gives the net backgrounds displayed
in Fig.~\ref{linsignal} for the SM Higgs boson with mass of 120 GeV. 
These background levels include the
expected luminosity from CAIN in the $J_z=0$ and $J_z=\pm 2$
initial two-photon configurations for 80\% electron
beam polarization. As for the case of circularly polarized laser beams,
the background is overwhelmingly dominated by the $J_z=\pm 2$
backgrounds.  QCD NLO and resummation
corrections to the $J_z=0$ two-jet cross section
in the case of linearly polarized laser beams are expected
to be much more modest in size than in the case of circularly polarized
beams~\cite{Jikia:2000rk}. But, even if these corrections were to increase 
the $J_z=0$ backgrounds by
as much as a factor of 10, the background level would be only
of order $3\%$ larger than that we have employed.

 \begin{figure}[h!]
\centerline{\psfig{file=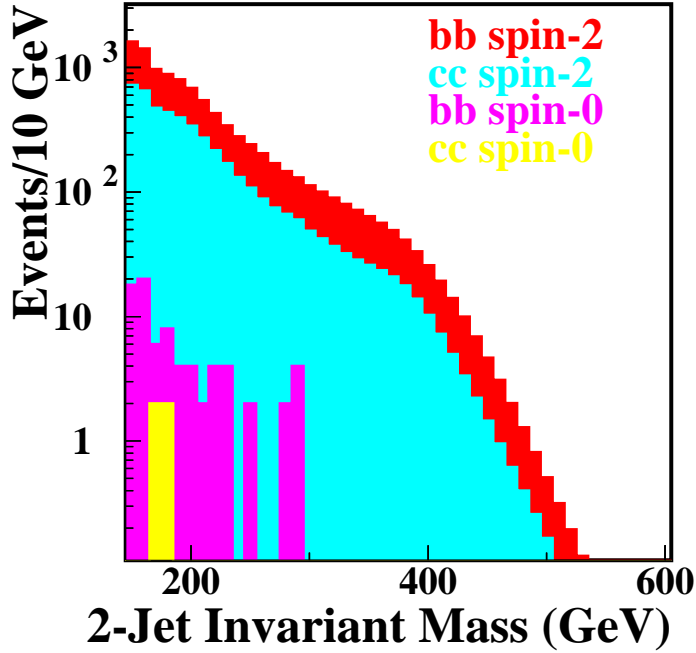,width=10cm}}
\vspace*{-.15in}
\caption[0]{For $\rts=630\gev$ and type-I (broad spectrum)
polarization configuration 
operation, we plot, as a function of 2-jet mass (in GeV),
the $J_z=0$ and $J_z=2$ backgrounds (events per $10^7$ sec
year per bin) for $c\anti c (g)$
and $b\anti b (g)$ production as obtained from PYTHIA/JETSET/ROOT
after all cuts. Colliding photon
luminosities and polarizations employed are those
predicted by the CAIN Monte Carlo assuming 100\% circular
polarization ($P_c=P_c'=+1$) for the laser beams and 
80\% polarization ($\lam_e=\lam_e'=0.4$) for the electron beams.}
  \label{i630}
\end{figure}

\begin{figure}[h!]
\centerline{\psfig{file=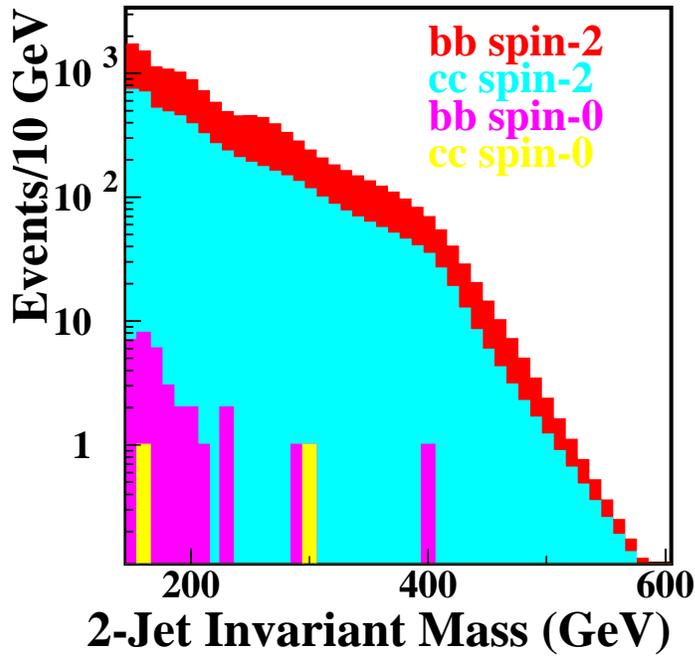,width=10cm}}
\vspace*{-.15in}
\caption[0]{As in Fig.~\ref{i630} except for type-II (peaked spectrum)
polarization configuration ($P_c=P_c'=-1$, $\lam_e=\lam_e'=0.4$).
}
  \label{ii630}
\end{figure}

Figs.~\ref{i630} and \ref{ii630} give the $J_z=0$ and $J_z=\pm2$
backgrounds incorporated in the $\rts=630\gev$ Figs.~\ref{signalploti}
and \ref{signalplotii} in the cases of type-I (broad spectrum)
and type-II (peaked spectrum) operation, respectively;
these are the results after all cuts, including the 2-jet cuts.
We see that even a factor of 10
QCD correction to the $J_z=0$ portion
of the background would result in at most a 10\% correction to the total
background. 
Let us compare this situation to \cite{Muhlleitner:2001kw} (see
\cite{Muhlleitner:2000jj} for details). There, the background 
is dominated by the $J_z=0$ contribution and QCD corrections
are essential for obtaining an appropriate background estimate.  
Although our signal cross section in the `without SUSY'
case is very close in value to that is
plotted in Fig.~2 of \cite{Muhlleitner:2001kw} (if
we convert our $I_\sigma$ to the cross section definition 
implicit in their figure), their background 
(obtained, we believe, assuming 100\% electron
polarization, $\lam_e=\lam_e'=0.5$ --- see comment below
Eq.~(3.61) in association with Fig.~3.12 of \cite{Muhlleitner:2000jj}) 
is much smaller than
their signal.  This is in sharp contrast to the background
level we obtain in the CAIN simulation with 80\% polarization 
($\lam_e=\lam_e'=0.4$), which background is quite comparable to our 
typical signal.

Another theoretical issue concerns the suppression factors
associated with imposing 2-jet cuts on the signal.
Our approach has been to generate the signal at tree-level
but in the context of PANDORA/PYTHIA/JETSET which allows for the generation
of extra jets via final state radiation.
The imposition of 2-jet cuts will then give rise to a 
suppression factor as computed in the context of PYTHIA/JETSET, which
factor is expected to be quite similar to that obtained 
from the analytic approaches but will also take into account experimental
issues related to jet definition, detector resolutions  and so forth.
In this regard, it is useful to compare to
Ref.~\cite{jikia}, which follows an approach very similar to ours.
Their Fig.~1 shows that before cuts about 75\%
of the Higgs events have more than 2 jets (using $y_{\rm cut}=0.02$).
For the same Higgs mass, we obtain almost exactly the same result.
Further, we find that this same percentage applies also for Higgs
masses in the $300-500\gev$ range.  A corresponding
result is not given in \cite{jikia} after imposing
their cuts. In our case, after cuts, especially the 
back-to-back and $|\cos\theta^*|<0.5$ cuts,
we find that roughly 90\% (95\%) of the events are 2-jet
for Higgs masses of $120\gev$ ($400\gev$). 

The final theoretical issue that requires discussion is the possible
importance of interference between signal and background. Here,
we refer to several discussions in \cite{Muhlleitner:2000jj}.  
First, as shown in their Eq.~(3.22), we note that the interference
cross section only involves the $J_z=0$ part of the background.
Since the Higgs bosons being considered are essentially produced
on-shell, and since after cuts 
the $J_z=0$ backgrounds are reduced to a level much smaller than 
the signal cross section, such interference will be small.
For example, \cite{Muhlleitner:2000jj} (see below Eq.~(3.64)) 
finds that the interference cross section 
is typically of order 1/100 to 1/1000 of the signal
cross section after imposing cuts similar those we consider.

\bigskip
\centerline{\bf Grant Support}
\bigskip

This work was supported in part by the U.S. Department of Energy
contract No. DE-FG03-91ER40674 and under the auspices of the U.S. Department of
Energy by the University of California, Lawrence Livermore National
Laboratory under Contract No.W-7405-Eng.48.

\clearpage


\end{document}